\documentclass[twocolumn,tighten,times]{aastex63}

\hypersetup{colorlinks=true,linkcolor=blue,citecolor=blue,urlcolor=blue,}
\usepackage{verbatim} % TODO: REMOVE PACKAGE
\usepackage{color,soul} % allows highlighting, remove when finished

\usepackage{lineno}
\shorttitle{\sc {\MgII} Absorber Evolution}
\shortauthors{\sc Churchill \etal}

\input{iondefs.sty}

\newcommand{\NX}{{\cal X}_{W_i}}
\newcommand{\NXz}{{\cal X}_{W_i}(z)}

\begin{document}

\title{13 Billion Years of {\MgII} Absorber Evolution}

\author[0000-0002-9125-8159]{Christopher W. Churchill}
\affiliation{Department of Astronomy, New Mexico State University, Las Cruces, NM 88003, USA}

\author[0009-0006-4626-832X]{Asif Abbas}
\affiliation{Department of Astronomy, New Mexico State University, Las Cruces, NM 88003, USA}

\author[0000-0003-1362-9302]{Glenn G. Kacprzak}
\affiliation{Centre for Astrophysics and Supercomputing, Swinburne University of Technology, Hawthorn, Victoria 3122, Australia}
%\affiliation{ARC Centre of Excellence for All Sky Astrophysics in 3 Dimensions (ASTRO 3D), Australia}

\author[0000-0003-2377-8352]{Nikole M. Nielsen}
\affiliation{Homer L. Dodge Department of Physics and Astronomy, The University of Oklahoma, 440 W. Brooks St., Norman, OK 73019, USA}
\affiliation{Centre for Astrophysics and Supercomputing, Swinburne University of Technology, Hawthorn, Victoria 3122, Australia}

%
% there is a 250 word count for the abstract; this version is 250 words!
% so if you add a word, you must remove a word! haha
%
\begin{abstract}
Applying ``apportioned integrals," we use $d{\cal N}/dX$ measurements to determine the {\MgII} absorber equivalent width distribution function for $W_r \geq 0.03$~{\AA} and $0 \leq z \leq 7$.  Adopting a Schechter distribution, ${f(z,W)dW \!=\! \Phi^*(W/W^*)^{\alpha} e^{-W/W^*} dW/W^*}$, we present the normalization, $\Phi^*(z)$, the characteristic equivalent width, $W^*(z)$, and the weak-end slope, $\alpha(z)$, as smooth functions of redshift. Measurements of $d{\cal N}/dX$ are robust for $z<4$ but less so at $z>4$ for weaker absorbers ($W_r \leq 0.3$~{\AA}). We bracketed two data-driven scenarios: from $z\sim7$ to $z\sim4$, $dN/dX$ of weak absorbers is (1) constant, or (2) decreasing. For scenario \#1, the evolution of $\Phi^*(z)$, $W^*(z)$, and $\alpha(z)$ show that in the post-reionization universe, weak systems are nonevolving while the incidence of the strongest systems increases until Cosmic Noon; following Cosmic Noon, the strongest absorbers slowly evolve away while the incidence of weak absorbers rapidly increases. For scenario \#2, the parameter evolution is such that, in the post-reionization universe, weak systems evolve away while the incidence of the strongest systems increases until Cosmic Noon; following Cosmic Noon, the behavior tracks the same as scenario \#1. We argue in favor of scenario \#2 based on corroborating {\OI}, {\CII}, and {\SiII} measurements at $z>4$.  Our results provide a unified, quantitative description for {\MgII} absorber evolution spanning 13 billion years of cosmic time and offer deeper insights into  galactic baryon cycle physics. They also highlight the need for deep $z>5$ {\MgII} surveys and have implications for detectability of a {\MgII} forest at $z>7$.
\end{abstract}

%the strongest absorbers slowly evolve away while the incidence of the weakest absorbers rapidly increases. 
%$\Phi^*(z)$ decreases while $W^*(z)$ increases and $\alpha(z)$ steepens, all monotonically;  For scenario (2), $\Phi(z)$ maximizes and $\alpha(z)$ is flattest at $z\sim4$. From $z\sim7$ to $z\sim4$, $\Phi(z)$ and $W^*(z)$ increases while $\alpha(z)$ becomes shallower and then for $z\sim4$ to $z\sim0$ $\Phi^*(z)$ decreases, $W^*(z)$ increases, and $\alpha(z)$ steepens.

\keywords{Galaxy evolution (594), Quasar absorption line spectroscopy (1317), Circumgalactic medium (1879)}
%\facility{Keck:I (HIRES)}
%\facility{VLT: (UVES)}

\section{Introduction}
\label{sec:intro}

% MgII  (>0.02) (>1.0)
% OI > 0.05
% CIV  (>0.05) (>0.6)

By quantitatively characterizing the global properties of various populations of quasar absorption lines systems, we gain insights into the astrophysics governing the evolution of the universe.  When we focus on absorption from a given ion, for example, {\OI}, {\MgII}, {\SiIV}, {\CIV}, or {\OVI}, we can chart and compare the cosmic mass densities of the individual ions as a function of cosmic time \citep[][]{bergeron05, shull12, noterdaeme12, dodorico13, dodorico22, boksy15, rao17, becker19, davies23, sebastian23, abbas24} and, in the case of {\HI} and {\HeII} absorbers, measure the thermal and ionization history of the universe \citep[][]{carilli04, fechner07, mcquinn09, becker13, syphers14, worseck19, bosman21, hu22, fan23}.  We can also measure the build up of metals and dust and chart the chemical enrichment history of the universe \citep[e.g.,][]{wolfe95, jtl1996, ferrera00, menard08, rafelski12, fumagalli16, lehner16, lehner22, peroux20, roman-duval22, davies23}.

It required several decades of surveys focused on distinct absorber populations to build our first census and characterization of the gaseous universe. These early surveys included the population of neutral {\HI} absorbers \citep[e.g.,][]{sargent80, hu95, lu96, weymann98}, and the abundant lithium-like low-ionization {\MgII} absorbers \citep[][]{lanzetta87, ssb88, caulet89, pb90, ss92, churchill99}, intermediate-ionization {\CIV} absorbers \citep[][]{sbs88, pb94, rauch96, cooksey13}, and, following the launch of the {\it Hubble Space Telescope\/} ({\it HST}), high-ionization {\OVI} absorbers \citep[][]{burles96, danforth08, tripp08, danforth16}. 

As high-resolution optical quasar spectra accumulated in the archives, surveys of {\MgII} and {\CIV} absorbers could be pushed to much higher sensitivities, allowing the gas kinematics to be studied and the properties of weaker absorbers to be characterized \citep[e.g.,][]{narayanan05, evans11, boksy15, mathes17, hasan20}. With the advent of sensitive infrared spectrographs on large telescopes, the redshifts over which these populations of absorbers can be surveyed have effectively been extended to the epoch of the first quasars, pushing the current frontier out to $z \!\sim\! 7$ \citep[e.g.,][]{matejek12, dodorico13, dodorico22, chen.s17, codor17,  christiansen23, davies23, sebastian23}. This has allowed metal-line absorption populations to be studied well past the Cosmic Noon period ($2 \!\leq\! z\!\leq\! 3$), when galaxy evolutionary processes, such as star formation, stellar driven winds, and accretion from the intergalactic medium (IGM), were at their most active \cite[e.g.,][]{forster20} and into the epoch of reionization ($z\! > \! 5.3$), when ionizing photons from the first galaxies transformed intergalactic space from a neutral medium into a highly ionized plasma \citep[e.g.,][]{quinn16}.

Since the first unbiased surveys, key characterizations of various absorber populations have included their redshift path density, $d{\cal N}/dz$, and their equivalent width and column density distribution functions. The quantity $d{\cal N}/dz$ is the number of absorbers per unit of redshift path surveyed. The equivalent width distribution is quantified as $f(W) \!=\! d^2\!{\cal N}/dzdW$, the number of absorbers per unit redshift path per unit of equivalent width, $W$.  Similarly, the column density distribution is quantified as $f(N) \!=\! d^2\!{\cal N}/dzdN$, the number of absorbers per unit redshift path per unit column density.  The column density distribution is more difficult to measure, as it requires that the absorption lines are resolved in high-resolution spectra and modeled using techniques such as Voigt profile decomposition \citep[e.g.,][]{carswell14}.  In comparison, measuring the equivalent width distribution is straightforward because $W$ is resolution invariant and can be measured by directly summing pixel flux decrements \citep[e.g.,][]{lanzetta87, schneider93}.  
However, in quasar spectra, it is well known that spurious blends from absorption lines not associated with the line being measured can result in added uncertainty in the measure equivalent width.

Even if a population of absorbers does not evolve with cosmic time, its $d{\cal N}/dz$, $f(W)$, and $f(N)$ will change with redshift as a function of the expansion of the universe. To remove redshift dependence, \citet{bahcall-peebles69} formulated a quantity known as the ``absorption path'', $X(z)$, defined such that $dX/dz \!=\! (1\!+\!z)^2 / \{ \Omega_m(1\!+\!z)^3 \!+\! \Omega_\Lambda \}^{1/2}$, where $\Omega_m$ and $\Omega_\Lambda$ are the present-epoch energy densities of matter and dark energy. The absorption path accounts for the radial and transverse components of cosmic  expansion, such that a non-evolving population of absorbers has a constant absorption path density, $d{\cal N}/dX \!=\! (c/H_0) n_0 \sigma_0$, where $n_0$ is the mean cosmic spatial number density of the absorber population and $\sigma_0$ is the mean cross section of the absorbing gas structure.  Hereafter, we will refer to the absorption path as the co-moving line-of-sight path. 

In terms of the co-moving line-of-sight path, the equivalent width distribution, $f(W) \!=\! d^2{\cal N}/dXdW$, and the column density distribution, $f(N) \!=\! d^2{\cal N}/dXdN$, will remain constant with redshift if the absorber population is not evolving. The equivalent width distribution is related to $d{\cal N}/dX$ via 
\begin{equation}
(d{\cal N}/dX)_z 
= \frac{c}{H_0}n_0(z)\sigma(z)
= \int _{0}^{\infty}  \!\!\! f(z,W) \, dW  \, ,
\label{eq:whatallthefussisabout}
\end{equation}
where we have added the notation to account for possible redshift evolution in the absorber population. Through this relation, we see that, at a given redshift, the co-moving line-of-sight path density is the area under $f(z,W)$. A similar relationship can be written for the column density distribution. As such, we see that $d{\cal N}/dX$ measures the product of the cosmic number density, $n_0(z)$, and cross section, $\sigma(z)$, at a given redshift, whereas $f(z,W)$ provides the distribution of the gas-structure absorption strengths.  The absorption strength depends on the gas metallicity, the ionization, the kinematics, the physical size, and the gas density, so $f(z,W)$ places global constraints on the convolved distribution of all these physical properties of the absorbing structures as a function of redshift. 

The equivalent width distribution of {\MgII} absorbers has probably been investigated over a larger redshift range and more thoroughly than any other metal-line absorber population \citep[e.g.,][]{bergeron84, lanzetta87, tytler87, ssb88, caulet89, pb90, ss92, churchill99, nestor05, nestor06, prochter06, narayanan07, ggk-cwc11, matejek12, seyffert13, zhu13, raghunathan2016, chen.s17, mathes17, bosman17, codor17, christiansen23, sebastian23}. This population of absorber is of particular interest because (1) it is a well-established tracer of the baryon cycle of galaxies \citep[e.g.,][]{bb91, sdp94, chen10-mgii, nielsen13-magiicat2, peroux20}, and (2) the cosmic incidence, $(d{\cal N}/dX)_z$, of the strongest absorbers (i.e., those with rest-frame equivalent widths $W_r \!\geq\! 1.0$~{\AA}) evolves in direct step with the global galactic star formation of the universe \citep{zhu13, matejek12, chen.s17}.  Even with a substantial body of work and focus on the evolution of {\MgII} absorbers, no unified comprehensive formulation of the functional form and redshift evolution of the {\MgII} equivalent width distribution has been specified. In view of recent and more sensitive measurements covering $z\!>\!2$ \citep[the cosmic period between reionization and Cosmic Noon, e.g.,][]{bosman17, codor17, christiansen23, davies23, sebastian23}, such a synthesis, or a first step toward such a synthesis, may now be at hand for redshifts up $z\sim 7$.

In this paper, we develop and apply a method to constrain the functional form and redshift evolution of the {\MgII} absorber equivalent width distribution function directly from the measured $(d{\cal N}/dX)_z$, as expressed in Eq.~\ref{eq:whatallthefussisabout}.  In Section~\ref{sec:ahistory}, we provide the historical development of {\MgII} observations and findings.  In Section~\ref{sec:currentpic}, we summarize the current measurements and draw inferences about how they inform us about the evolution of {\MgII} absorbers.  The formalism of our method is presented in Section~\ref{sec:modeling} and, in Section~\ref{sec:results}, we test and implement our method and present our constraints on the evolution of $f(z,W_r)$. In Section~\ref{sec:discussion}, we discuss our findings and, in Section~\ref{sec:conclude}, we provide our conclusions.  When a cosmological model is invoked, we adopted the cosmological parameters $\Omega_m = 0.31$, $\Omega_\Lambda = 0.69$, and $H_0 = 67.7$~{\kms}~Mpc$^{-1}$ \citep{planck18}.

\section{The Magnesium Chronicles}
\label{sec:ahistory}

First-generation studies of quasar spectra discovered that absorption lines from fine-structure {\MgIIdblt} doublets are common and that it was likely they arose in gas structures intervening to the background quasars  \citep[e.g.,][]{kinman67, carswell75, burbidge77}. 
Over the next quarter century, blind and unbiased surveys yielded deeper insights into the characteristic properties of {\MgII} absorbers \citep[e.g.,][]{bergeron84, lanzetta87, tytler87, ssb88, caulet89, pb90, ss92}. The survey of \citet{ss92} observed 103 quasars covering the redshift range $0.3 \!\leq\! z \!\leq\! 2.2$ over which {\MgII} absorbers could be detected to a minimum detection threshold of $W_r \!=\! 0.3$~{\AA}. They fitted the equivalent width distribution to both a power-law, of the form 
\begin{equation}
f(W)dW =\Phi^* W^\alpha dW \, , 
\end{equation}
and an exponential function of the form,
\begin{equation}
f(W) \, dW = \Phi^* \, e ^{-(W/W^\ast)}  dW/W^\ast
\, ,
\label{eq:exponentialfunction}
\end{equation}
where $\alpha$ is the power-law index, $W^*$ is the characteristic equivalent width, and $\Phi^*$ is the normalization. 
%They obtained $\alpha \simeq -1.7$ for the power law, and $W^* \simeq 0.66$~{\AA} for the exponential distribution, but 
They could not confidently distinguish between the two functional forms. However, they reported that the median rest-frame equivalent width $W_r$ at $z \!\sim\! 1.8$ is $\sim\! 40$\% larger than at $z \!\sim\! 0.6$ with confidence level of 97\%. This finding, and the clear measurement that the redshift evolution of $d{\cal N}/dz$ became more rapid as the mean equivalent width of the sample was increased, strengthened the findings of \citet{lanzetta87} and \citet{caulet89} that the equivalent width distribution evolves with redshift such that fewer stronger absorbers are present at later cosmic times compared to $z \!\sim\! 2$.

% SS92 W*=0.66 <z>=1.12
% LTW87 W*=0.88 <z>=1.62
% Tytler87 alpha=-2.2
% LTW87 alpha=-1.99
% Caulet89
% Petijean & Bergeron 1990

Also consistent with the other surveys at the time, \citet{ss92} suggested there is a deficiency of weak {\MgII} absorption lines and estimated that at least $\sim\! 80$\% of all {\MgII} absorbers have  $W_r \!\geq\! 0.3$~{\AA}. There was a straightforward argument for this prediction. First, photoionization modeling of low-ionization gas predicted {\MgII} absorption would arise almost exclusively in structures that are optically thick to hydrogen ionizing photons.  These structures, known as Lyman Limit systems (LLSs), are characterized by having neutral hydrogen column densities of $N({\HI}) \!\geq\! 10^{17.2}$~cm$^{-2}$. Second, for $z \!<\! 2$ (the extent of surveys of that time), the cosmic redshift path density of {\MgII} absorbers with $W_r \!\geq\! 0.3$~{\AA} and of LLSs are nearly equivalent, meaning that, statistically, it could be assumed {\MgII} absorbers and LLSs were one-in-the same gas structures. Consequently, a sharp down turn in the {\MgII} equivalent width distribution below $W_r \!=\! 0.3$~{\AA} was predicted in deeper surveys sensitive to smaller equivalent widths. 

However, tentative signs that $W_r \!<\! 0.3$~{\AA} {\MgII} absorbers may dominate the population were reported by \citet{womble95} and \citet{tripp97}. Shortly thereafter, \citet{churchill99} surveyed 30 HIRES/Keck spectra for these weak {\MgII} absorbers in the redshift range $0.4 \!\leq\! z \!\leq\! 1.4$.  They showed that weak {\MgII} absorbers comprise $\sim\! 65$\% of the total {\MgII} absorber population and that their redshift path density was a factor of $\sim\! 4$ greater than that of LLSs.  This implies that $\sim\! 75$\% of weak {\MgII} absorbers must arise in sub-LLS environments, i.e., those with $N({\HI}) \!<\! 10^{17.2}$~cm$^{-2}$. They demonstrated that the equivalent width distribution continued to increase to small $W_r$, following a power law, $f(W) \!\sim\! W^{-1}$, with no indication of a turnover in the distribution down to  $W_r \!=\! 0.02$~{\AA}. 

Extending {\MgII} absorber studies to $z \leq 0.15$ using ultraviolet spectra,  \citet{churchill01} measured $d{\cal N}/dz$ for {\MgII} absorbers with $W_r \!>\! 0.3$~{\AA}.  Comparing this measurement to the $0.3 \!\leq\! z\leq 2.2$ measurements of \citet{ss92}, it was deduced that $f(W_r)$ for $W_r \!>\! 0.3$~{\AA} did not strongly evolve from $z \!\sim\! 0.3$ to $z \!\sim\! 0.05$; however, the results were not well constrained. \citet{narayanan05} conducted an ultraviolet survey for weak absorbers at $z \!\leq\! 0.3$ and, compared to the $0.4 \!\leq\! z \!\leq\! 1.4$ measurements of \citet{churchill99}, found no evidence for evolution in $d{\cal N}/dz$ nor the power-law slope of $f(W_r)$ for $W_r \!<\! 0.3$~{\AA} from $z \!\sim\! 0.9$ to the present epoch. 

Using Early Data Release (DR) optical spectra from the Sloan Digital Sky Survey (SDSS), \citet{nestor05} undertook a {\MgII} absorber survey covering $0.4 \!\leq\! z \!\leq\! 2.3$. They established that $f(W_r)$ follows an exponential function for $W_r \!>\! 0.3$~{\AA}, ruling out a power-law function for this equivalent width range. They also clearly demonstrated that the characteristic equivalent width, $W^*$, evolves. It was measured to decrease from $W^* \!\simeq\! 0.8$~{\AA} at $z \!\sim\! 1.5$ to $W^* \!\simeq\! 0.6$~{\AA} at $z \!\sim\! 0.5$. In a separate study, \citet{nestor06} confirmed the upturn in $f(W_r)$ for $W_r \!<\! 0.3$~{\AA} relative to an extrapolation of the exponential distribution fitted to $W_r \!>\! 0.3$~{\AA} absorbers. The upturn was consistent with the power-law fit of \citet{churchill99}. Using DR3 SDSS spectra, \citet{prochter06} examined $f(W_r)$ over $0.35 \!\leq\! z \!\leq\! 2.0$ for $W_r \!\geq\! 1.0$~{\AA}. They suggested that the function $f(W_r)dW \!=\! \Phi^* W^\alpha e^{-W} dW$ provides a qualitatively better description of the equivalent width distribution than does a power-law distribution.  Using DR5 SDSS spectra, \citet{lundgren09} measured $d{\cal N}/dz$ for multiple finite ranges of $W_r$, but did not quantitatively examine the shape of $f(W_r)$.

% Prochter d{\cal N}/dX is for Wr > 1
% Wr > 1.0 
% 1.0 < Wr < 1.4 
% Wr > 1.8 

Using HIRES/Keck and UVES/VLT spectra, \citet{narayanan07} extended the search for weak {\MgII} absorbers to redshift $z \!\sim\! 2.4$. They found that $d{\cal N}/dz$ increases by a factor of roughly $2.5$ from $z \!\sim\! 2.4$ to $z \!\sim\! 1.2$. This indicated fewer weak {\MgII} absorbers during Cosmic Noon than at later cosmic times. For $z \!\sim\! 0.9$, \citet{narayanan07} confirmed that $f(W_r)$ for $W_r \!<\! 0.3$~{\AA} was a power-law distribution consistent with that found by \citet{churchill99}. However, for $z \!\sim\! 2$, they argued that the observed $f(W_r)$ of weak absorbers favored an extension of the exponential distribution fitted by \citet{nestor05} down to $W_r \!\sim\! 0.02$~{\AA}. That is, the higher redshift data sampling Cosmic Noon exhibited only a slight overabundance with respect to the extrapolation of the exponential function fitted to $W_r \!\geq\! 0.3$~{\AA} by \citet{nestor05}, and did not follow as steep a power-law distribution below $W_r \!<\! 0.3$~{\AA} as was observed for $z \!\sim\! 0.9$.

Noting the power-law behavior for the distribution of the smallest equivalent widths and the exponential behavior for larger equivalent widths, \citet{ggk-cwc11} suggested $f(W_r)$ might be best-described by a Schechter function \citep{schechter76}, 
\begin{equation}
f(W) \, dW \!=\! \Phi^* (W/W^\ast)^{\alpha} \, 
e ^{-(W/W^\ast)}  dW/W^\ast
\, ,
\label{eq:schechterfunction}
\end{equation}
over the full measured range of $W_r$, where $\Phi^*$ is the normalization, $W^*$ is the exponential scale length (or characteristic equivalent width), and $\alpha$ is the ``weak-end'' power-law slope. They combined the measurements of \citet{ss92} and \cite{nestor05} for absorbers with $W_r \!\geq\! 0.3$~{\AA} covering $0.3 \!\leq\! z \!\leq\! 2.3$, with the weak absorbers measured by \citet{churchill99} and \citet{narayanan07} limited to $0.4 \!\leq\! z \!\leq\! 1.4$, and obtained the fitted values $\alpha \!\simeq\! -0.9$ and $W^{\ast} \!\simeq\! 1.7$~{\AA}.  The combined data constitute a heterogeneous sample across redshift, rendering the fit results an admixture of $f(W_r)$ at different cosmic times. Indeed, the fit omitted the weak absorbers for $z \!>\! 1.4$ from \citet{narayanan07}, which would have yielded a flatter ``weak-end'' power-law slope and a larger $W^*$.  The \citet{ggk-cwc11} study, however, clearly showed that a Schechter function is a viable functional form for $f(W_r)$.

In the 2010s, surveys for {\MgII} were extended beyond ${z \!\sim\! 2.4}$ with the advent of infrared spectrographs on 8--10 meter class telescopes. Using the FIRE/Magellan facility, \citet{matejek12} conducted the first blind survey for high-redshift {\MgII} absorbers ($1.9 \leq z \leq 6.3$) to a threshold of $W_r = 0.3$~{\AA}. Incorporating their measurements with those of the $0.3 \leq z \leq 2.3$ measurements of \citet{nestor05}, they found that, for absorbers in the equivalent width range  $0.3 \leq W_r < 1.0$~{\AA}, the redshift path density $d{\cal N}/dz$ is not evolving from $z \sim 6$ to $z \sim 0.4$.  For the strongest absorbers, i.e., those with $W_r \geq 1.0$~{\AA}, \citet{matejek12} found that $d{\cal N}/dX$ changes (evolves) quite strongly. It increases by a factor of $\sim\! 3$ from $z\!\sim\! 6$ to Cosmic Noon, where it peaks, and then declines by a factor of $\sim\! 1.5$ by $z \!\sim\! 0.4$.  In summary, the $d{\cal N}/dX$ of {\MgII} absorbers with $0.3 \!\leq\! W_r \!<\! 1.0 $~{\AA} do not evolve from $0.4 \!\leq\! z \!\leq\! 6.4$, whereas $d{\cal N}/dX$ of the strongest absorbers with $W_r \!\geq\! 1$~{\AA} rises from $z \!\sim\! 6$ to Cosmic Noon then declines toward the present epoch. We will call this type of evolution ``Type A'' evolution.\footnote{As many astrophysical quantities \citep[for example, the global star formation density of the universe,][]{madau14}, evolve in this manner, i.e., increase from early times at higher redshifts, peak during Cosmic Noon ($2 \!\leq\! z \!\leq\! 3$), and then decline toward the present epoch, and as this evolution has what we might call a classic ``A'' shape, we will hereafter intermittently invoke the shorthand term ``Type~A'' when referring to this type of evolution. Conversely, when referring to evolution of the opposite sense, i.e., that which decreases from early times at higher redshifts, minimizes during Cosmic Noon, and then increases toward the present epoch,  we will intermittently invoke the term ``Type~V''.} \citet{matejek12} modeled $f(W_r)$ as an exponential function for $W_r \!>\! 0.3$~{\AA}. They found clear redshift evolution of $W^*$, which mirrors the classic Type~A evolution of $d{\cal N}/dX$ for $W_r \!\geq\! 1.0$~{\AA} absorbers in that it peaks during Cosmic Noon. 

By DR7, the number of SDSS optical quasar spectra that covered $0.4 \!\leq\! z \!\leq\! 2.3$ for {\MgII} absorbers with $W_r > 0.3$~{\AA} increased to $\sim\! 100,\!000$. \citet{zhu13} and \citet{seyffert13} conducted independent surveys and studied the $\sim\! 40,\!000$ {\MgII} absorbers found in these spectra.  These studies dramatically improved the statistics of $d{\cal N}/dz$ and $d{\cal N}/dX$ for strong {\MgII} absorbers at these redshifts. Assuming an exponential function for $f(W_r)$, both studies also confirmed the decrease in $W^*$ reported by \citet{nestor05} from Cosmic Noon to $z \!\sim\! 0.5$.  In fact, \citet{zhu13} were able to measure the redshift where $W^*$ maximized. They found that $W^*$ increases from about $W^* \!\simeq\! 0.65$~{\AA} at $z \!\sim\! 2$ to a peak value of $W^* \!\simeq\! 0.75$~{\AA} at $z \!\sim\! 1.5$  and then decreases to $W^* \!\simeq\! 0.5$~{\AA} by $z \!\sim\! 0.5$. Using DR12 SDSS spectra, \citet{raghunathan2016} searched $\sim\! 260,\!000$ background quasars and found $\sim\! 40,\!000$ {\MgII} absorbers.  They found significant evolution of the systems with $W_r \geq 1.2$~{\AA} for $z\!<\!1$, consistent with other previous studies.

In 2017, several studies of {\MgII} absorbers appeared in the literature.  \citet{mathes17} studied some 1200 {\MgII} absorbers detected in $\sim\! 600$ HIRES/Keck and UVES/VLT quasar spectra covering the range $0.2 \!\leq\! z \!\leq\! 2.6$ to a detection threshold of $W_r \!=\! 0.02$~{\AA}. Using FIRE/Magellan to survey 100 quasars over the redshift range $3.6 \!\leq\! z \!\leq\! 7.1$, \citet{chen.s17} studied some $280$ high-redshift {\MgII} absorbers.
%with $W_r \!\geq\! 0.3$~{\AA}. 
Surveying four X-shooter/VLT quasar spectra, \citet{codor17} studied some $50$ {\MgII} absorbers in the redshift range $1.9 \!\leq\! z \!\leq\! 6.4$.
%, including 10 weak absorbers having $0.12 \!\leq\! W_r \!<\! 0.3$~{\AA}. 
\citet{bosman17} surveyed an X-shooter spectrum of the $z = 7.1$ quasar ULAS J1120+0641 
%and found a surprising number of weak {\MgII} absorbers at 
across the redshift range $5.9 \!\leq\! z \!\leq\! 7.0$ and found seven {\MgII} absorbers. 

A key contribution of the \citet{chen.s17} survey was a tightening in the statistics on $d{\cal N}/dz$, $d{\cal N}/dX$, and $f(W_r)$ for $W_r \!>\! 0.3$~{\AA} at $z \!>\! 2.5$.  Assuming an exponential function for  $f(W_r)$, they reported that $W^*$ increases from $W^* \!\simeq\! 0.3$~{\AA} at $z \!\sim\! 6.3$ to $W^* \!\simeq\! 0.8$~{\AA} at $z \!\sim\! 2.5$.  It was now firmly established that $f(W_r)$ evolved such that both $d{\cal N}/dX$ of $W_r \!\geq\! 1.0$~{\AA} {\MgII} absorbers and the magnitude of $W^*$ peak around $z \!\sim\! 2$, indicating that the strongest {\MgII} absorbers are most frequent during Cosmic Noon. Both $W^*$ and $d{\cal N}/dX$ of the strongest {\MgII} absorbers exhibit Type~A evolution.

Key results of \citet{mathes17} were improved statistics on $d{\cal N}/dz$, $d{\cal N}/dX$, and $f(W_r)$ for the weakest {\MgII} absorbers over the redshift range $0.2 \!\leq\! z \!\leq\! 2.6$.  A Schechter function was required for the best fit to $f(W_r)$ and the weak-end power-law slope was seen to evolve from $\alpha \!\simeq\! -0.8$ at $z \!\sim\! 2$ to $\alpha \!\simeq\! -1.1$ at $z \!\sim\! 0.5$. This indicates a greater relative number of weak systems at low redshifts compared to Cosmic Noon.  \citet{mathes17} substantially improved the measurements of $d{\cal N}/dX$ of weak absorbers for $z\!<\!2.6$.  The quasi-linear increase of $d{\cal N}/dX$ with cosmic time from $z \!\sim\! 2.4$ to $z\!\sim\!0.4$ corroborates evolution in $f(W_r)$; there is a greater cosmic incidence of weak {\MgII} absorbers at the present epoch and a lower incidence (by a factor of $\sim\! 2$) during Cosmic Noon.  Extrapolating the $d{\cal N}/dX$ for the weak absorbers to higher redshift led to the prediction that no weak {\MgII} absorbers existed at $z \!\geq\! 3.3$.  That prediction was proved false. Ten weak {\MgII} absorbers in the redshift range $2 \!\leq\! z \!\leq\! 6$ were found by \citet{codor17} and three weak {\MgII} absorbers were found in the redshift range $6 \!\leq\! z \!\leq\! 7$ by \citet{bosman17}. However, these two surveys are based on only four quasar sight lines \citep{codor17} and one sight line \citep{bosman17}, respectively, and may suffer from cosmic variance.

For these weak absorbers, the $d{\cal N}/dX$ measurements of \citet{narayanan07} and  \citet{mathes17} overlap with the measurements of \citet{codor17} at $2.0 \!\leq\! z \!\leq\! 2.4$, and these works are consistent with one another across this range.  The interesting result is that $d{\cal N}/dX$ is a factor of a few  {\it larger\/} at $z \!\sim\! 6.5$ than at $z \!\sim\! 2$. That is, the combined $d{\cal N}/dX$ measurements from the $z \!<\! 2.4$ optical and $z \!>\! 2.0$ infrared surveys suggested Type~V evolution for weak {\MgII} absorbers with their minimum cosmic incidence occurring at Cosmic Noon. 

Both \citet{bosman17} and \citet{codor17} pointed out that, at the highest redshifts, the weak absorbers are present in numbers greater than are predicted by an exponential equivalent width distribution, $f(W_r)$, and that the data are more consistent with a power-law slope for $W_r \!<\! 0.3$~{\AA}.  Fitting a Schechter function to their full sample, \citet{bosman17} showed that the weak-end slope of $f(W_r)$ at $z\!\sim\!6$ is within the $2~\sigma$ uncertainties of that found by \citet{mathes17} for $z \!\leq\! 2$. 

The NIRSpec instrument on the {\it James Webb Space Telescope\/} ({\it JWST\/}) allows studies of {\MgII} to extend to $z\!>\!15$ (assuming bright quasars will be found up to those redshifts!).  In a theoretical study, \citet{hennawi21} used hydrodynamic cosmological simulations and conducted a modest-sized mock NIRSpec survey to demonstrate that, for a sufficiently neutral and metal enriched IGM, {\it JWST\/} should be able to detect a ``forest'' of weak {\MgII} absorbers with $d{\cal N}/dX \!\sim\! 0.1$ at $z \!\sim\! 7.5$ for a Schechter equivalent width distribution function with estimated parameters $(\Phi^*, W^*, \alpha) = (0.44, 1.0~\hbox{\AA}, -0.8)$.\footnote{Quoted $d{\cal N}/dX$ and $\Phi^*$ have been converted from quoted $d{\cal N}/dz = 5.2$ and $N^* = 2.35$ using $dX/dz = 5.3$ at $z=7.5$}  These Schechter parameters are based on the combined $d{\cal N}/dX$ measurements of \citet{chen.s17} and \citet{bosman17} at $z \!\sim\! 6.4$, where the weak-end slope is constrained by the \citet{bosman17} measurements. A ground-based survey of 10 $z \!\geq\! 6.8$ quasars observed with the less sensitive MOSFIRE and NIRES spectrographs on Keck and X-shooter on VLT yielded no {\MgII} forest lines over the redshift range $5.9 \leq z \leq 7.4$  \citep{tie23}. 

\citet{christiansen23} studied NIRSpec spectra of four quasars sensitive to $W_r \geq 0.3$~{\AA} over the redshift range $2.3 \!\leq\! z \!\leq\! 7.5$ and found 29 {\MgII} absorbers, including two at the currently highest redshifts, $z \!=\! 7.37$ and $z \!=\! 7.44$.  They measure $d{\cal N}/dX$ for two populations, those with $W_r \geq 0.3$~{\AA} and those with $W_r \geq 1.0$~{\AA}. Within their relatively large uncertainties, their measured $d{\cal N}/dX$ are consistent with those of \citet{chen.s17}. In particular, they were able to confirm no evidence for evolution in the co-moving line-of-sight path density of {\MgII} absorbers with $W_r \!\geq\! 0.3$~{\AA}.

\citet{davies23-survey} and \citet{sebastian23} conducted what is presently the deepest $z\!\geq 2$ survey of {\MgII} absorbers using X-shooter spectra from the XQR-30 survey \citep{dodorico23}. They detected 131 weak absorbers and were sensitive to a 50\% completeness down to $W_r \!=\! 0.03$~{\AA}.  Over $2.0 \!\leq\! z \!\leq\! 6.4 $, their measured $d{\cal N}/dX$ for the strongest absorbers, corroborates the evolution measured by \citet{chen.s17}.  
For this very same redshift range, $d{\cal N}/dX$ for the weakest absorbers was found to be flat with redshift, indicating no evolution in this population from Cosmic Noon up to $z \sim 6.4$. \cite{sebastian23} showed that $f(W_r)$ at $2.0 \!\leq\! z \!\leq\! 6.4 $ is consistent with a Schechter distribution function, and obtain $\alpha \simeq -0.7$ and $W^* \simeq 1.4$~{\AA} (see Eq.~\ref{eq:schechterfunction}). 

Fixing $W^*$, \cite{sebastian23} found tentative evidence that the weak-end slope becomes shallower as redshift decreases, meaning that a greater co-moving line-of-sight path density of weak {\MgII} absorbers would be predicted in the higher redshift range of their survey ($4.1 \!\leq\! z \!\leq\! 6.4$) compared to the pre-Cosmic Noon and Cosmic Noon redshift regime ($1.9 \!\leq\! z \!\leq\! 4.1$). This suggestive, yet tentative evolution is interesting because it may be indicating that the cosmic incidence of weak absorbers for $z \sim 6$ is declining from higher redshifts. Indeed, the measurement of \citet{bosman17} at $z\sim 6.5$ is a factor of $2.5\times$ above that measured at $z\sim 6.2$ and a factor of $4\times$ above that measured at $z\sim 5.2$ by \cite{sebastian23}, respectively. Taken at face value, this trend at the highest redshift indicates that weak {\MgII} absorbers exhibit Type V evolution with a minimum cosmic incidence at Cosmic Noon. The error bars for the two $z>6$ data points slightly overlap, so Type V evolution remains tentative.

\section{From Cosmic Incidence to Evolution}
\label{sec:currentpic}

%tttttttttttttttttttttttttttttttttttttttttttttttttttttttttttttttt
%\begin{deluxetable}{ccl}
%\tablewidth{0pt}
%\tablecaption{Adopted $d{\cal N}/dX$ Measurements \label{tab:dNdXdata}}
%\tablehead{
%\colhead{$W_r$ Bin} &
%\colhead{Redshift} &
%\colhead{Referenced} \\[-6pt]
%\colhead{(\AA)} &
%\colhead{Range} &
%\colhead{Work}
%}
%\startdata
%$0.03-0.3$ & $0.0-0.3$ & \citet{narayanan05} \\[-4pt]
%           & $0.3-2.4$ & \citet{narayanan07} \\[-4pt]
%           & $0.4-2.5$ & \citet{mathes17} \\[-4pt]
%           & $1.9-6.4$ & \citet{sebastian23} \\%[-4pt]
%           & $2.0-5.4$ & \citet{codor17} \\[-4pt]
%           & $6.1-6.7$ & \citet{bosman17} \\
%$0.3-0.6$  & $0.0-0.2$ & \citet{churchill01} \\[-4pt]
%           & $0.4-2.3$ & \citet{nestor05} \\[-4pt]
%           & $1.9-7.1$ & \citet{chen.s17} \\
%$0.6-1.0$  & $0.0-0.2$ & \citet{churchill01} \\[-4pt]
%           & $0.4-2.3$ & \citet{nestor05} \\[-4pt]
%           &  & + \citet{seyffert13} \\[-4pt]
%           &  & + \citet{zhu13} \\[-4pt]  % Fig 13
%           & $1.9-7.1$ & \citet{chen.s17} \\
%$> 1.0$    & $0.0-0.2$ & \citet{churchill01} \\[-4pt]
%           & $0.4-2.3$ & \citet{zhu13} \\[-4pt]  % Fig 13
%           &  & + \citet{seyffert13} \\[-4pt]
%           & $1.9-6.4$ & \citet{sebastian23} \\[-4pt]
%           & $1.9-7.1$ & \citet{chen.s17} \\
%\enddata 
%\end{deluxetable}
%tttttttttttttttttttttttttttttttttttttttttttttttttttttttttttttttt

%tttttttttttttttttttttttttttttttttttttttttttttttttttttttttttttttt
\begin{deluxetable*}{cclcc}
\tablewidth{0pt}
\tablecaption{Adopted $d{\cal N}/dX$ Measurements \label{tab:dNdXdata}}
\tablehead{
\colhead{$W_r$ Bin} &
\colhead{Redshift} &
\colhead{Referenced} &
\colhead{Resolution} &
\colhead{Completeness} \\[-6pt]
\colhead{(\AA)} &
\colhead{Range} &
\colhead{Work} &
\colhead{($R=\lambda/\Delta \lambda$)} &
\colhead{to minimum $W_r$} 
}
\startdata
$W_r \in [0.03,0.3)$ & $0.0-0.3$ & \citet{narayanan05} & 45,000 & 91\% \\[-4pt]
           & $0.3-2.4$ & \citet{narayanan07} & 45,000 & 92\% \\[-4pt]
           & $0.4-2.5$ & \citet{mathes17} & 45,000 & 99\% \\[-4pt]
           & $1.9-6.4$ & \citet{sebastian23} & 17,000 & 50\%  \\[-4pt]
%           & $2.0-5.4$ & \citet{codor17} \\[-4pt]
           & $6.1-6.7$ & \citet{bosman17} & 10,000 & 60\% \\
$W_r \in [0.3,0.6)$  & $0.0-0.2$ & \citet{churchill01} & 1330 & 16\% \\[-4pt]
           & $0.4-2.3$ & \citet{nestor05} & 1560--2650 & 5\% \\[-4pt]
           & $0.4-2.3$ & \citet{seyffert13} & 1560--2650 &  60\% \\[-4pt]
           & $1.9-7.1$ & \citet{chen.s17} & 6000--8000 &  60\%\\
$W_r \in [0.6,1.0)$  & $0.0-0.2$ & \citet{churchill01} & 1330 &  80\%\\[-4pt]
           & $0.4-2.3$ & \citet{nestor05} & 1560--2650 &  26\% \\[-4pt]
           &  & + \citet{seyffert13} & 1560--2650 & 75\% \\[-4pt]
           &  & + \citet{zhu13} & 1560--2650 &  48\% \\[-4pt]  % Fig 13
           & $1.9-7.1$ & \citet{chen.s17} & 6000--8000 &  90\% \\
$W_r \geq 1.0$    & $0.0-0.2$ & \citet{churchill01} & 1330 & 95\% \\[-4pt]
           & $0.4-2.3$ & \citet{zhu13} & 1560--2650 & 75\% \\[-4pt]  % Fig 13
           &  & + \citet{seyffert13} & 1560--2650 &  85\% \\[-4pt]
           & $1.9-6.4$ & \citet{sebastian23} & 17,000 & 100\% \\[-4pt]
           & $1.9-7.1$ & \citet{chen.s17} & 6000--8000 & 100\%  \\
\enddata 
\end{deluxetable*}
%tttttttttttttttttttttttttttttttttttttttttttttttttttttttttttttttt

\begin{figure}[thb]
\centering
\includegraphics[width=0.98\columnwidth]{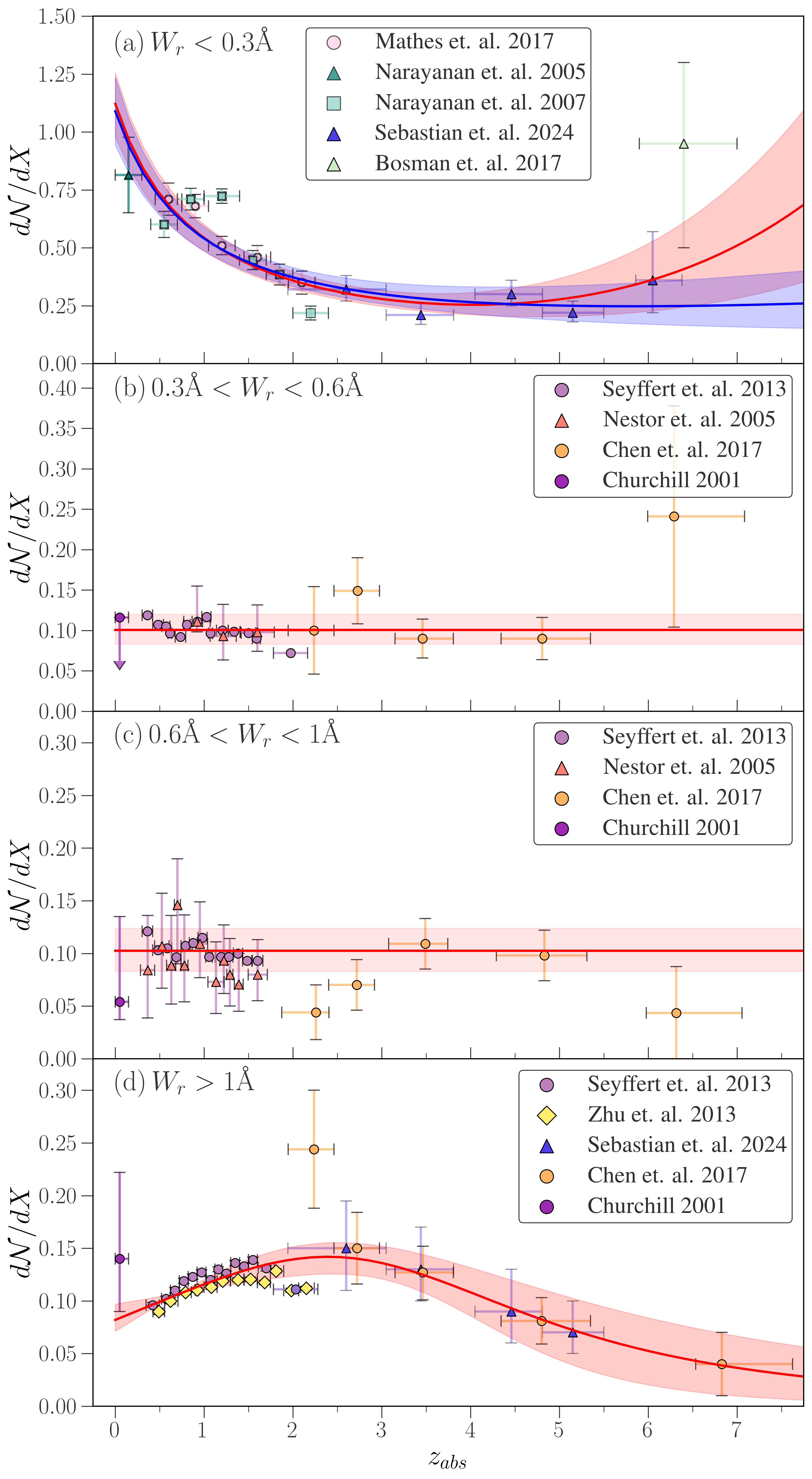}
%\captionsetup{justification=raggedright}
%\captionsetup{justification=justified, singlelinecheck=off} 
\caption{Adopted $d{\cal N}/dX$ measurements for $0 \!\leq\! z \!\leq\! 7$. 
(a) $W_r \!\in\! (0.03,0.3]$~{\AA}. 
(b) $W_r \!\in\! (0.3,0.6]$~{\AA}. 
(c) $W_r \!\in\! (0.6,1.0]$~{\AA}.
(d) $W_r \!\in \!(1.0,\infty)$~{\AA}. 
Curves are fits to the data (described in Section~\ref{sec:modeling}) and shaded regions represent $1~\sigma$ envelopes.  For $W_r \!<\! 0.3$~{\AA} two models have been fitted, one omitting the \citet{bosman17} measurement (S24, blue curve), which shows no evolution for $z>4.5$,  and one including the Bosman {\etal} point (S24+B17, red curve), which indicates a decreasing incidence from $z\sim 7$ to $z\sim 4.5$.
}
\label{fig:MgIIdNdX}
\end{figure}

We compiled selected $d{\cal N}/dX$ measurements from the literature. The adopted works are summarized in Table~\ref{tab:dNdXdata} and their measured values are illustrated in Figure~\ref{fig:MgIIdNdX}. Historically, the majority of surveys have published {\MgII} absorber path densities in the four rest-frame equivalent width ranges $W_1 \in (0.03, 0.3]$, $W_2 \in (0.3, 0.6]$, $W_3 \in (0.6, 1.0]$, and  $W_4 \in (1.0, \infty)$~{\AA} and we adopted these measurements because, when combined, they provide the greatest redundancy and therefore the greatest statistical leverage. For cases where a survey published $d{\cal N}/dz$ only, we converted the measurement to $d{\cal N}/dX$ using the correction factor $\Delta z/\Delta X$, where $\Delta z = (z^+ \!-\! z^-)$ is the redshift interval of the measurement and $\Delta X \!=\! X(z^+)\!-\!X(z^-)$ is the corresponding  ``absorption path'' interval, where we used $X(z) \!=\! 
(2/3\Omega_m)[
\{\Omega_m(1\!+\!z)^3 \!+\!\Omega_\Lambda \}^{1/2} -1 ]$. 

For UV surveys covering $z \!\leq\! 0.3$, we adopted the $d{\cal N}/dX$ measurements of \citet{narayanan05} and \citet{churchill01}.  The former provide data for the weak absorbers and the latter for those with $W_r \!\geq\! 0.3$~{\AA}.  For optical surveys covering $0.4 \!\leq\! z \!\leq\! 2.3 $, we adopted the measurements from the SDSS surveys of \citet{nestor05}, \citet{seyffert13} and \citet{zhu13}, which cover $W_r \!\geq\! 0.3$~{\AA}. For the weak absorbers in this redshift range, we adopted the measurements of \citet{narayanan07} and \citet{mathes17}. For $2.0 \!\leq\! z \!\leq\! 7$, we adopted the IR measurements of \citet{bosman17}, \citet{chen.s17}, and \citet{sebastian23}.  Whereas the work of \citet{chen.s17} is derived from FIRE spectra and cover only $W_r \!\geq\! 0.3$~{\AA}, the works of \citet{bosman17} and \citet{sebastian23} use X-shooter and cover $W_r \!\geq\! 0.03$~{\AA}.

%\begin{equation}
%\frac{\Delta z}{\Delta X} = 
%\frac{(z^+ - z^-)}{X(z^+)-X(z^-)} \, ,
%\end{equation}

%ffffffffffffffffffffffffffffffffffffffffffffffff
%ffffffffffffffffffffffffffffffffffffffffffffffff

\subsection{The Cosmic Incidence}
\label{sec:thedNdXdata}

As illustrated in Figure~\ref{fig:MgIIdNdX}(a), the $d{\cal N}/dX$ measurements of \citet{sebastian23} suggest that $d{\cal N}/dX$ of the weak population of absorbers does not evolve from $z \!\sim\! 6$ to Cosmic Noon. However, inclusion of the \citet{bosman17} measurement at $z\sim 6.4$, and accounting for the tentative evolution reported by \citet{sebastian23} for $z>4$, one could argue that the nature of the evolution at the highest measured redshifts remains an open question. Combining with $d{\cal N}/dX$ measurements of \citet{narayanan05, narayanan07} and \citet{mathes17}, the default picture is that the cosmic incidence of the weak population remains steady from $z \!\sim\! 7$ down through Cosmic Noon and then increases to the present epoch.  

As illustrated in Figure~\ref{fig:MgIIdNdX}(b), for intermediate absorbers in the range $0.3 \!\leq\! W_r \!<\! 0.6$~{\AA}, the measurements of \citet{churchill01} \citet{nestor05}, \citet{seyffert13}, \citet{zhu13}, and \citet{chen.s17} suggest no evolution in $d{\cal N}/dX$ from $z \!\sim\! 7$ to the present epoch.  As illustrated in Figure~\ref{fig:MgIIdNdX}(c), for intermediate absorbers in the range $0.6 \!\leq\! W_r \!<\! 1.0$~{\AA}, a similar lack of evolution is observed. As illustrated in Figure~\ref{fig:MgIIdNdX}(d), for the strongest {\MgII} absorbers, $d{\cal N}/dX$ clearly exhibits classic Type~A evolution \citep{nestor05, seyffert13, zhu13, chen.s17, sebastian23}. That is, the population increases from $z \!\sim\! 7$ to Cosmic Noon where it peaks and then decreases toward the present epoch. 

\subsection{Inferring Evolution}
\label{sec:thefWimplications}

In accordance with Eq.~\ref{eq:whatallthefussisabout}, the differing redshift evolution of $d{\cal N}/dX$ across finite equivalent width ranges is  informing us that the equivalent width distribution is evolving. First, since $d{\cal N}/dX$ of the weak absorbers is increasing from Cosmic Noon to the present epoch \citep[e.g.,][]{narayanan07, mathes17} and $d{\cal N}/dX$ of the strongest absorbers is decreasing from Cosmic Noon to the present epoch \citep[e.g.,][]{nestor05, seyffert13, zhu13, raghunathan2016}, we can immediately infer that the {\it relative\/} frequency of the weakest absorbers  has been increasing and the {\it relative\/} frequency of the strongest absorbers has been decreasing since Cosmic Noon.  If, for example, $f(z,W_r)$ is modeled as a Schechter function for which the weak-end power-law, $\alpha$, evolves with redshift, then we can infer that the value of $\alpha$ would become more negative (the distribution would become steeper) from Cosmic Noon to the present epoch.  This inference is consistent with the tentative findings of \citet{narayanan07} and \citet{mathes17}, both of whom suggest that the weak-end power-law slope steepens from Cosmic Noon to $z \!\sim\! 0.3$ from their direct measurements of the distribution of equivalent widths.  

For cosmic times prior and leading up to Cosmic Noon, $d{\cal N}/dX$ for the weakest absorbers may be constant, as per the measurements of \citet{sebastian23}. We can infer that the weak-end slope would remain fairly constant or slightly steepens toward Cosmic Noon (given that the strongest systems are increasing across this cosmic time).  Attempts to measure the weak-end slope directly from the measured equivalent widths of the detected absorbers have been unable to provided full clarity as to how weak absorbers evolve at these redshifts  \citep{bosman17, codor17, hennawi21, sebastian23}.  If, on the otherhand, weak absorbers exhibit Type V evolution, we would infer that the faint end slope would be steeper at $z>6$, would flatten toward Cosmic Noon, and then would steepen again toward the present epoch.

For the $d{\cal N}/dX$ of the strongest absorbers, both \citet{chen.s17} and \citet{sebastian23} suggest that the cosmic incidence of strong absorbers increases from $z \!\sim\! 7$ to Cosmic Noon. Both \citet{seyffert13} and \citet{zhu13} are in agreement that the cosmic incidence decreases from Cosmic Noon to the present epoch.  We might then infer that $W^*$ would increase from $z\!\sim\!7$ to Cosmic Noon and then decrease toward the present epoch (also exhibiting classic Type~A evolution).  As reported by \citet{chen.s17}, this is precisely what is determined {\it when one assumes an exponential equivalent width distribution function\/}. 

However, for a Schechter function formalism, it is less clear how $W^*$ is expected to evolve if $\alpha$ is also evolving.  
Consider the post-Cosmic Noon era, where we expect the weak-end slope to steepen toward the present epoch. Measurements of the evolution of the strongest absorbers in the post-Cosmic Noon era inform us that their cosmic incidence was higher at Cosmic Noon and decreased toward the present epoch. This decrease in the incidence of the strongest absorbers means that the {\it partial area\/} under the distribution function integrated over $W_r \!\ge\! 1$~{\AA} has decreased.  
Arithmetically, this decrease in the partial area can occur due to a decrease in the normalization of $f(z,W)$, even if accompanied by an incremental {\it increase\/} in $W^*$.  We will revisit this scenario below.

\section{Constraining Evolution}
\label{sec:modeling}

Our goal is to quantitatively characterize the redshift evolution in the {\MgII} absorber equivalent width distribution function using current measurements in the literature.  In principle, $f(z,W)$ can be measured directly by counting the number of measured absorbers in a given co-moving line-of-sight path interval as a function of equivalent width interval.  In practice, one must account for the  equivalent width detection sensitivity threshold of the survey and correct for the detection completeness as a function of redshift.  In order to conduct a meta analysis of existing {\MgII} absorber surveys, we would need to not only obtain the complete line lists and measured $W_r$ of all {\MgII} absorbers of all included samples, we would also need to properly combine the completeness functions, confidence levels, and redshift path coverage of each of the surveys. This would be a very challenging endeavor.  However, most published surveys also provide measurements of $d{\cal N}/dz$ and/or $d{\cal N}/dX$, which already take in to account the detection sensitivities and redshift path coverage of the surveys.  That is, $d{\cal N}/dz$ and $d{\cal N}/dX$ measurements can be directly compared between surveys.\footnote{This is strictly true for $d{\cal N}/dz$.  However, computing $d{\cal N}/dX$ requires a cosmological model and different surveys may adopt different models or cosmological parameters. Variations between $d{\cal N}/dX$ measurements due to choice of cosmological parameters in a $\Lambda$CDM model should typically be smaller than the experimental uncertainties.}  We will concentrate on $d{\cal N}/dX$ measurements and exploit the relationship between $d{\cal N}/dX$ and $f(z,W)$ given in Eq.~\ref{eq:whatallthefussisabout} to constrain evolution in $f(z,W)$. 

In practice, individual $d{\cal N}/dX$ values are measured in finite redshift ranges, $z \in (z^-,z^+)$ and rest-frame equivalent widths in bins $W_i\in (W_i^-,W_i^+)$. From the general relation given by Eq.~\ref{eq:whatallthefussisabout}, we can write
\begin{equation}
   (d{\cal N}/dX) _{z,W_i}  =
   \int _{W_i^{-}}^{W_i^{+}}
   \!\!\! f(z,W) \, dW  \, ,
\label{eq:dNdXfromfW}
\end{equation}
where the subscripts ``${z,W_i}$'' specify the redshift and equivalent width bin of the measured $d{\cal N}/dX$. The proper integral on right hand side of Eq.~\ref{eq:dNdXfromfW} represents an apportioned area under $f(z,W)$ equal to the measured $(d{\cal N}/dX)_{z,W_i}$ at this redshift for this equivalent width bin. Reversing this statement, a measured $(d{\cal N}/dX)_{z,W_i}$ equals the corresponding apportioned area under the equivalent width distribution.  Thus, in principle, a data set of measured $(d{\cal N}/dX)_{z,W_i}$ for which a range of $z$ and $W_i$ are thoroughly represented provides constraints on the entire shape and amplitude of $f(z,W)$.  Hereafter, we will refer to the application of these constraints as the apportioned integral method (AIM).

%ffffffffffffffffffffffffffffffffffffffffffffffffffffffffffffffff
\begin{figure}[htb]
\includegraphics[width=1.0\columnwidth]{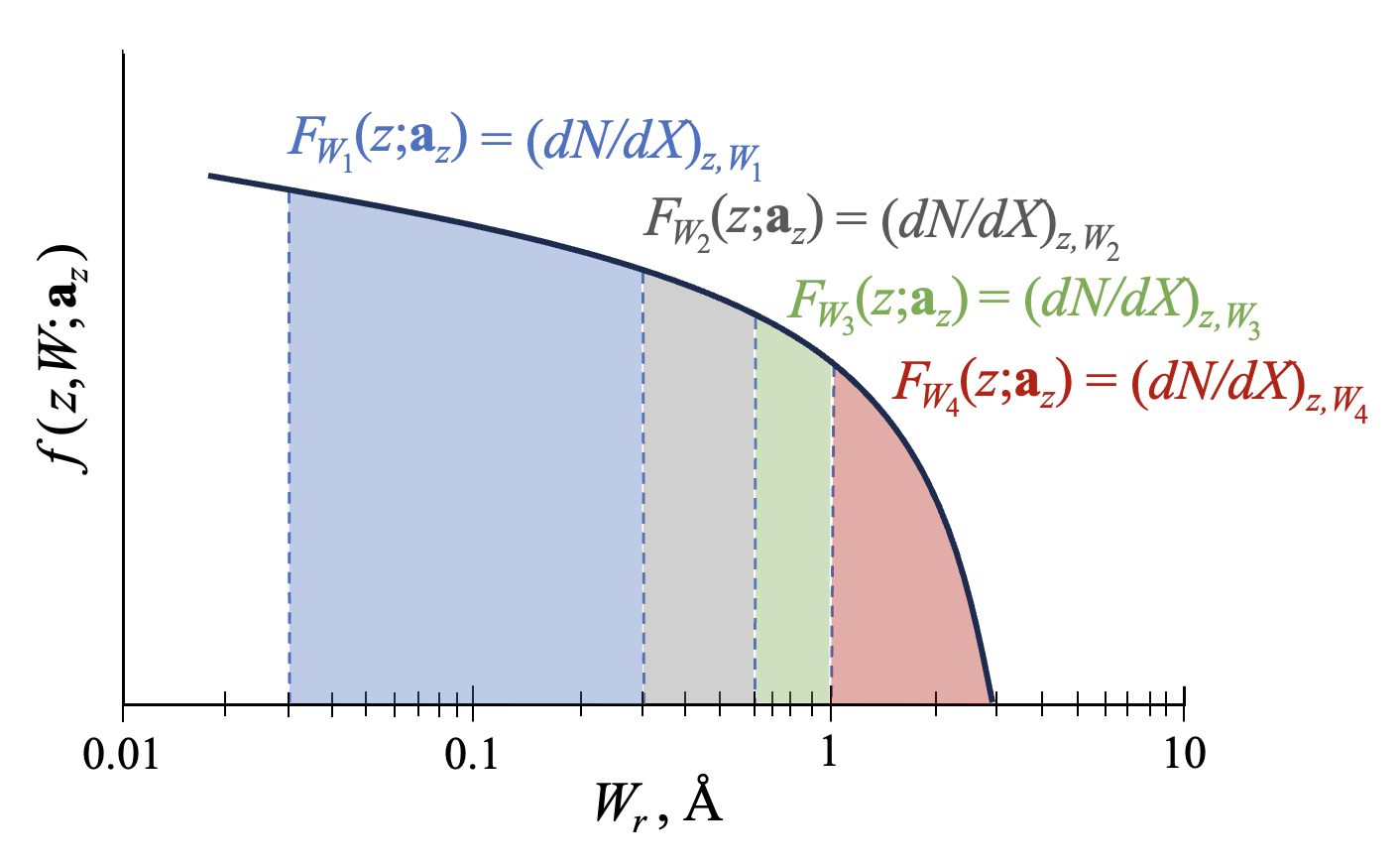}
%\fig{Figures/dNdX-Schematic.pdf} {0.49\textwidth}{} 
%\vspace{-25pt}
\caption{
A schematic of a parameterized equivalent width distribution function, $f(z,W;{\bf a}_z)$, for an arbitrary redshift.  The apportioned integrated areas under the curve, $F_{W_i}(z,{\bf a}_z)$, as given by Eq.~\ref{eq:theconstraintfunc}, are shown for the four adopted equivalent width bins, $W_1 \in (0.03,0.3]$~{\AA} (blue), $W_2 \in (0.3,0.6]$~{\AA} (gray), $W_3 \in (0.6,1.0]$~{\AA} (green), and $W_4 \in (1.0,\infty)$~{\AA} (red). If $f(z,W;{\bf a}_z)$ is an accurate description of $d{\cal N}/dX$ then each of these these apportioned areas should be equal to the corresponding measured $(d{\cal N}/dX)_{z,W_i}$.}
\label{fig:fWintegrals}
\end{figure}
%ffffffffffffffffffffffffffffffffffffffffffffffffffffffffffffffff

To model the evolution, we adopt a parameterized distribution, $f(z,W;{\bf a}_z)$, where ${\bf a}_z \!=\! {\bf a}(z)$ represents the vector of redshift-dependent parameters fitted to the data. In consideration of the works of \citet{ggk-cwc11}, \citet{mathes17}, \citet{bosman17}, and \citet{sebastian23}, we assume a Schechter function, 
\begin{equation}
\!\! f(z,W;{\bf a}_z)\, dW = a_{1} t^{a_3}  
e ^{-t} dt
\, ,
\label{eq:fittedEDF}
\end{equation}
where $t=W/a_2$, and where $a_1 = \Phi^*(z)$ is the normalization, $a_2 = W^*(z)$ is the characteristic equivalent width, and $a_3 = \alpha(z)$ is the weak-end slope at redshift $z$, respectively. We rewrite the right hand side of Eq.~\ref{eq:dNdXfromfW} (the apportioned area) in terms of the parameterized distribution function, giving
\begin{equation}
  F_{W_i}(z;{\bf a}_z) =  \int _{W_i^{-}}^{W_i^{+}}
   \!\!\! f(z,W;{\bf a}_z) \, dW  \, ,
\label{eq:theconstraintfunc}
\end{equation}
and, substituting, we obtain
\begin{equation}
(d{\cal N}/dX)_{z,W_i} = F_{W_i}(z;{\bf a}_z) \, , 
\label{eq:solvethis}
\end{equation}
where the left hand side is measured and the right hand side is the apportioned integrated area of the parameterize equivalent width distribution function.

In Figure~\ref{fig:fWintegrals}, we illustrate a plausible parameterized distribution function, $f(z,W;{\bf a}_z)$, for a given redshift. We show the four historically common equivalent width bins we have adopted for this work. The shaded regions represent the apportioned integrated areas under $f(z,W;{\bf a}_z)$ corresponding to each equivalent width bin as given by Eq.~\ref{eq:theconstraintfunc}. If $f(z,W;{\bf a}_z)$ accurately reflects the real-world  distribution function, then, in each equivalent width bin, the apportioned integrated area $F_{W_i}(z;{\bf a}_z)$ will equal the corresponding measured value of $(d{\cal N}/dX)_{z,W_i}$.  

A full quantified characterization of the redshift evolution of the equivalent width distribution requires that, {\it at each redshift}, we simultaneously satisfy four equations with three unknowns (four independent versions of Eq.~\ref{eq:solvethis}, one for each $W_i$ bin, and each with three free parameters, $\Phi^*(z)$, $W^*(z)$, and $\alpha(z)$). This constitutes an overdetermined system of independent nonlinear equations. These systems rarely have a single solution that simultaneously satisfies all equations.  A ``best'' solution is typically obtained using least squares minimization techniques. Furthermore, each measured $(d{\cal N}/dX)_{z,W_i}$ is the product of complex absorption line detection algorithms and redshift-dependent completeness corrections to estimate the redshift path.  These corrections are particularly important for the weakest absorbers, where completeness corrections can be large and vary with redshift. This means that measurement uncertainties will introduce scatter about a best solution.

From inspection of Figure~\ref{fig:MgIIdNdX}, we see that the measured $(d{\cal N}/dX)_{z,W_i}$ exhibit a non-trivial degree of scatter. First, there can be systematic offsets between published works, such as the measurements of \citet{seyffert13} and \citet{zhu13}.  Second, within a given study, there can be substantial scatter between adjacent redshift bins, for example the measurements of \citet{narayanan07} or the measurements of \citet{chen.s17}.  As our goal is to characterize the evolution in $f(z,W_r)$, we wish to minimize the effects of scatter by focusing on the trends apparent in the data.  Thus, we have fitted redshift-dependent functions to provide a smooth representation of the evolution of $(d{\cal N}/dX)_{z,W_i}$.

%tttttttttttttttttttttttttttttttttttttttttttttttttttttttttttttttt
\begin{deluxetable}{crr}[th]
\vglue 0.3in
\tablewidth{0pt}
\tablecaption{Weak Absorber Fit Parameters \label{tab:dNdXfits-Weak}}
\tablehead{
\multicolumn{3}{c}{$\NXz = N_1(1+z)^{\gamma_1} + N_2 (1+z)^{\gamma_2}$} \\
\colhead{Parameter} &
%\colhead{Value\tablenotemark{${\dagger}$}} &
\colhead{Value (S24)} &
\colhead{Value (S24+B17)} 
}
\startdata 
$N_1$        %& $1.008\pm0.096$ 
& $1.083\pm0.090$ & $1.122\pm0.138$ \\[-4pt]
$\gamma_1$   %& $ -0.913\pm0.024$ 
& $ -1.051\pm0.093$ & $-1.050\pm0.026$\\[-4pt]
$N_2$        %& $0.0016\pm0.0007$ 
& $0.006\pm0.007$ & $(3.7\pm2.5)\!\times\!10^{-5}$\\[-4pt]
$\gamma_2$   %& $3.086\pm0.025$ 
& $1.448\pm0.097$ & $4.450\pm0.026$
\enddata 
%\tablenotetext{\dagger}{w/ \citet{bosman17} and \citet{codor17}.}
%\tablenotetext{\ddagger}{w/ \citet{sebastian23}.}
\end{deluxetable}
%tttttttttttttttttttttttttttttttttttttttttttttttttttttttttttttttt

%tttttttttttttttttttttttttttttttttttttttttttttttttttttttttttttttt
\begin{deluxetable}{cr}[h!t]
\tablewidth{0pt}
%\vglue -0.3in
\tablecaption{Strong Absorber Fit Parameters \label{tab:dNdXfits-Strong}}
\tablehead{
\multicolumn{2}{c}{$\NXz = N_0(1+\beta z)/\{1+(z/z_0)^\gamma \}$} \\
\colhead{Parameter} &
\colhead{Value} 
}
\startdata
$N_0$     & $0.082\pm0.013$   \\[-4pt]
$\beta$   & $0.035\pm0.018$ \\[-4pt]
$z_0$     & $3.958\pm0.141$ \\[-4pt]
$\gamma$  & $3.624\pm1.415$  
\enddata 
\end{deluxetable}
%tttttttttttttttttttttttttttttttttttttttttttttttttttttttttttttttt

For weak absorbers ($W_r \!\in\! [0.03,0.3)$~{\AA}), we adopted a model for the co-moving line-of-sight path  density of the form
\begin{equation}
 \NXz
= N_1 (1+z)^{\gamma_1} 
+ N_2 (1+z)^{\gamma_2} \, ,
\label{eq:dNdx-fit-V}
\end{equation}
for which $n_0\sigma_0 = (H_0/c)(N_1\!+\!N_2)$ is the product of the spatial number density and average absorber cross section at the present epoch.  
%We applied this fitting function to the data in Figure~\ref{fig:MgIIdNdX}(a).  We simultaneously fitted the data of \citet{narayanan05, narayanan07} and \citet{mathes17} as these are consistent measurements. However, due to the tension between the measurements of \citet{bosman17}, \citet{codor17}, and \citet{sebastian23}, we fit the data of \citet{sebastian23} separately from those of \citet{bosman17} and \citet{codor17}.  This allowed us to examine the implications of possible evolution in the weak absorbers at $z\!>\!2$ as compared to a possible lack of evolution.
%N1 = 1.122 +/- 0.138
%g1 = -1.050 +/- 0.026
%N2 = 3.667e-5 +/- 2.5e-5
%g2 = 4.45 +/- 0.026

To account for Type~A evolution, which is seen for the strongest absorbers ($W_r \!\in\! (1.0,\infty)$~{\AA}), we adopted a model for the co-moving line-of-sight path density of the form \citep[][]{cole01},
\begin{equation}
 \NXz = \frac{N_0+\beta z}{1+(z/z_0)^\gamma}  \, ,
\label{eq:dNdx-fit-A}
\end{equation}
over the domain $0 \!\leq\! z \!\leq\! 7$. For this model, the product of the spatial number density and average absorber cross section at the present epoch is  $n_0\sigma_0 \!=\! (H_0/c)N_0$. We applied this fitting function to all the data presented in Figure~\ref{fig:MgIIdNdX}(d).

To perform the fits, we used the fitting code {\sc Fitmrq} \citep{press02-numrecipe}.  The fits and their $1\sigma$ uncertainties are shown in Figure~\ref{fig:MgIIdNdX}. The best-fit parameters are listed in Table~\ref{tab:dNdXfits-Weak} and Table~\ref{tab:dNdXfits-Strong}, respectively. 
For the intermediate absorbers ($W_r \!\in\! [0.3,0.6)$~{\AA} and $W_r \!\in\! [0.6,1.0)$~{\AA}), the $d{\cal N}/dX$ measurements are consistent with no evolution. We thus computed the variance-weighted mean values for these equivalent width bins. For $W_r \!\in\! [0.3,0.6)$~{\AA}, we obtained $\NXz \!=\! 1.005\pm0.004$ and for $W_r \!\in\! [0.6,1.0)$~{\AA} we obtained $\NXz \!=\! 1.024\pm0.004$.

\section{Tests, Applications, and Results}
\label{sec:results}

The apportioned integral method was implemented using least-squares minimization to obtain ${\bf a}_z$. The function we minimized is  \begin{equation}
    {\cal L}_z({\bf a}_z) = \sum_{i=1}
    \frac{[\NXz - F_{W_i}(z;{\bf a}_z) ]^2}{\sigma^2_{\NX}(z)} \, ,
\label{eq:likefunc}
\end{equation}
where ${\NXz}$ is the smoothed fitted value of $(d{\cal N}/dX)_{z,W_i}$, the $1\sigma$ uncertainty in $\NXz$ is $\sigma_{\NX}(z)$, and $F_{W_i}(z;{\bf a}_z)$ is the apportioned integrated area under the parameterized equivalent width distribution function at redshift $z$ in the equivalent width bin indexed by $i$.  We stepped through redshift from $z\!=\!0$ to $z\!=\!7$ in steps of $\Delta z \!=\! 0.05$. At each redshift, the sum is performed over the  equivalent width bins denoted by $W_i$.  The minimization is performed using {\sc lmfit} \citep{lmfit}, a non-linear least square minimization package.

\subsection{Testing Apportioned Integrals}

\begin{figure}[htb]
\centering
\includegraphics[width=0.45\textwidth]{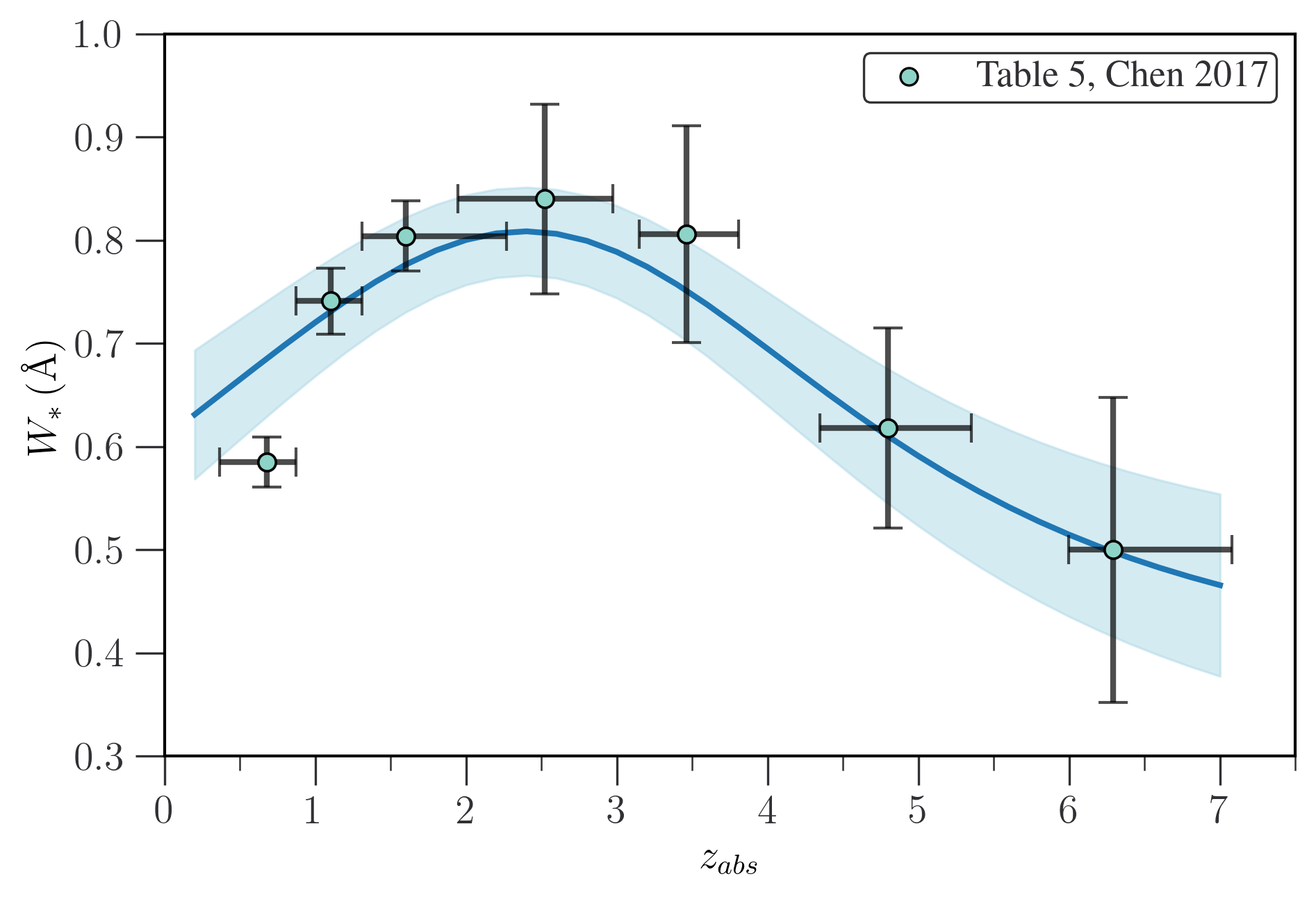} 
\caption{This is a test result from applying apportioned integrals to the redshift evolution of $W^*$ for $W_r \geq 0.3$~{\AA} absorbers assuming an exponential equivalent width distribution. The curve and shading are the predicted parameter $W^*(z)$ and its $\pm1\sigma$ uncertainty from the apportioned integral method (see Section~\ref{sec:modeling}) using the smoothed models of the $d{\cal N}/dX$ data presented in Figure~\ref{fig:MgIIdNdX}(b,c,d).  The apportioned integral method is highly consistent with the data determined from maximum likelihood least-squares fitting analysis applied directly to the measure equivalent widths \citep[see][]{nestor05, chen.s17}.  Note: the curve is not a fit to the data; it is an independent prediction from apportioned integrals.}
\label{fig:Wstarofz}
\end{figure}

For absorbers with $W_r \!\geq\! 0.3$~{\AA}, the redshift evolution in $W^*$ from  $z\!=\!6.4$ to $z\!=\!0.3$ has been determined using maximum-likelihood least-square fitting methods directly applied to the measured equivalent widths. For the regime $W_r \!\geq\! 0.3$~{\AA} an exponential distribution function, $f(W) \, dW = \Phi^* e^{-W/W^*} dW/W^*$, is adopted. \citet{chen.s17} showed that $W^*$ exhibited Type~A evolution, increasing from $W^* \!\sim\! 0.5$~{\AA} at $z \!\sim\! 6.4$ to $W^* \!\sim\! 0.8$~{\AA} near its peak at Cosmic Noon and then decreasing toward $z \sim 0.3$ where $W^* \!\sim\! 0.6$~{\AA}. In Figure~\ref{fig:Wstarofz}, we present the least-square results (data points) of \citet[][$z\!<\!2$]{nestor05} and \citet[][$z\!>\!2$, also see their Table~5]{chen.s17}.

%ffffffffffffffffffffffffffffffffffffffffffffffffffffffffffffffff
%ffffffffffffffffffffffffffffffffffffffffffffffffffffffffffffffff

%ffffffffffffffffffffffffffffffffffffffffffffffffffffffffffffffff
\begin{figure*}[bth]
\centering
\includegraphics[width=0.98\textwidth]{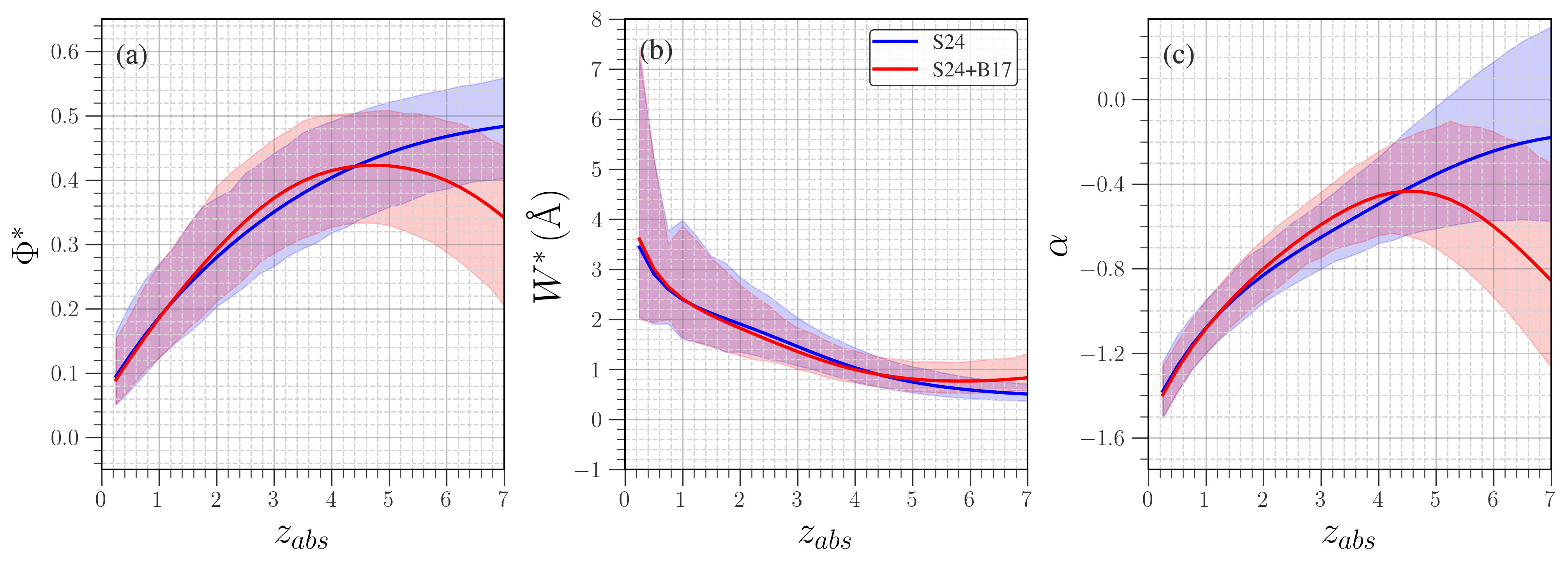} 
\caption{Redshift evolution of the fitted parameters for a Schechter equivalent width distribution function, (a) $\Phi^*(z)$, (b) $W^*(z)$, and (c) $\alpha(z)$,  using the apportioned integral method applied to the smoothed $d{\cal N}/dX$ data presented in Figure~\ref{fig:MgIIdNdX}. 
Two models of $d{\cal N}/dX$ are investigated, the S24 model and the S24+B17 model. The best-fit parameters for the equivalent width distribution are given by the solid curves and were obtained using the least-square minimization code {\sc lmfit}.  The $1\sigma$ uncertainties in the best-fit parameters, shown as the shaded regions, were estimated using the MCMC code {\sc emcee~v3} to better characterize the posterior uncertainty distributions.}
\label{fig:results}
\end{figure*}
%ffffffffffffffffffffffffffffffffffffffffffffffffffffffffffffffff

To test the apportioned integral method, we applied it to the $d{\cal N}/dX$ data for $W_r\!\geq\!0.3$~{\AA} (see Figure~\ref{fig:MgIIdNdX}(b,c,d)), omitting $d{\cal N}/dX$ for $W_r \!<\! 0.3$~{\AA}. We assumed an exponential distribution function.  Applying the apportion integral method, we then obtained $\Phi^*(z)$ and $W^*(z)$ by minimizing Eq.~\ref{eq:likefunc}. Our results for $W^*(z)$ are presented in Figure~\ref{fig:Wstarofz}. The curve and shaded region (representing the $1\sigma$ uncertainties) is $W^*(z)$ determined from the apportioned integral method.  Note that the curve is not a fit to the $W^*$ data of \citet{nestor05} and \citet{chen.s17}; it is an independent measurement. The apportioned integral method is in good agreement and captures the essence of the redshift evolution of $W^*$, with no more than a minor discrepancy as $z \rightarrow 0$. We did not undertake a comparison with the normalization, $\Phi^*(z)$, which is denoted $N^*$ by \citet{nestor05} and \citet{chen.s17}, because the normalization criterion differs between those authors. \citet{nestor05} normalize to $d{\cal N}/dz$ and \citet{chen.s17} normalize to the number of absorbers in their survey.  Fortunately, the characteristic equivalent width as determined by \citet{nestor05} and \citet{chen.s17} using maximum likelihood methods is independent of the convention adopted for the normalization. 

It is important to emphasize that the apportioned integral method and the maximum likelihood least-squares fitting method are entirely different.  The former minimizes the differences between $(d{\cal N}/dX)_{z,W_i}$ and the corresponding apportioned areas under the parameterized $f(z,W_r)$ curve. The latter involves a maximum likelihood fit (moderated through the detection completeness function) directly applied to the measured equivalent widths in the absorber sample for the assumed equivalent width distribution function. We embrace agreement  between these very different approaches as validation of the apportioned integral method.  

\subsection{Applying Apportioned Integrals}
\label{sec:AIM}

Having demonstrated that the apportioned integral method yields results commensurate with maximum likelihood least-squares fitting methods, we extended our analysis to include the weak {\MgII} absorbers (see Figure~\ref{fig:MgIIdNdX}(a)) under the assumption that $f(z,W)$ is a Schechter function. The Schechter function is able to capture the power-law distribution in the weak absorber regime.  Given the open question of Type V evolution for weak absorbers (discussed at the very end of Section~\ref{sec:ahistory}),  we independently applied the method to the two smoothed fitted models of $d{\cal N}/dX$ shown in Figure~\ref{fig:MgIIdNdX}(a). These models are denoted S24 and S24+B17, respectively. The former excludes the measurement of \citet{bosman17} at $z \sim 6.4$, whereas the latter includes it.
%Due the tension at $z\!>\!2$ between the \citet{sebastian23} measurements and those of \citet{bosman17} and \citet{codor17}, where the former suggest no evolution in the weak absorber $d{\cal N}/dX$ for $z\!>\!2$ and the latter suggest higher incidence at $z \!\sim\! 7$ decreasing toward Cosmic Noon (i.e., Type~V evolution), we independently applied the method to the two smoothed fitted models to $d{\cal N}/dX$ shown in Figure~\ref{fig:MgIIdNdX}(a).

In Figure~\ref{fig:results}, we present our results for the best-fitted parameters, i.e., the normalization, $\Phi^*(z)$, the characteristic equivalent width, $W^*(z)$, and the weak-end power-law slope, $\alpha(z)$, as a function of redshift. 
%Figure~\ref{fig:results}(a,b,c) shows the evolution in the fitted parameters for the scenario in which the cosmic incidence of weak {\MgII} absorbers declines from $z \!\sim\! 7$ to Cosmic Noon. Figure~\ref{fig:results}(d,e,f) shows the evolution in the fitted parameters for the scenario in which the cosmic incidence of weak {\MgII} absorbers does not evolve across this high redshift range.  
%Given that the cosmic incidence of weak {\MgII} absorbers appears to not evolve from $z \!\sim\! 7$ to Cosmic Noon, 
For the S24 model of the measured cosmic incidence, we find that, from $z\!\sim\!7$ to the present epoch, (1) the normalization monotonically declines from $\Phi^* \!\sim\! 0.5$ to $\Phi^* \!\sim\!0.1$, (2) the characteristic equivalent width monotonically increases from $W^* \!\sim\! 0.6$~{\AA} to $W^*\!\sim\! 3.2$~{\AA}, and (3) the weak-end slope monotonically steepens from $\alpha \!\sim\! -0.2$ to $\alpha \!\sim\! -1.4$.

For $z < 4.5$, the equivalent width distribution fitted parameters derived from the S24+B17 model are very similar to those determined for the S24 model. However, at $z\sim 4.5$, the normalization peaks at $\Phi^* \!\sim\! 0.4$ and the weak-end slope is flattest with $\alpha \!\sim\! -0.45$.  The departure of the two models occurs from $z=7$ to $z\sim4.5$, where the normalization increases from $\Phi^* \!\sim\! 0.35$ to $\Phi^* \!\sim\!0.4$ and the weak-end slope shallows from $\alpha \!\sim\! -0.8$ to $\alpha \!\sim\! -0.45$.  Interestingly, the evolution of $W^*$ is similar for both models of $d{\cal N}/dX$.

%For $z \!<\! 2$, the post-Cosmic Noon era, the two scenarios yield virtually identical redshift evolution and distribution functions.  
%The normalization decreases from $\Phi^* \!\sim\! 0.5$ at $z\!=\!7$ to $\Phi^* \!\sim\! 0.1$ at $z\!=\!7$. The characteristic equivalent width increases from $W^* \!\sim\! 0.6$~{\AA} to $W^* \!\sim\! 3.2$~{\AA} over this redshift regime.  The weak-end slope becomes steeper, evolving from $\alpha \!\sim\! -0.2$ to $\alpha \!\sim\! -1.4$.

%For $z \!>\! 2$, the pre-Cosmic Noon era, the two scenarios yield quite different redshift evolution and distribution functions.  If the incidence of weak {\MgII} absorbers declines from $z \!\sim\! 7$ to Cosmic Noon, we see that (1) the normalization exhibits Type~A evolution, peaking at Cosmic Noon, (2) the characteristic equivalent width is consistent with little to no evolution up to Cosmic Noon, and (3) the weak-end slope also exhibits Type~A evolution in that it is steepest at $z\!\sim\!7$, flattens toward Cosmic Noon, and then steepens again toward the present epoch. 

\subsection{Characterizing Parameter Covariance}

Although the {\sc lmfit} least-squares minimization algorithm returns uncertainties in the fitted parameters, these uncertainties are determined from the absolute values of the diagonal elements of the covariance matrix and do not include off-diagonal dependencies between parameters.  In order to characterize the posterior uncertainty distributions for the fitted parameters and assess the covariant relationships between parameters, we employed Markov-Chain Monte Carlo (MCMC) modeling.  We used the Python ensemble sampling toolkit for affine-invariant MCMC called {\sc emcee~v3} \citep{Foreman_Mackey_2019}.  We employed a likelihood function given by the logarithm of Eq.~\ref{eq:likefunc}, but with fixed $\sigma^2_{\NX}(z)\!=\!1$. We adopted Gaussian priors based on the best-fit values and standard deviations from the least-squares solution.\footnote{To clarify, we are not using the least-squares solution as a prior to obtain a solution from the MCMC method; we are using MCMC only to explore the covariance of the uncertainties in the best-fitted parameters.}  We ran {\sc emcee~v3} as a function of redshift and adopted the two-sided 68\% confidence levels as the $1\sigma$ confidence interval of the fitted parameters.  These are shown as the shaded regions in Figure~\ref{fig:results}.

%ffffffffffffffffffffffffffffffffffffffffffffffffffffffffffffffff
\begin{figure}[bt]
\centering
\includegraphics[width=0.46\textwidth]{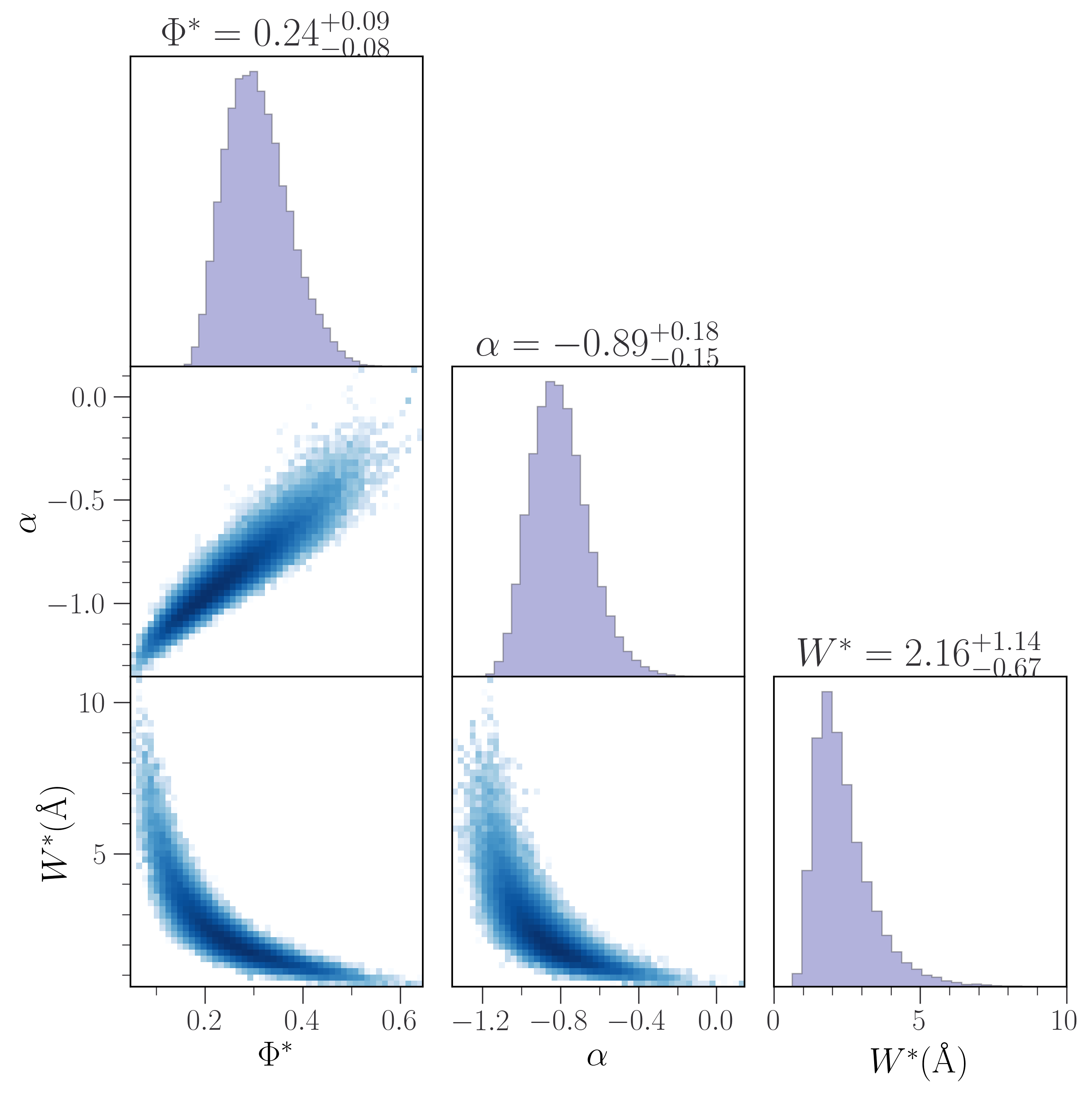} 
\caption{The 2D covariance between the fitted parameters and the 1D posterior uncertainty distribution from an MCMC analysis for $z=1.8$.  %This example is for the scenario of high-redshift weak {\MgII} absorber evolution (shown in Figure~\ref{fig:results}(a,b,c)), as per \citet{bosman17} and \citet{codor17}.  
Gaussian priors were assumed based on the mean values and standard deviations adopted from the maximum-likelihood least-squares solution. The final adopted parameters and their uncertainties for $z=1.8$ are presented above the respective panels.}
\label{fig:MCMCcorner}
\end{figure}
%ffffffffffffffffffffffffffffffffffffffffffffffffffffffffffffffff

In Figure~\ref{fig:MCMCcorner}, we show an example of the posterior bivariate uncertainty distributions for $\Phi^*(z)$, $W^*(z)$, and $\alpha(z)$ at $z=1.8$, corresponding to the evolution curves in Figures~\ref{fig:results}. The covariance between the weak-end slope and normalization is linear, in that a shallower (steeper) weak-end slope tends to be associated with a larger (smaller) normalization.  This strong anti-correlation is due to conservation of the total area under the equivalent width distribution function.  Area is increased for a steeper slope, $\alpha$, so normalization, $\Phi^*$, correspondingly decreases and, in the regime of the fitted parameters, the relationship is linear. The covariance between the characteristic $W^*$ and both $\alpha$ and $\Phi^*$ follow a quasi-inverse linear relationship, i.e., $y \!\propto\! 1/x$.
The value of $W^*$ is larger (smaller) for steeper (flatter) weak-end distribution function.  The same applies for the covariance between $W^*$ and $\Phi^*$, as the value of $W^*$ is larger (smaller) for smaller (larger) normalization.

These covariant relationships, which are specific to a Schechter function, explain the relationships between the Schechter parameters as they evolve with redshift (as manifest in Figure~\ref{fig:results}).  That is, at any given redshift, the best-fitted values of the fitted parameters are highly reflective of their covariant dependencies. This provides some insight into why, for example, at redshifts where the weak-end slope is steepest, the normalization is smallest, and vice versa.  The covariance also explains, for example, why the uncertainties in $W^*$ are largest when the weak-end slope is steepest and the normalization is smallest (as can be seen in the 2D distributions for $W^*$ vs.\ $\Phi^*$ and $W^*$ vs.\ $\alpha$).

\subsection{The Distribution Functions}

\begin{figure}[tb]
\centering
\includegraphics[width=0.45\textwidth]{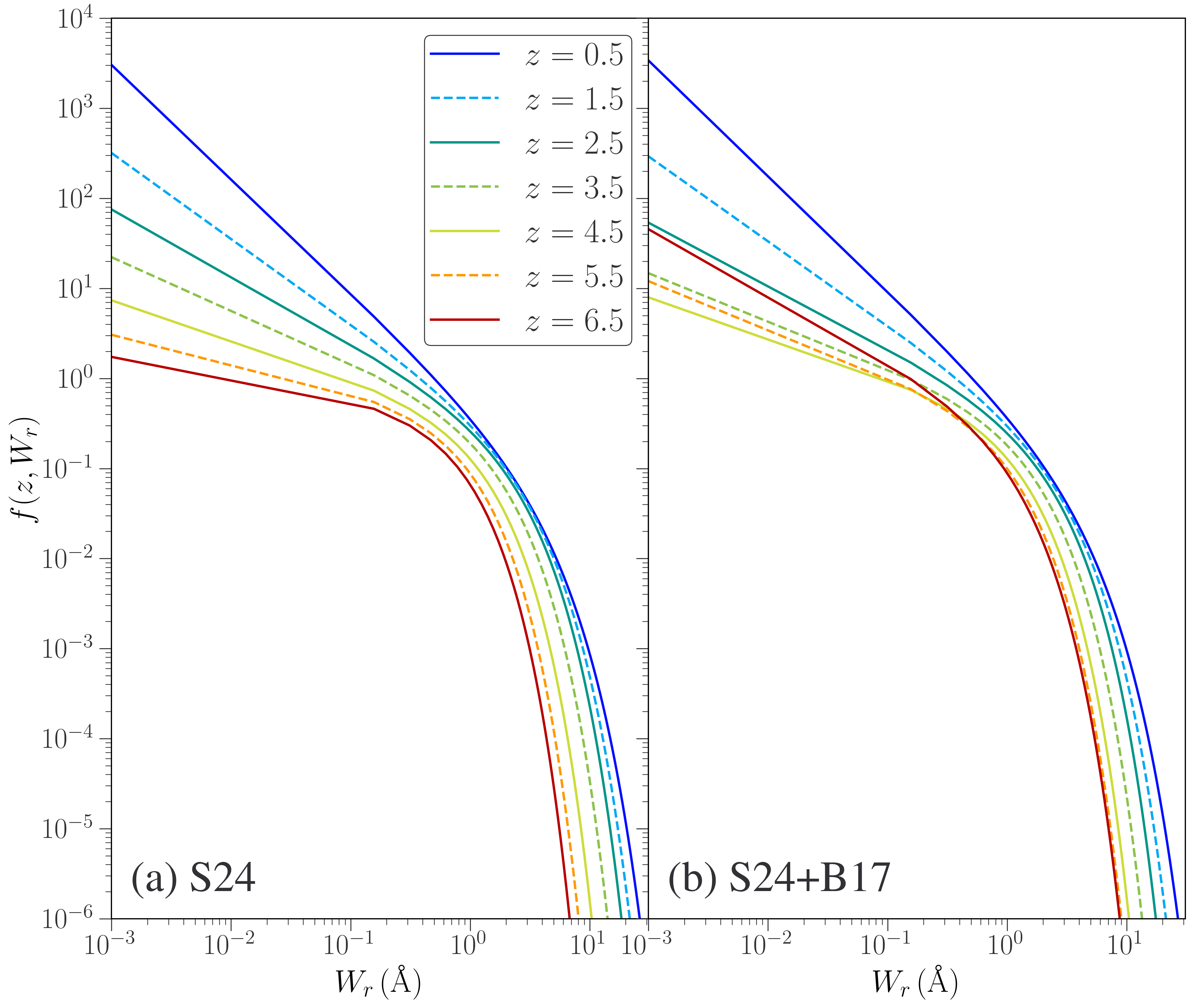} 
\caption{The {\MgII} Schechter equivalent width distribution functions, $f(z,W_r;{\bf a}_z)$, obtained using the apportioned integral method applied to the smoothed $d{\cal N}/dX$ measurements. These distribution functions are generated using the parameters listed in Table~\ref{tab:avector}. The curves are colored blue to red with increasing redshift over the range $0.5 \!\leq\! z \!\leq\! 6.5$ in intervals of $\Delta z=1$. Note that the curves cross as $W_r \!\rightarrow\! \infty$~{\AA}. (a) The S24 model. (b) The S24+B17 model. 
% (a)  The $f(z,W_r;{\bf a}_z)$ for the scenario of high-redshift weak {\MgII} absorber evolution as per \citet{bosman17} and \citet{codor17}.  (b)  
% These $f(z,W_r;{\bf a}_z)$ curves correspond to  of no evolution in the high-redshift weak {\MgII} absorbers as per \citet{sebastian23}.
}
\label{fig:TheSchechterfuncs}
\end{figure}

In Table~\ref{tab:avector}, we list the best-fit parameters that are illustrated in Figure~\ref{fig:results} over the redshift range $0.25 \!\leq\! z \!\leq\! 7.00 $ in steps of $\Delta z = 0.25$.  %We tabulated the parameters for both scenarios of high-redshift weak absorber evolution.
In Figure~\ref{fig:TheSchechterfuncs}, we present the best-fit {\MgII} equivalent width distribution functions determined from application of the apportioned integral method assuming Schechter functions.  The evolution of the functions is shown over the redshift range $0.5 \!\leq\! z \!\leq\! 6.5 $ in steps of $\Delta z = 1$.  The curves are colored according to redshift.  

%Figure~\ref{fig:TheSchechterfuncs}(a) shows the distribution functions for the scenario of high-redshift weak {\MgII} absorber evolution as per \citet{bosman17} and \citet{codor17}. Note that the steepest weak-end slopes are at $z \!=\!0.5$ and $z\!=\!6.5$ and the shallowest slopes are at $z\!=\!2.5$ and $z\!=\!3.5$.  This indicates that weak absorbers have their highest incidence during reionization, diminish during Comic noon, and then increase in incidence toward the present epoch. This is opposite to the strongest absorbers, which increase from reionization to their peak during Comic Noon, following which they diminish toward the present epoch.

%ffffffffffffffffffffffffffffffffffffffffffffffffffffffffffffffff
%ffffffffffffffffffffffffffffffffffffffffffffffffffffffffffffffff

%Figure~\ref{fig:TheSchechterfuncs}(b) shows distribution functions for  the scenario of no evolution in the high-redshift weak {\MgII} absorbers as per \citet{sebastian23}. 
%Though we refer to this as a no evolution scenario, 

%tttttttttttttttttttttttttttttttttttttttttttttttttttttttttttttttt
\begin{deluxetable*}{ccccccc}
\tablewidth{0pt}
\tablecaption{Fitted Parameters from Apportioned Integrals\label{tab:avector}}
\tablehead{
\multicolumn{7}{c}{}\\[-10pt]
\multicolumn{7}{c}{$f(z,W_r;{\bf a}_z) = (\Phi^*/W^*) (W/W^*)^{\alpha} \, e^{-W/W^*}$ } \\[5pt]
& \multicolumn{3}{c}{S24} & \multicolumn{3}{c}{S24+B17} \\[-6pt]
& \multicolumn{3}{c}{---------------------------------------------} 
& \multicolumn{3}{c}{---------------------------------------------}\\[-6pt]
\colhead{Redshift} &
\colhead{$\Phi^*(z)$} &
\colhead{$W^*(z)$,~{\AA}} &
\colhead{$\alpha(z)$} &
%\colhead{Redshift} &
\colhead{$\Phi^*(z)$} &
\colhead{$W^*(z)$,~{\AA}} &
\colhead{$\alpha(z)$} 
}
\startdata
${0.25}$ % -------------------------------------------
& $0.09^{+0.07}_{-0.04}$ & $3.44^{+3.79}_{-1.39}$ & $-1.38^{+0.14}_{-0.13}$ % s23
& $0.09^{+0.06}_{-0.04}$ & $3.61^{+3.83}_{-1.60}$ & $-1.40^{+0.14}_{-0.11}$ % s23 + b17
\\[-0.5pt]
${0.50}$ % -------------------------------------------
& $0.13^{+0.08}_{-0.06}$ & $2.92^{+2.31}_{-1.01}$ & $-1.26^{+0.15}_{-0.12}$ % s23
& $0.12^{+0.07}_{-0.05}$ & $3.01^{+2.23}_{-1.07}$ & $-1.28^{+0.13}_{-0.11}$ % s23 + b17
\\[-0.5pt]
${0.75}$ % -------------------------------------------
& $0.16^{+0.09}_{-0.06}$ & $2.60^{+1.17}_{-0.69}$ & $-1.16^{+0.14}_{-0.12}$ % s23
& $0.15^{+0.08}_{-0.05}$ & $2.65^{+0.86}_{-0.65}$ & $-1.17^{+0.13}_{-0.10}$ % s23 + b17
\\[-0.5pt]
${1.00}$ % -------------------------------------------
& $0.19^{+0.08}_{-0.06}$ & $2.40^{+1.59}_{-0.80}$ & $-1.08^{+0.13}_{-0.12}$ % s23
& $0.19^{+0.08}_{-0.06}$ & $2.41^{+1.45}_{-0.77}$ & $-1.08^{+0.13}_{-0.12}$ % s23 + b17
\\[-0.5pt]
${1.25}$ % -------------------------------------------
& $0.21^{+0.08}_{-0.07}$ & $2.24^{+1.41}_{-0.71}$ & $-1.01^{+0.12}_{-0.13}$ % s23
& $0.21^{+0.08}_{-0.07}$ & $2.23^{+1.34}_{-0.69}$ & $-1.00^{+0.13}_{-0.12}$ % s23 + b17
\\[-0.5pt]
${1.50}$ % -------------------------------------------
& $0.24^{+0.09}_{-0.07}$ & $2.12^{+1.14}_{-0.65}$ & $-0.94^{+0.14}_{-0.14}$ % s23
& $0.24^{+0.09}_{-0.07}$ & $2.08^{+1.14}_{-0.63}$ & $-0.93^{+0.14}_{-0.13}$ % s23 + b17
\\[-0.5pt]
${1.75}$ % -------------------------------------------
& $0.26^{+0.10}_{-0.08}$ & $2.01^{+1.12}_{-0.67}$ & $-0.88^{+0.15}_{-0.14}$ % s23
& $0.27^{+0.09}_{-0.08}$ & $1.95^{+1.02}_{-0.59}$ & $-0.86^{+0.14}_{-0.13}$ % s23 + b17
\\[-0.5pt]
${2.00}$ % -------------------------------------------
& $0.28^{+0.09}_{-0.08}$ & $1.91^{+0.94}_{-0.56}$ & $-0.83^{+0.14}_{-0.13}$ % s23
& $0.29^{+0.10}_{-0.08}$ & $1.82^{+0.86}_{-0.55}$ & $-0.80^{+0.14}_{-0.14}$ % s23 + b17
\\[-0.5pt]
${2.25}$ % -------------------------------------------
& $0.30^{+0.08}_{-0.08}$ & $1.80^{+0.84}_{-0.51}$ & $-0.78^{+0.14}_{-0.14}$ % s23
& $0.31^{+0.10}_{-0.09}$ & $1.70^{+0.76}_{-0.51}$ & $-0.74^{+0.15}_{-0.15}$ % s23 + b17
\\[-0.5pt]
${2.50}$ % -------------------------------------------
& $0.32^{+0.09}_{-0.08}$ & $1.69^{+0.78}_{-0.48}$ & $-0.74^{+0.15}_{-0.14}$ % s23
& $0.34^{+0.09}_{-0.09}$ & $1.58^{+0.69}_{-0.42}$ & $-0.69^{+0.14}_{-0.14}$ % s23 + b17
\\[-0.5pt]
${2.75}$ % -------------------------------------------
& $0.33^{+0.09}_{-0.08}$ & $1.57^{+0.67}_{-0.42}$ & $-0.69^{+0.15}_{-0.15}$ % s23
& $0.36^{+0.09}_{-0.08}$ & $1.47^{+0.55}_{-0.36}$ & $-0.64^{+0.15}_{-0.14}$ % s23 + b17
\\[-0.5pt]
${3.00}$ % -------------------------------------------
& $0.35^{+0.09}_{-0.08}$ & $1.46^{+0.57}_{-0.39}$ & $-0.65^{+0.16}_{-0.15}$ % s23
& $0.37^{+0.09}_{-0.09}$ & $1.36^{+0.47}_{-0.35}$ & $-0.59^{+0.15}_{-0.15}$ % s23 + b17
\\[-0.5pt]
${3.25}$ % -------------------------------------------
& $0.37^{+0.09}_{-0.08}$ & $1.34^{+0.51}_{-0.34}$ & $-0.61^{+0.18}_{-0.15}$ % s23
& $0.39^{+0.09}_{-0.09}$ & $1.25^{+0.47}_{-0.31}$ & $-0.55^{+0.16}_{-0.16}$ % s23 + b17
\\[-0.5pt]
${3.50}$ % -------------------------------------------
& $0.38^{+0.09}_{-0.09}$ & $1.23^{+0.46}_{-0.32}$ & $-0.57^{+0.19}_{-0.17}$ % s23
& $0.40^{+0.09}_{-0.09}$ & $1.16^{+0.42}_{-0.31}$ & $-0.51^{+0.18}_{-0.17}$ % s23 + b17
\\[-0.5pt]
${3.75}$ % -------------------------------------------
& $0.39^{+0.09}_{-0.09}$ & $1.13^{+0.43}_{-0.30}$ & $-0.53^{+0.21}_{-0.18}$ % s23
& $0.41^{+0.09}_{-0.09}$ & $1.07^{+0.37}_{-0.29}$ & $-0.48^{+0.19}_{-0.19}$ % s23 + b17
\\[-0.5pt]
${4.00}$ % -------------------------------------------
& $0.40^{+0.09}_{-0.09}$ & $1.03^{+0.40}_{-0.28}$ & $-0.50^{+0.22}_{-0.18}$ % s23
& $0.41^{+0.09}_{-0.09}$ & $1.00^{+0.37}_{-0.26}$ & $-0.46^{+0.21}_{-0.19}$ % s23 + b17
\\[-0.5pt]
${4.25}$ % -------------------------------------------
& $0.41^{+0.08}_{-0.09}$ & $0.95^{+0.37}_{-0.26}$ & $-0.46^{+0.25}_{-0.21}$ % s23
& $0.42^{+0.08}_{-0.09}$ & $0.93^{+0.36}_{-0.26}$ & $-0.44^{+0.23}_{-0.19}$ % s23 + b17
\\[-0.5pt]
${4.50}$ % -------------------------------------------
& $0.42^{+0.08}_{-0.09}$ & $0.87^{+0.34}_{-0.24}$ & $-0.42^{+0.28}_{-0.22}$ % s23
& $0.42^{+0.08}_{-0.09}$ & $0.88^{+0.35}_{-0.24}$ & $-0.43^{+0.25}_{-0.21}$ % s23 + b17
\\[-0.5pt]
${4.75}$ % -------------------------------------------
& $0.43^{+0.08}_{-0.09}$ & $0.81^{+0.33}_{-0.23}$ & $-0.39^{+0.30}_{-0.24}$ % s23
& $0.42^{+0.08}_{-0.09}$ & $0.84^{+0.36}_{-0.24}$ & $-0.44^{+0.29}_{-0.23}$ % s23 + b17
\\[-0.5pt]
${5.00}$ % -------------------------------------------
& $0.44^{+0.08}_{-0.08}$ & $0.75^{+0.31}_{-0.21}$ & $-0.35^{+0.32}_{-0.26}$ % s23
& $0.42^{+0.09}_{-0.09}$ & $0.81^{+0.35}_{-0.25}$ & $-0.45^{+0.32}_{-0.25}$ % s23 + b17
\\[-0.5pt]
${5.25}$ % -------------------------------------------
& $0.45^{+0.08}_{-0.09}$ & $0.70^{+0.31}_{-0.20}$ & $-0.32^{+0.34}_{-0.27}$ % s23
& $0.42^{+0.08}_{-0.10}$ & $0.79^{+0.37}_{-0.24}$ & $-0.47^{+0.37}_{-0.28}$ % s23 + b17
\\[-0.5pt]
${5.50}$ % -------------------------------------------
& $0.46^{+0.08}_{-0.08}$ & $0.66^{+0.28}_{-0.19}$ & $-0.29^{+0.37}_{-0.29}$ % s23
& $0.41^{+0.09}_{-0.10}$ & $0.77^{+0.38}_{-0.24}$ & $-0.51^{+0.38}_{-0.29}$ % s23 + b17
\\[-0.5pt]
${5.75}$ % -------------------------------------------
& $0.46^{+0.07}_{-0.08}$ & $0.62^{+0.26}_{-0.18}$ & $-0.27^{+0.40}_{-0.31}$ % s23
& $0.41^{+0.09}_{-0.10}$ & $0.77^{+0.38}_{-0.24}$ & $-0.55^{+0.42}_{-0.32}$ % s23 + b17
\\[-0.5pt]
${6.00}$ % -------------------------------------------
& $0.47^{+0.07}_{-0.08}$ & $0.59^{+0.25}_{-0.17}$ & $-0.24^{+0.42}_{-0.33}$ % s23
& $0.40^{+0.09}_{-0.11}$ & $0.77^{+0.39}_{-0.24}$ & $-0.60^{+0.45}_{-0.34}$ % s23 + b17
\\[-0.5pt]
${6.25}$ % -------------------------------------------
& $0.47^{+0.07}_{-0.08}$ & $0.56^{+0.24}_{-0.16}$ & $-0.22^{+0.45}_{-0.35}$ % s23
& $0.39^{+0.10}_{-0.12}$ & $0.78^{+0.42}_{-0.24}$ & $-0.66^{+0.47}_{-0.35}$ % s23 + b17
\\[-0.5pt]
${6.50}$ % -------------------------------------------
& $0.48^{+0.07}_{-0.08}$ & $0.54^{+0.22}_{-0.15}$ & $-0.21^{+0.48}_{-0.36}$ % s23
& $0.37^{+0.10}_{-0.12}$ & $0.79^{+0.42}_{-0.24}$ & $-0.72^{+0.52}_{-0.37}$ % s23 + b17
\\[-0.5pt]
${6.75}$ % -------------------------------------------
& $0.48^{+0.07}_{-0.08}$ & $0.52^{+0.21}_{-0.14}$ & $-0.19^{+0.50}_{-0.38}$ % s23
& $0.36^{+0.10}_{-0.13}$ & $0.81^{+0.42}_{-0.23}$ & $-0.79^{+0.52}_{-0.39}$ % s23 + b17
\\[-0.5pt]
${7.00}$ % -------------------------------------------
& $0.48^{+0.08}_{-0.08}$ & $0.51^{+0.21}_{-0.14}$ & $-0.18^{+0.52}_{-0.39}$ % s23
& $0.34^{+0.11}_{-0.14}$ & $0.83^{+0.49}_{-0.25}$ & $-0.86^{+0.55}_{-0.41}$ % s23 + b17
\\[2pt]
\enddata 
\end{deluxetable*}
%tttttttttttttttttttttttttttttttttttttttttttttttttttttttttttttttt

For both the S24 and S24+B17 models, the steepening of the weak-end slopes with decreasing redshift for $z< 4.5$ robustly indicates that the incidence of weak {\MgII} absorbers ($W_r < 0.3$~{\AA}) increases monotonically with the passage of cosmic time across Cosmic Noon through to the current epoch.  
For $z>4.5$, the S24+B17 model yields evolution such that there is a slowly decreasing incidence of weak absorbers as the universe emerges from  reionization, whereas the S24 model indicates little-to-no evolution in the weak absorber incidence as the universe emerges from reionization.   

These distribution functions are also consistent with the peak in the cosmic incidence of  strong {\MgII} absorbers ($W_r \!\geq\! 1$~{\AA}) at Cosmic Noon (Type A evolution). At first glance, this might seem counter intuitive given that the areas under the plotted curves for $W_r \!\geq\! 1$~{\AA} {\it visually appear\/} to increase monotonically with decreasing redshift. However, the apportioned integrated area under the distribution function is over the domain $1 \!\leq\! W_r \!\leq\! \infty$~{\AA} at each redshift. The peak of the cosmic incidence of the strongest absorbers at Cosmic Noon derives from the trend between normalization and the characteristic equivalent width as a function of redshift. The combined behavior, when the integration is carried out to infinity, yields the peak at Cosmic Noon.

%The constraints placed on the apportioned integral method by the measured $d{\cal N}/dX$ of the strong absorbers as a function of redshift are identical for both scenarios of high-redshift weak absorber evolution. However, the very different $d{\cal N}/dX$ values of the weak absorbers for the evolution and no-evolution scenarios resulted in very different fitted Schechter parameters for $z\!>\!2$.  This is true not only for the weak-end slope, $\alpha(z)$, but also for the normalization and characteristic equivalent width, $\Phi^*(z)$ and $W^*(z)$, respectively. The very different evolution in these parameters for $z\!>\!2$ can be see in Figure~\ref{fig:results}.  A consequence of these different parameters is that, even for strong absorbers ($W_r \!\geq\! 1$~{\AA}), the amplitude and shapes of the Schechter distribution functions differ between the scenarios of high-redshift weak absorber evolution. This can be seen in Figure~\ref{fig:TheSchechterfuncs}.

%ffffffffffffffffffffffffffffffffffffffffffffffffffffffffffffffff
\begin{figure}[thb]
\centering
\includegraphics[width=0.46\textwidth]{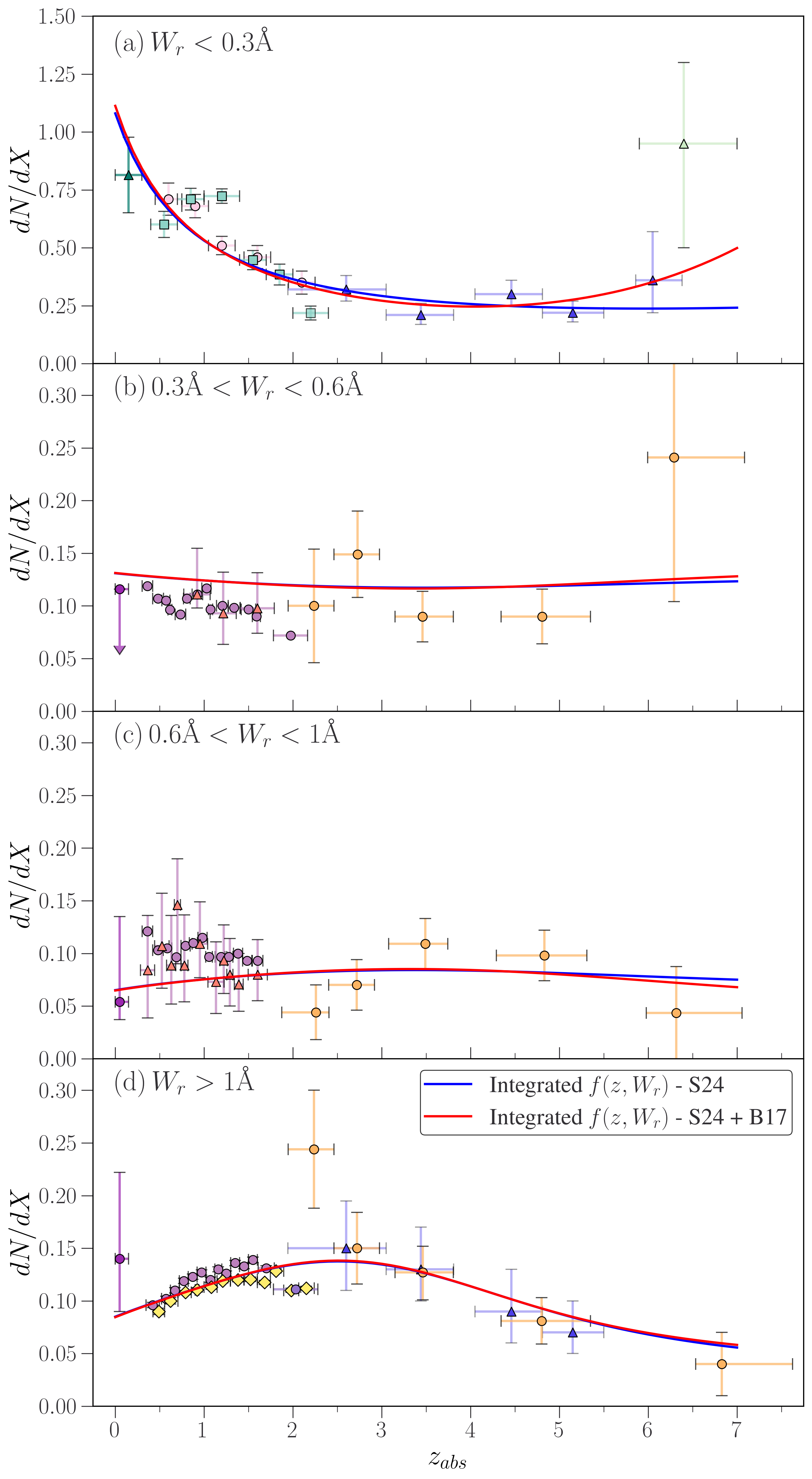} 
\caption{The self-consistency test of the apportioned integral method.  The curves are the ``predicted'' $d{\cal N}/dX$ obtained by computing the apportioned integrals (Eqs.~\ref{eq:theconstraintfunc} and \ref{eq:solvethis}) of the parameterized equivalent width distribution functions, $f(z,W_r;{\bf a}_z)$.  The data points are identical to the data in Figure~\ref{fig:MgIIdNdX}. These curves are not fits to the $d{\cal N}/dX$ measurements; they are derived by integrating apportioned integrals of the parameterized equivalent width distribution functions presented in Table~\ref{sec:ahistory} that were obtained using the apportioned integral method.}
\label{fig:PredictiondNdX}
\end{figure}
%ffffffffffffffffffffffffffffffffffffffffffffffffffffffffffffffff

\subsection{Integrating the Parameterized Distribution Functions}

To demonstrate the distribution functions tabulated in Table~\ref{tab:avector} recover (and can predict) $d{\cal N}/dX$ measurements, we computed the apportioned integrals (Eqs.~\ref{eq:theconstraintfunc} and \ref{eq:solvethis})  under the parameterized distributions as a function of redshift (some of which are shown in Figure~\ref{fig:TheSchechterfuncs}) and compared the ``predicted'' $d{\cal N}/dX$ to the measured data.  As can be seen in  Figure~\ref{fig:PredictiondNdX}, the best-fitted equivalent width distribution functions provide a good description of the data for both the S24 and S24+B17 models.  

This is an important test because it establishes that the apportioned integral method (1) yields distribution functions that are consistent with and can reproduce the measured $d{\cal N}/dX$ data,  (2) is flexible enough to yield such functions even for different manifestations of $d{\cal N}/dX$ evolution, and (3) can be deployed for developing models of equivalent width distribution functions capable of predicting $d{\cal N}/dX$ values for any desired equivalent range and redshift.  This demonstration does not prove that the parameterized distribution functions and their evolution reflect the true underlying evolution of the distribution of {\MgII} absorbers, only that the apportioned integral method yields distributions that succeed in describing the measured $d{\cal N}/dX$ data.  The primary utility of the method then is to provide working models that provide a complete and self-consistent quantitative description of the evolution of the equivalent widths.  However, given the parameterized distribution functions determined using the apportion integral method, it is of interest to consider how they fit into theoretical models and the observed universe of absorbers.

\section{Discussion}
\label{sec:discussion}

Measurements of global characterization of astrophysical quantities are key for developing and refining the physics applied in hydrodynamic cosmological simulations that model the evolution of galaxies \citep[e.g.,][]{ford13, churchill15,  oppenheimer18, kacprzak19, peeples19, appleby21}, the intergalactic medium \citep[e.g.,][]{dave99, richter06, tepper-garcia11, oppenheimer12, nelson18}, and reionization astrophysics \citep[e.g.,][]{oppenheimer09, doughty18, finlator18, gnedin22, puchwein23}. 

For galaxies, these characterizations include the cosmic star formation rate density \citep[][]{madau14}, the stellar mass to halo mass function \citep[e.g.,][]{behroozi13}, the luminosity function of galaxies \citep[e.g.,][]{bouwens21}, and the main-sequence of star forming galaxies \citep[e.g.,][]{popesso23}.  For absorbers, these include the mass densities, the redshift and co-moving line of sight path densities, and the equivalent width and column density distributions of various populations absorbers \citep[e.g.,][]{danforth16, mathes17, dodorico22,  davies23-survey}. For example, distribution functions of absorber strengths have been shown to place meaningful constraints for discriminating between competing models of galactic outflows \citep[e.g.,][]{oppenheimer06, bird16, rahmati16}. It is our hope that the apportioned integral method may find application in providing more accurate constraints on the redshift evolution of equivalent width and column density distributions from the large body of path density measurements obtained by individual absorber surveys.  

\subsection{A Brief Muse on {\MgII} Absorber Evolution}

The characteristics of the population of {\MgII} absorbers is of particular interest because of the well-established association with the circumgalactic medium (CGM) at low redshifts \citep[][]{bb91, sdp94, churchill05, zibetti05, bouche06, chen10-mgii, kacprzak10,  nielsen13-magiicat2, dutta20, huang21}.  Though direct association between {\MgII} absorbers and galaxies is not firmly established for $z\!\geq\!2$, \citep[however, see][]{nielsen20, nielsen22}, an association with galaxies from the era of the earliest galaxies across Cosmic Noon is expected \citep[e.g.,][]{keating14}. One observational clue is that the path density of strong {\MgII} absorbers ($W_r \geq 1.0$~{\AA}) evolves with redshift in direct step with the cosmic star formation rate from $z \!\sim\! 6$ to $z \!\sim\! 0$, which increases from the epoch of reionization, peaks during Cosmic Noon, and declines to the present epoch \citep{matejek12, zhu13, chen.s17}, i.e., classic Type~A evolution.  We would also note that a substantial fraction of the strongest of these absorbers are hypothesized to reflect stellar driven winds \citep[e.g.,][]{bond01-winds, bond01-bubbles, rubin10, nestor11}. However, the strong absorbers alone do not provide a complete picture of the universe of {\MgII} absorbing gas structures. 

Interestingly, the path density of intermediate strength {\MgII} absorbers (i.e., $0.3 \!\leq\! W_r \!<\! 1.0$~{\AA}) is consistent with little-to-no cosmic evolution from reionization through Cosmic Noon to the present epoch. This leads one to question whether the array of localized creation and destruction mechanisms in the halos of galaxies \citep[e.g.,][]{maller04, faucher-giguere23, tan23} balance in some kind of cosmic non-evolving equilibrium even though there are dramatic changes in the star formation rates, ionization conditions, galaxy merging rates and structure growth across cosmic time. This would be a remarkable situation, as these absorbers are believed to arise in low-entropy, dynamically turbulent, multi-phase environments closely coupled to the stochastic processes of galaxy evolution \citep[e.g.,][]{mccourt18, keller20, esmerian21, lochhaas21, pandya23}. This would imply a chaotic equilibrium conspiring to create a steady-state cosmic incidence of {\MgII} absorbers confined to this equivalent width range in a universe where host-galaxies systematically evolve in their global stellar and halo mass properties \citep[e.g.,][]{behroozi13, forster20, lyu23}.

As for the weak {\MgII} absorbers (i.e., those with $W_r \!<\! 0.3$~{\AA}), their path density evolution is well characterized for $z \!<\! 4$. From Cosmic Noon to the present epoch their cosmic incidence increases, while over the same cosmic time period the incidence of stronger systems is decreasing. This could imply that the strong absorbing structures are fragmenting into weak absorbing structures \citep[e.g.,][]{mccourt18}. Evolution in the kinematics of strong ($W_r \geq 1$~{\AA}) and intermediate ($0.3 \leq W_r <1$~{\AA}) absorbers is not inconsistent with a scenario in which some intermediate absorbers represent ``transition'' gas structures between strong and weak absorbers in a post-Cosmic Noon universe \citep[][]{churchill03, mshar07}.  However, this process would need to occur such that the cosmic incidence of intermediate absorbers  remains constant. Alternatively, weak absorber evolution could be spatially independent of or physically decoupled from strong absorber evolution in that the balance between the creation and destruction rates of gas structures that yield weak absorption is favoring increasingly higher creation rates as the universe evolves post-Cosmic Noon. This implies that weak absorbers arise in astrophysical environments that differ from those of strong absorbers \citep[e.g.,][]{churchill00-archiveI, churchill00-archiveII, churchill03, rigby02, narayanan08}.

Both scenarios could, in principle, manifest as evolution of the post-Cosmic Noon equivalent width distribution function precisely as we have determined using the apportioned integral method, i.e., the weak-end slope, $\alpha$, continually gets steeper toward the present epoch while the cosmic incidence of strong absorbers declines. Remarkably, the evolution in the characteristic $W^*$ is increasing, providing additional insight into the evolution of the intermediate and strongest absorbers. This evolution of $W^*$ indicates that structures giving rise to strong absorbers tend to have larger equivalent widths as the present epoch is approached while their lowered incidence is a consequence of the decreasing normalization, $\Phi^*$.  That is, the {\it evolution of the equivalent width distribution as measured herein suggests there are fewer of these ``larger'' structures as the post-Cosmic Noon universe evolves, but that they progressively become stronger absorbers on average}. 

%For $z \!>\! 2.5$, i.e., from reionization through Cosmic Noon, the independent measurements of $d{\cal N}/dX$ for the weakest absorbers are not in full agreement. If weak absorber evolution is marked by a decreasing incidence from reionization to Cosmic Noon, as per \citet{bosman17} and \citet{codor17}, and accounting for the increasing incidence of this population from Cosmic Noon to the present epoch \citep{narayanan07, mathes17}, we would be seeing that the weak population of {\MgII} evolves {\it opposite\/} to that of the strong {\MgII} absorbers \citep{matejek12, zhu13, chen.s17}.  That is weak absorbers exhibit Type~V evolution while strong systems exhibit Type~A evolution. In this case, the equivalent width distribution is characterized by a steep slope during reionization that steadily flattens to Cosmic Noon and then steadily gets progressively steeper towards the present epoch until it is as steep as it was during reionization. This scenario implies mechanisms that create weak {\MgII} absorber structures were in place at $z \!\sim\! 7$, following which they steadily diminished as the universe became more active, and then steadily increased as the universe became more quiescent. That the minimum in the weak population is aligned with the maximum in the global star formation and activity in galaxies marked by the Cosmic Noon period \citep[][]{vandevoort11, behroozi13, madau14, forster20} is highly suggestive of a causal connection. 

\subsection{High Redshift Weak {\MgII} Absorbers}

%On the other hand, 
If weak absorbers over the range $z\!\sim\! 6.5$ to $z \!\sim\! 2.5$ are characterized by the S24 model, then (1) the cosmic incidence of these structures does not vary substantially from the epoch of reionization to Cosmic Noon, and (2) the weak-end slope, $\alpha(z)$, remains relative shallow throughout the pre-Cosmic Noon universe while progressively increasing the rate at which it is steepening throughout the post-Cosmic Noon universe. This evolutionary scenario would imply that the balance between astrophysical mechanisms that create and destroy weak absorbing structures would not induce strong evolution until after the universe passed through Cosmic Noon. 

\begin{figure}[htb]
\centering
\includegraphics[width=0.48\textwidth]{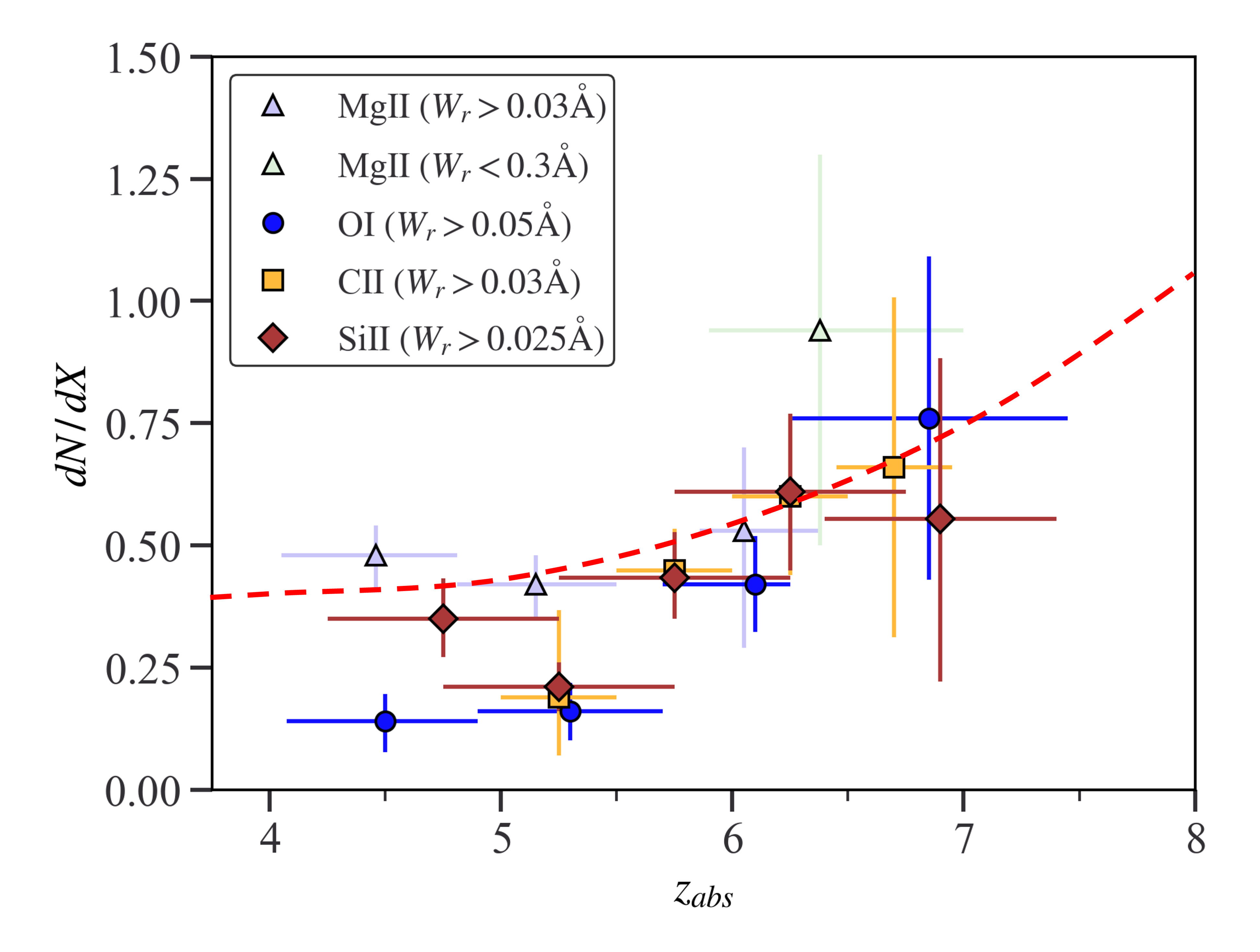} 
\caption{Measured $d{\cal N}/dX$ at $4.5 \leq z \leq 7$ for the low-ionization ions {\OI} for $W_r \geq 0.05$~{\AA}, and {\CII}, {\SiII}, and {\MgII} for $W_r \geq 0.03$~{\AA}.  The {\OI} data are from \citet{becker19}, the {\CII} and {\SiII} are from the calculations of \citet{christiansen23} using the data of \citet{davies23-survey}, and the {\MgII} data are from \citet{sebastian23} using the data of \citet{davies23-survey}. Also shown is the $z=6.4$ data point from \citet{bosman17} for $W_r<0.3$~{\AA}.  The dashed curve is the S24+B17 model from Eq.~\ref{eq:dNdx-fit-V} scaled to the {\MgII} measurements of \citet{sebastian23}.  All low-ionization ions show a trend in which their cosmic incidence declines as the universe emerges from the epoch of reionization.}
\label{fig:dNdXhighz}
\end{figure}

% see their Fig.~12
However, there are reasons to embrace the scenario of Type~V evolution in weak {\MgII} absorbers, as loosely represented by the S24+B17 model. A decreasing $d{\cal N}/dX$ from $z\sim7$ to $z \sim 4.5$ for weak {\MgII} would be consistent with the observed path densities evolution at $z \!>\!4.5$ for weak absorbers of other low ions, such as {\OI}, {\CII}, and {\SiII},  \citep[see][]{christiansen23}. As shown in Figure~\ref{fig:dNdXhighz}, the path density of {\OI} absorbers with minimum threshold $W_r \!>\! 0.05${\AA} \citep{becker19, sebastian23} and of weak {\CII} and {\SiII} absorbers with minimum threshold $W_r \!>\! 0.03${\AA} \citep{davies23-survey, christiansen23, sebastian23}, when taken together, all suggest a globally declining trend in the incidence of low-ionization ions from $z\sim 7$ to $z\sim 4.5$. Within uncertainties, the cosmic incidence of {\MgII} absorbers with minimum threshold $W_r \!>\! 0.03${\AA} \citep{sebastian23} hint at the same declining trend. Although the \citet{bosman17} measurement of weak {\MgII} at $z\sim 6.4$ is quite uncertain, it is also suggestive of a decreasing cosmic incidence of low-ionization metals from $z\sim 7$ to $z\sim 4.5$. 
%It is interesting to note that the lack of evolution in weak {\MgII} absorbers as measured by \citet{sebastian23} is derived from the survey results of \citet{davies23-survey}, for which weak {\CII} and {\SiII} measurements do indicate evolution \citep[][]{christiansen23}. 

%ffffffffffffffffffffffffffffffffffffffffffffffffffffffffffffffff
%ffffffffffffffffffffffffffffffffffffffffffffffffffffffffffffffff

Conversely, the weak absorber evolution in $d{\cal N}/dX$ measured for the higher ions {\CIV} and {\SiIV} are observed to {\it increase\/} from $z \!\sim\! 7$ to $z \!\sim\! 4.5$ \citep{dodorico22, davies23-survey}. Considering additional observations and theoretically-based arguments \citep{finlator16, becker19, cooper19, bosman21, bosman22, christiansen23}, the consensus is that the body of evidence indicates changes in the ionization state of the IGM at these redshifts and not changes in gas-phase metallicity.  More specifically, models suggest that the CGM and IGM at $5 \leq z \leq 7$ had a softer, more inhomogeneous ionizing background relative to $z \!<\! 5$ in the pre-Cosmic Noon universe \citep[e.g.,][]{kulkarni19, faucher-giguere20, puchwein23}.  Under photoionization conditions, the Mg$^+$ ion is created when $I \!=\! h\nu \!\geq\! 7.65$~eV photons ionize Mg$^0$ and is destroyed into Mg$^{+2}$ via ionization by $I \!\geq\! 15.04$~eV photons.  These ionization thresholds bracket that of neutral hydrogen, which is ionized starting at $I \!=\! 13.59$~eV. Neutral oxygen is ionized at $I \!=\! 13.61$~eV. Singly ionized carbon (C$^+$) is created at $I \!=\! 11.26$ and ionized at $I \!=\! 24.38$~eV. Singly ionized silicon (Si$^+$) is created at $I \!=\! 8.15$ and ionized at $I \!=\! 16.35$~eV.  Naive expectations is that the path density evolution of these low-ionization ions should trace one another (which is indeed seen for {\OI}, {\CII}, and {\SiII}).

In particular, given that both Si$^+$ and Mg$^+$ ions are $\alpha$-group elements formed via core-collapse supernovae and have very similar ionization potentials, it would be even more securely expected that their evolution would closely track. However, their evolution could differ if their photoionization rates and/or recombination rates differ.  For example, the dielectronic recombination rate for Mg$^+$ can be 5-10 times higher than for Si$^+$ \citep[e.g.,][]{verner96, badnell03, draine11}. The slightly lower ionization potential of Mg$^{+}$ means that in optically thin gas it can ionize away in lower density gas than does Si$^{+}$; but in optically thick and/or higher density gas the ionization of the two ions is in lock step \citep[e.g.,][]{oppenheimer18}.

Expectations for $z \!\sim\! 6$ are that, for low-ionization ions such as {\OI} and {\MgII} with $W_r \!\sim\!0.1$~{\AA}, the absorption arises in overdensities on the order of 80--100 \citep[e.g.,][]{keating14, becker19}. This places the detection of these absorbers (and presumably weak {\SiII} as well) firmly in the regime of the high redshift CGM \citep[e.g.,][]{simcoe12}.  
The nature and observational consequences of the background ionizing spectrum during reionization are not highly certain \citep[e.g.,][]{finlator16, finlator18, puchwein23} and constraints on the mean free path of ionizing photons and on the ionization fraction of hydrogen from the observed evolution of {\Lya} optical depths indicate a broadly inhomogeneous state for the higher-redshift IGM \citep[e.g.,][]{bosman21, bosman22}.  

Given that the {\SiII} and {\MgII} column densities behave virtually identically as a function of impact parameter and their CGM covering fractions are virtually identical in low-redshift simulated galaxies \citep[e.g.,][]{oppenheimer18}, if $d{\cal N}/dX$ of {\MgII} does not trace that of {\SiII}, we would need to infer that the relative ionization balance of the Mg$^+$ and Si$^{+}$ ions is substantially different during reionization than in the post-Cosmic Noon universe; that is, their {\it relative\/} column densities, extent around galaxies, and CGM covering fractions would be very different during the epoch of reionization.  However, if such stark differences were manifest in the relative evolution of {\SiII} and {\MgII} due to ionization, such differences would be expected to be even more accentuated between {\SiII} and {\CII} due to the much higher ionization potential of the C$^{+}$ ion. The current data do not appear to support that. 

We believe that a lack of evolution for weak {\MgII} absorbers as the universe emerges from the epoch of reionization is difficult to reconcile with observations and theoretical expectations. In light of the above observational and theoretical considerations, we expect the co-moving line of sight path density of weak {\MgII} absorbers should evolve in lock step with the observed decrease in the co-moving line of sight path density of weak {\CII} and {\SiII} as the universe emerges from the epoch of reionization.
Such evolution is tentatively suggested by the elevated weak {\MgII} incidence measured by \citet{bosman17} at $z \sim 6.4$. This would favor the S24+B17 parametrized model of $f(z,W_r;{\bf a}_z)$ presented in Figure~\ref{fig:results}, Figure~\ref{fig:TheSchechterfuncs}(b) and Table~\ref{tab:avector}.

\subsection{What About a {\MgII} Forest?}

Whether weak {\MgII} absorbers exhibit Type V evolution, and therefore elevated cosmic incidence at $z \geq 7$, may have implications for the existence and detection of a {\MgII} forest in the IGM at $z \!\sim\! 7.5$  \citep{oh02, hennawi21}.  If weak absorbers exhibit Type~V evolution and the equivalent width distribution function evolves as shown for the S24+B17 model in Figure~\ref{fig:TheSchechterfuncs}(b), then conditions for weak {\MgII} in the high-redshift CGM would hint at a ubiquity of metal enrichment and lend confidence that a {\MgII} forest, if it exists, should be detectable \citep{hennawi21}. 

Though this forest was not detected in recent ground-based spectra \citep{tie23}, it is predicted to be detected in the much more sensitive spectra from {\it JWST}/NIRSpec observations \citep[see][]{hennawi21, tie23}.  On the other hand, if the evolution reflects little-to-no evolution in the cosmic incidence for $z \geq 7$ as characterized by the S24 model and shown in Figure~\ref{fig:TheSchechterfuncs}(a), then a detectable weak {\MgII} forest may be more challenging than currently predicted.  In their unsuccessful search for the {\MgII} forest, \citet{tie23} were able to identify several CGM weak {\MgII} absorbers and to determine an upper limit of $\hbox{[Mg/H]} < -3.7$ (95\% confidence) for the IGM at their median redshift, $z \!=\! 6.5$.

\section{Conclusion}
\label{sec:conclude}

We conducted a meta-analysis of the co-moving line-of-sight path density, $d{\cal N}/dX$, of {\MgII} absorbers with the goal of characterizing the shape and evolution of the {\MgII}~$\lambda 2796$ equivalent width distribution function for $W_r \geq 0.03$~{\AA} over the redshift range $0 \!\leq\! z \!\leq\! 7$. We implemented a novel method in which we integrated the apportioned areas under parameterized distribution functions and equated these areas to the measure $d{\cal N}/dX$ in given equivalent width ranges as a function of redshift. We call this the apportioned integral method (AIM).  Applying this method, we found that a Schechter function provides a good description of the distribution functions at all redshifts and measured the full characterization by obtaining the maximum likelihood parameters, i.e., the normalization $\Phi^*(z)$, the characteristic equivalent width, $W^*(z)$, and the weak-end power-law slope, $\alpha(z)$, as a function of redshift.

Due to uncertainty in the measured $d{\cal N}/dX$ of weak absorbers ($W_r < 0.3$~{\AA}) for $z \sim 6$, we studied two evolutionary scenarios. In the first (called the S24 model), the weak {\MgII} cosmic incidence is non-evolving from reionization to Cosmic Noon and then increases to the present epoch. In the second (called the S24+B17 model), the weak {\MgII} cosmic incidence exhibits Type~V evolution; the incidence is high at $z \!\sim\! 7$ during reionization, declines to its minimum at $z \!\sim\! 2.5$ during Cosmic noon, and then increases again to the present epoch ($z \!\sim\! 0$). For both scenarios, we determined $\Phi^*(z)$, $W^*(z)$, and $\alpha(z)$ by applying the apportioned integral method.  We characterized the covariance in the maximum likelihood parameters and estimated their correlated uncertainties using Markov-Chain Monte Carlo (MCMC) modeling.  We plotted the redshift evolution of the parameters in Figure~\ref{fig:results}. We tabulated the best-fit parameters and their uncertainties in Table~\ref{tab:avector} as a function of redshift from $z=0.25$ to $z=7.00$ in steps of $z=0.25$.  In Figure~\ref{fig:TheSchechterfuncs}, we overplotted the best-fit {\MgII} equivalent width distribution functions for redshifts $0.5 \!\leq\! z \!\leq\! 6.5$ in steps of $\Delta z \!=\! 1$.  

We consider the evolution in our best-fit equivalent width distribution from Cosmic Noon to the present epcoh to be robust, as there is no tension in the $d{\cal N}/dX$ measurements following Cosmic Noon.  For the pre-Cosmic Noon era the two adopted models of the cosmic incidence (S24 and S24+B17)  yield quite different evolution.  If weak {\MgII} absorbers evolve away from $z \!\sim\! 7$ to Cosmic Noon, the normalization $\Phi^*$ exhibits Type~A evolution and $\alpha$ also evolves such that it is relatively steep during reionization, flattens toward Cosmic Noon, and then steepens again toward the present epoch. If  weak {\MgII} absorbers do not evolve from $z \!\sim\! 7$ to Cosmic Noon, $\Phi^*$ steadily declines, $W^*$ steadily increases, and $\alpha $ steadily steepens from reionization to the present epoch.

We performed two validation tests of the apportioned integral method. In the first, we showed that the method successfully reproduces the measured Type~A evolution of $W^*(z)$ over the redshift range $0.3 \!\leq\! z \!\leq\! 6.4$ for $W_r \!>\! 0.3$~{\AA} assuming an exponential equivalent width distribution function \citep[see][]{chen.s17}.  In the second, we showed that the apportioned areas under the maximum likelihood parameterized Schechter equivalent width distribution functions successfully reproduce the measured $d{\cal N}/dX$ data for the full redshift range $0 \!\leq\! z \!\leq\! 7$. This demonstrated that the distribution functions provide an accurate representation of the evolution of {\MgII} absorbers consistent with the measured cosmic path densities and that the parameterized distribution functions can be employed to predict $d{\cal N}/dX$ for any equivalent width minimum threshold or finite equivalent width bin for $W_r \geq 0.03$~{\AA} at any redshift in the range $0 \leq z \leq 7$.

The apportioned integral method can be easily applied to any population of absorber for which path densities are measured across multiple finite equivalent width ranges and/or a series of different minimum equivalent width thresholds.  Furthermore, the apportioned integral method has the potential to measure the equivalent width distribution function with high precision. First, equivalent width measurements are invariant to spectral resolution, so the measurements are independent of the specifications of the instrument.  Second, path density measurements already account for the survey sizes and survey-specific detection completeness and redshift-dependent sensitivity functions. Third, the method is highly flexible in that an apportioned integral can be uniquely specified for any equivalent width range or minimum threshold, even redundant, or partially overlapping ranges or threshold. Simply put, the more surveys and the deeper their equivalent width detection thresholds, the better.  All that is required are trustworthy path density measurements with small uncertainty measurements. 

This work highlights the need for robust surveys of {\MgII} absorbers at $z>5$, especially those that can achieve detection thresholds down to $W_r \simeq 0.03$~{\AA} and less than 30\% one-sided uncertainties in $d{\cal N}/dX$.   As path density measurements continue to accumulate, we can continually improve results from the apportioned integral method for {\OI} {\CII}, and {\SiII} absorbers.  The apportioned integral method could easily be applied to column density distribution functions using $dN/dX$ measurements parsed by column density bins and/or minimum thresholds. For example, the relative redshift evolution of the distribution functions of {\CIII} to {\CIV} and {\CIV} to {\SiIV} could provide constraints on the evolution of the ultraviolet ionizing background radiation \citep[e.g.,][]{songaila96, songaila98, songaila05, giroux97, schaye03, simcoe04, vasiliev14, boksy15, dodorico22} during the epoch of {\HeII} reionization at $3 \!\leq\! z \!\leq\! 5$ \citep[e.g.,][]{nath96, fechner06, fechner07, bolton09, mcquinn09, schmidt17}.

With increased sensitivities to weaker absorbers and the proper choice of comparative absorber populations, the method may be very helpful at discriminating various ionization backgrounds and their degrees of inhomogeneity in theoretical treatments of reionization for $z \!\sim\! 7$ to $z \!\sim\! 5$ \citep[e.g.,][]{doughty18}.  If $d{\cal N}/dX$ of {\FeII} is ever measured, the comparative redshift evolution of the distribution functions may be useful for tracking global average [$\alpha$/Fe] abundance patterns as function of redshift \citep[e.g.,][]{rodriguez12, zou21}. 
In future work, we plan to apply the apportioned integral method to constrain the shape and redshift evolution of the {\CIV} equivalent width distribution function, for which there exists path density measurements virtually as complete as those for {\MgII} absorbers.

\begin{center}
ACKNOWLEDGMENTS
\end{center}

% The authors appreciate the comments from the anonymous referee, % which helped improve the manuscript. 
We thank Alma Sebastian for discussions about their measurements and for sharing them prior to publication.  A.A. acknowledges the support through a William Webber Fellowship administered by the Department of Astronomy at New Mexico State University. Parts of this research were supported by the Australian Research Council Centre of Excellence for All Sky Astrophysics in 3 Dimensions (ASTRO 3D), through project number CE170100013.
 
%\newpage

\bibliographystyle{aasjournal}  
\bibliography{main}

\begin{thebibliography}{}
\expandafter\ifx\csname natexlab\endcsname\relax\def\natexlab#1{#1}\fi
\providecommand{\url}[1]{\href{#1}{#1}}
\providecommand{\dodoi}[1]{doi:~\href{http://doi.org/#1}{\nolinkurl{#1}}}
\providecommand{\doeprint}[1]{\href{http://ascl.net/#1}{\nolinkurl{http://ascl.net/#1}}}
\providecommand{\doarXiv}[1]{\href{https://arxiv.org/abs/#1}{\nolinkurl{https://arxiv.org/abs/#1}}}

\bibitem[{{Abbas} {et~al.}(2024){Abbas}, {Churchill}, {Kacprzak}, {Lidman},
  {Guatelli}, \& {Bellstedt}}]{abbas24}
{Abbas}, A., {Churchill}, C.~W., {Kacprzak}, G.~G., {et~al.} 2024, \apj, 966,
  242, \dodoi{10.3847/1538-4357/ad35cc}

\bibitem[{{Appleby} {et~al.}(2021){Appleby}, {Dav{\'e}}, {Sorini},
  {Storey-Fisher}, \& {Smith}}]{appleby21}
{Appleby}, S., {Dav{\'e}}, R., {Sorini}, D., {Storey-Fisher}, K., \& {Smith},
  B. 2021, \mnras, 507, 2383, \dodoi{10.1093/mnras/stab2310}

\bibitem[{{Badnell} {et~al.}(2003){Badnell}, {O'Mullane}, {Summers}, {Altun},
  {Bautista}, {Colgan}, {Gorczyca}, {Mitnik}, {Pindzola}, \&
  {Zatsarinny}}]{badnell03}
{Badnell}, N.~R., {O'Mullane}, M.~G., {Summers}, H.~P., {et~al.} 2003, \aap,
  406, 1151, \dodoi{10.1051/0004-6361:20030816}

\bibitem[{{Bahcall} \& {Peebles}(1969)}]{bahcall-peebles69}
{Bahcall}, J.~N., \& {Peebles}, P.~J.~E. 1969, \apjl, 156, L7,
  \dodoi{10.1086/180337}

\bibitem[{{Becker} \& {Bolton}(2013)}]{becker13}
{Becker}, G.~D., \& {Bolton}, J.~S. 2013, \mnras, 436, 1023,
  \dodoi{10.1093/mnras/stt1610}

\bibitem[{{Becker} {et~al.}(2019){Becker}, {Pettini}, {Rafelski}, {D'Odorico},
  {Boera}, {Christensen}, {Cupani}, {Ellison}, {Farina}, {Fumagalli},
  {L{\'o}pez}, {Neeleman}, {Ryan-Weber}, \& {Worseck}}]{becker19}
{Becker}, G.~D., {Pettini}, M., {Rafelski}, M., {et~al.} 2019, \apj, 883, 163,
  \dodoi{10.3847/1538-4357/ab3eb5}

\bibitem[{{Behroozi} {et~al.}(2013){Behroozi}, {Wechsler}, \&
  {Conroy}}]{behroozi13}
{Behroozi}, P.~S., {Wechsler}, R.~H., \& {Conroy}, C. 2013, \apj, 770, 57,
  \dodoi{10.1088/0004-637X/770/1/57}

\bibitem[{{Bergeron} \& {Boisse}(1984)}]{bergeron84}
{Bergeron}, J., \& {Boisse}, P. 1984, \aap, 133, 374

\bibitem[{{Bergeron} \& {Boiss{\'e}}(1991)}]{bb91}
{Bergeron}, J., \& {Boiss{\'e}}, P. 1991, \aap, 243, 344

\bibitem[{{Bergeron} \& {Herbert-Fort}(2005)}]{bergeron05}
{Bergeron}, J., \& {Herbert-Fort}, S. 2005, in IAU Colloq. 199: Probing
  Galaxies through Quasar Absorption Lines, ed. P.~{Williams}, C.-G. {Shu}, \&
  B.~{Menard}, 265--280, \dodoi{10.1017/S1743921305002693}

\bibitem[{{Bird} {et~al.}(2016){Bird}, {Rubin}, {Suresh}, \&
  {Hernquist}}]{bird16}
{Bird}, S., {Rubin}, K. H.~R., {Suresh}, J., \& {Hernquist}, L. 2016, \mnras,
  462, 307, \dodoi{10.1093/mnras/stw1582}

\bibitem[{{Boksenberg} \& {Sargent}(2015)}]{boksy15}
{Boksenberg}, A., \& {Sargent}, W. L.~W. 2015, \apjs, 218, 7,
  \dodoi{10.1088/0067-0049/218/1/7}

\bibitem[{{Bolton} {et~al.}(2009){Bolton}, {Oh}, \& {Furlanetto}}]{bolton09}
{Bolton}, J.~S., {Oh}, S.~P., \& {Furlanetto}, S.~R. 2009, \mnras, 396, 2405,
  \dodoi{10.1111/j.1365-2966.2009.14914.x}

\bibitem[{{Bond} {et~al.}(2001{\natexlab{a}}){Bond}, {Churchill}, {Charlton},
  \& {Vogt}}]{bond01-winds}
{Bond}, N.~A., {Churchill}, C.~W., {Charlton}, J.~C., \& {Vogt}, S.~S.
  2001{\natexlab{a}}, \apj, 562, 641, \dodoi{10.1086/323876}

\bibitem[{{Bond} {et~al.}(2001{\natexlab{b}}){Bond}, {Churchill}, {Charlton},
  \& {Vogt}}]{bond01-bubbles}
---. 2001{\natexlab{b}}, \apj, 557, 761, \dodoi{10.1086/321689}

\bibitem[{{Bosman}(2021)}]{bosman21}
{Bosman}, S. E.~I. 2021, arXiv e-prints, arXiv:2108.12446,
  \dodoi{10.48550/arXiv.2108.12446}

\bibitem[{{Bosman} {et~al.}(2017){Bosman}, {Becker}, {Haehnelt}, {Hewett},
  {McMahon}, {Mortlock}, {Simpson}, \& {Venemans}}]{bosman17}
{Bosman}, S. E.~I., {Becker}, G.~D., {Haehnelt}, M.~G., {et~al.} 2017, \mnras,
  470, 1919, \dodoi{10.1093/mnras/stx1305}

\bibitem[{{Bosman} {et~al.}(2022){Bosman}, {Davies}, {Becker}, {Keating},
  {Davies}, {Zhu}, {Eilers}, {D'Odorico}, {Bian}, {Bischetti}, {Cristiani},
  {Fan}, {Farina}, {Haehnelt}, {Hennawi}, {Kulkarni}, {Mesinger}, {Meyer},
  {Onoue}, {Pallottini}, {Qin}, {Ryan-Weber}, {Schindler}, {Walter}, {Wang}, \&
  {Yang}}]{bosman22}
{Bosman}, S. E.~I., {Davies}, F.~B., {Becker}, G.~D., {et~al.} 2022, \mnras,
  514, 55, \dodoi{10.1093/mnras/stac1046}

\bibitem[{{Bouch{\'e}} {et~al.}(2006){Bouch{\'e}}, {Murphy}, {P{\'e}roux},
  {Csabai}, \& {Wild}}]{bouche06}
{Bouch{\'e}}, N., {Murphy}, M.~T., {P{\'e}roux}, C., {Csabai}, I., \& {Wild},
  V. 2006, \mnras, 371, 495, \dodoi{10.1111/j.1365-2966.2006.10685.x}

\bibitem[{{Bouwens} {et~al.}(2021){Bouwens}, {Oesch}, {Stefanon},
  {Illingworth}, {Labb{\'e}}, {Reddy}, {Atek}, {Montes}, {Naidu},
  {Nanayakkara}, {Nelson}, \& {Wilkins}}]{bouwens21}
{Bouwens}, R.~J., {Oesch}, P.~A., {Stefanon}, M., {et~al.} 2021, \aj, 162, 47,
  \dodoi{10.3847/1538-3881/abf83e}

\bibitem[{{Burbidge} {et~al.}(1977){Burbidge}, {Smith}, {Weymann}, \&
  {Williams}}]{burbidge77}
{Burbidge}, E.~M., {Smith}, H.~E., {Weymann}, R.~J., \& {Williams}, R.~E. 1977,
  \apj, 218, 1, \dodoi{10.1086/155651}

\bibitem[{{Burles} \& {Tytler}(1996)}]{burles96}
{Burles}, S., \& {Tytler}, D. 1996, \apj, 460, 584, \dodoi{10.1086/176994}

\bibitem[{{Carilli} {et~al.}(2004){Carilli}, {Gnedin}, {Furlanetto}, \&
  {Owen}}]{carilli04}
{Carilli}, C.~L., {Gnedin}, N., {Furlanetto}, S., \& {Owen}, F. 2004, \nar, 48,
  1053, \dodoi{10.1016/j.newar.2004.09.027}

\bibitem[{{Carswell} {et~al.}(1975){Carswell}, {Hilliard}, {Strittmatter},
  {Taylor}, \& {Weymann}}]{carswell75}
{Carswell}, R.~F., {Hilliard}, R.~L., {Strittmatter}, P.~A., {Taylor}, D.~J.,
  \& {Weymann}, R.~J. 1975, \apj, 196, 351, \dodoi{10.1086/153418}

\bibitem[{{Carswell} \& {Webb}(2014)}]{carswell14}
{Carswell}, R.~F., \& {Webb}, J.~K. 2014, {VPFIT: Voigt profile fitting
  program}.
\newblock \doeprint{1408.015}

\bibitem[{{Caulet}(1989)}]{caulet89}
{Caulet}, A. 1989, \apj, 340, 90, \dodoi{10.1086/167378}

\bibitem[{{Chen} {et~al.}(2010){Chen}, {Helsby}, {Gauthier}, {Shectman},
  {Thompson}, \& {Tinker}}]{chen10-mgii}
{Chen}, H.-W., {Helsby}, J.~E., {Gauthier}, J.-R., {et~al.} 2010, \apj, 714,
  1521, \dodoi{10.1088/0004-637X/714/2/1521}

\bibitem[{{Chen} {et~al.}(2017){Chen}, {Simcoe}, {Torrey}, {Ba{\~n}ados},
  {Cooksey}, {Cooper}, {Furesz}, {Matejek}, {Miller}, {Turner}, {Venemans},
  {Decarli}, {Farina}, {Mazzucchelli}, \& {Walter}}]{chen.s17}
{Chen}, S.-F.~S., {Simcoe}, R.~A., {Torrey}, P., {et~al.} 2017, \apj, 850, 188,
  \dodoi{10.3847/1538-4357/aa9707}

\bibitem[{{Christensen} {et~al.}(2023){Christensen}, {Jakobsen}, {Willott},
  {Arribas}, {Bunker}, {Charlot}, {Maiolino}, {Marshall}, {Perna}, \&
  {{\"U}bler}}]{christiansen23}
{Christensen}, L., {Jakobsen}, P., {Willott}, C., {et~al.} 2023, \aap, 680,
  A82, \dodoi{10.1051/0004-6361/202347943}

\bibitem[{{Churchill}(2001)}]{churchill01}
{Churchill}, C.~W. 2001, \apj, 560, 92, \dodoi{10.1086/322512}

\bibitem[{{Churchill} {et~al.}(2005){Churchill}, {Kacprzak}, \&
  {Steidel}}]{churchill05}
{Churchill}, C.~W., {Kacprzak}, G.~G., \& {Steidel}, C.~C. 2005, in IAU Colloq.
  199: Probing Galaxies through Quasar Absorption Lines, ed. P.~{Williams},
  C.-G. {Shu}, \& B.~{Menard}, 24--41, \dodoi{10.1017/S1743921305002401}

\bibitem[{{Churchill} {et~al.}(2000{\natexlab{a}}){Churchill}, {Mellon},
  {Charlton}, {Jannuzi}, {Kirhakos}, {Steidel}, \&
  {Schneider}}]{churchill00-archiveI}
{Churchill}, C.~W., {Mellon}, R.~R., {Charlton}, J.~C., {et~al.}
  2000{\natexlab{a}}, \apjs, 130, 91, \dodoi{10.1086/317343}

\bibitem[{{Churchill} {et~al.}(2000{\natexlab{b}}){Churchill}, {Mellon},
  {Charlton}, {Jannuzi}, {Kirhakos}, {Steidel}, \&
  {Schneider}}]{churchill00-archiveII}
---. 2000{\natexlab{b}}, \apj, 543, 577, \dodoi{10.1086/317120}

\bibitem[{{Churchill} {et~al.}(1999){Churchill}, {Rigby}, {Charlton}, \&
  {Vogt}}]{churchill99}
{Churchill}, C.~W., {Rigby}, J.~R., {Charlton}, J.~C., \& {Vogt}, S.~S. 1999,
  \apjs, 120, 51, \dodoi{10.1086/313168}

\bibitem[{{Churchill} {et~al.}(2015){Churchill}, {Vander Vliet},
  {Trujillo-Gomez}, {Kacprzak}, \& {Klypin}}]{churchill15}
{Churchill}, C.~W., {Vander Vliet}, J.~R., {Trujillo-Gomez}, S., {Kacprzak},
  G.~G., \& {Klypin}, A. 2015, \apj, 802, 10,
  \dodoi{10.1088/0004-637X/802/1/10}

\bibitem[{{Churchill} {et~al.}(2003){Churchill}, {Vogt}, \&
  {Charlton}}]{churchill03}
{Churchill}, C.~W., {Vogt}, S.~S., \& {Charlton}, J.~C. 2003, \aj, 125, 98,
  \dodoi{10.1086/345513}

\bibitem[{{Codoreanu} {et~al.}(2017){Codoreanu}, {Ryan-Weber}, {Crighton},
  {Becker}, {Pettini}, {Madau}, \& {Venemans}}]{codor17}
{Codoreanu}, A., {Ryan-Weber}, E.~V., {Crighton}, N. H.~M., {et~al.} 2017,
  \mnras, 472, 1023, \dodoi{10.1093/mnras/stx1985}

\bibitem[{{Cole} {et~al.}(2001){Cole}, {Norberg}, {Baugh}, {Frenk},
  {Bland-Hawthorn}, {Bridges}, {Cannon}, {Colless}, {Collins}, {Couch},
  {Cross}, {Dalton}, {De Propris}, {Driver}, {Efstathiou}, {Ellis},
  {Glazebrook}, {Jackson}, {Lahav}, {Lewis}, {Lumsden}, {Maddox}, {Madgwick},
  {Peacock}, {Peterson}, {Sutherland}, \& {Taylor}}]{cole01}
{Cole}, S., {Norberg}, P., {Baugh}, C.~M., {et~al.} 2001, \mnras, 326, 255,
  \dodoi{10.1046/j.1365-8711.2001.04591.x}

\bibitem[{{Cooksey} {et~al.}(2013){Cooksey}, {Kao}, {Simcoe}, {O'Meara}, \&
  {Prochaska}}]{cooksey13}
{Cooksey}, K.~L., {Kao}, M.~M., {Simcoe}, R.~A., {O'Meara}, J.~M., \&
  {Prochaska}, J.~X. 2013, \apj, 763, 37, \dodoi{10.1088/0004-637X/763/1/37}

\bibitem[{{Cooper} {et~al.}(2019){Cooper}, {Simcoe}, {Cooksey}, {Bordoloi},
  {Miller}, {Furesz}, {Turner}, \& {Ba{\~n}ados}}]{cooper19}
{Cooper}, T.~J., {Simcoe}, R.~A., {Cooksey}, K.~L., {et~al.} 2019, \apj, 882,
  77, \dodoi{10.3847/1538-4357/ab3402}

\bibitem[{{Danforth} \& {Shull}(2008)}]{danforth08}
{Danforth}, C.~W., \& {Shull}, J.~M. 2008, \apj, 679, 194,
  \dodoi{10.1086/587127}

\bibitem[{{Danforth} {et~al.}(2016){Danforth}, {Keeney}, {Tilton}, {Shull},
  {Stocke}, {Stevans}, {Pieri}, {Savage}, {France}, {Syphers}, {Smith},
  {Green}, {Froning}, {Penton}, \& {Osterman}}]{danforth16}
{Danforth}, C.~W., {Keeney}, B.~A., {Tilton}, E.~M., {et~al.} 2016, \apj, 817,
  111, \dodoi{10.3847/0004-637X/817/2/111}

\bibitem[{{Dav{\'e}} {et~al.}(1999){Dav{\'e}}, {Hernquist}, {Katz}, \&
  {Weinberg}}]{dave99}
{Dav{\'e}}, R., {Hernquist}, L., {Katz}, N., \& {Weinberg}, D.~H. 1999, \apj,
  511, 521, \dodoi{10.1086/306722}

\bibitem[{{Davies} {et~al.}(2023{\natexlab{a}}){Davies}, {Ryan-Weber},
  {D'Odorico}, {Bosman}, {Meyer}, {Becker}, {Cupani}, {Keating}, {Bischetti},
  {Davies}, {Eilers}, {Farina}, {Haehnelt}, {Pallottini}, \& {Zhu}}]{davies23}
{Davies}, R.~L., {Ryan-Weber}, E., {D'Odorico}, V., {et~al.}
  2023{\natexlab{a}}, \mnras, 521, 314, \dodoi{10.1093/mnras/stad294}

\bibitem[{{Davies} {et~al.}(2023{\natexlab{b}}){Davies}, {Ryan-Weber},
  {D'Odorico}, {Bosman}, {Meyer}, {Becker}, {Cupani}, {Bischetti}, {Sebastian},
  {Eilers}, {Farina}, {Wang}, {Yang}, \& {Zhu}}]{davies23-survey}
---. 2023{\natexlab{b}}, \mnras, 521, 289, \dodoi{10.1093/mnras/stac3662}

\bibitem[{{D'Odorico} {et~al.}(2013){D'Odorico}, {Cupani}, {Cristiani},
  {Maiolino}, {Molaro}, {Nonino}, {Centuri{\'o}n}, {Cimatti}, {di Serego
  Alighieri}, {Fiore}, {Fontana}, {Gallerani}, {Giallongo}, {Mannucci},
  {Marconi}, {Pentericci}, {Viel}, \& {Vladilo}}]{dodorico13}
{D'Odorico}, V., {Cupani}, G., {Cristiani}, S., {et~al.} 2013, \mnras, 435,
  1198, \dodoi{10.1093/mnras/stt1365}

\bibitem[{{D'Odorico} {et~al.}(2022){D'Odorico}, {Finlator}, {Cristiani},
  {Cupani}, {Perrotta}, {Calura}, {C{\`e}nturion}, {Becker}, {Berg}, {Lopez},
  {Ellison}, \& {Pomante}}]{dodorico22}
{D'Odorico}, V., {Finlator}, K., {Cristiani}, S., {et~al.} 2022, \mnras, 512,
  2389, \dodoi{10.1093/mnras/stac545}

\bibitem[{{D'Odorico} {et~al.}(2023){D'Odorico}, {Ba{\~n}ados}, {Becker},
  {Bischetti}, {Bosman}, {Cupani}, {Davies}, {Farina}, {Ferrara}, {Feruglio},
  {Mazzucchelli}, {Ryan-Weber}, {Schindler}, {Sodini}, {Venemans}, {Walter},
  {Chen}, {Lai}, {Zhu}, {Bian}, {Campo}, {Carniani}, {Cristiani}, {Davies},
  {Decarli}, {Drake}, {Eilers}, {Fan}, {Gaikwad}, {Gallerani}, {Greig},
  {Haehnelt}, {Hennawi}, {Keating}, {Kulkarni}, {Mesinger}, {Meyer},
  {Neeleman}, {Onoue}, {Pallottini}, {Qin}, {Rojas-Ruiz}, {Satyavolu},
  {Sebastian}, {Tripodi}, {Wang}, {Wolfson}, {Yang}, \&
  {Zanchettin}}]{dodorico23}
{D'Odorico}, V., {Ba{\~n}ados}, E., {Becker}, G.~D., {et~al.} 2023, \mnras,
  523, 1399, \dodoi{10.1093/mnras/stad1468}

\bibitem[{{Doughty} {et~al.}(2018){Doughty}, {Finlator}, {Oppenheimer},
  {Dav{\'e}}, \& {Zackrisson}}]{doughty18}
{Doughty}, C., {Finlator}, K., {Oppenheimer}, B.~D., {Dav{\'e}}, R., \&
  {Zackrisson}, E. 2018, \mnras, 475, 4717, \dodoi{10.1093/mnras/sty156}

\bibitem[{{Draine}(2011)}]{draine11}
{Draine}, B.~T. 2011, {Physics of the Interstellar and Intergalactic Medium}

\bibitem[{{Dutta} {et~al.}(2020){Dutta}, {Fumagalli}, {Fossati}, {Lofthouse},
  {Prochaska}, {Arrigoni Battaia}, {Bielby}, {Cantalupo}, {Cooke}, {Murphy}, \&
  {O'Meara}}]{dutta20}
{Dutta}, R., {Fumagalli}, M., {Fossati}, M., {et~al.} 2020, \mnras, 499, 5022,
  \dodoi{10.1093/mnras/staa3147}

\bibitem[{{Esmerian} {et~al.}(2021){Esmerian}, {Kravtsov}, {Hafen},
  {Faucher-Gigu{\`e}re}, {Quataert}, {Stern}, {Kere{\v{s}}}, \&
  {Wetzel}}]{esmerian21}
{Esmerian}, C.~J., {Kravtsov}, A.~V., {Hafen}, Z., {et~al.} 2021, \mnras, 505,
  1841, \dodoi{10.1093/mnras/stab1281}

\bibitem[{{Evans}(2011)}]{evans11}
{Evans}, J.~L. 2011, PhD thesis, New Mexico State University

\bibitem[{{Fan} {et~al.}(2023){Fan}, {Ba{\~n}ados}, \& {Simcoe}}]{fan23}
{Fan}, X., {Ba{\~n}ados}, E., \& {Simcoe}, R.~A. 2023, \araa, 61, 373,
  \dodoi{10.1146/annurev-astro-052920-102455}

\bibitem[{{Faucher-Gigu{\`e}re}(2020)}]{faucher-giguere20}
{Faucher-Gigu{\`e}re}, C.-A. 2020, \mnras, 493, 1614,
  \dodoi{10.1093/mnras/staa302}

\bibitem[{{Faucher-Gigu{\`e}re} \& {Oh}(2023)}]{faucher-giguere23}
{Faucher-Gigu{\`e}re}, C.-A., \& {Oh}, S.~P. 2023, \araa, 61, 131,
  \dodoi{10.1146/annurev-astro-052920-125203}

\bibitem[{{Fechner} \& {Reimers}(2007)}]{fechner07}
{Fechner}, C., \& {Reimers}, D. 2007, \aap, 463, 69,
  \dodoi{10.1051/0004-6361:20066566}

\bibitem[{{Fechner} {et~al.}(2006){Fechner}, {Reimers}, {Kriss}, {Baade},
  {Blair}, {Giroux}, {Green}, {Moos}, {Morton}, {Scott}, {Shull}, {Simcoe},
  {Songaila}, \& {Zheng}}]{fechner06}
{Fechner}, C., {Reimers}, D., {Kriss}, G.~A., {et~al.} 2006, \aap, 455, 91,
  \dodoi{10.1051/0004-6361:20064950}

\bibitem[{{Ferrara} {et~al.}(2000){Ferrara}, {Pettini}, \&
  {Shchekinov}}]{ferrera00}
{Ferrara}, A., {Pettini}, M., \& {Shchekinov}, Y. 2000, \mnras, 319, 539,
  \dodoi{10.1046/j.1365-8711.2000.03857.x}

\bibitem[{{Finlator} {et~al.}(2018){Finlator}, {Keating}, {Oppenheimer},
  {Dav{\'e}}, \& {Zackrisson}}]{finlator18}
{Finlator}, K., {Keating}, L., {Oppenheimer}, B.~D., {Dav{\'e}}, R., \&
  {Zackrisson}, E. 2018, \mnras, 480, 2628, \dodoi{10.1093/mnras/sty1949}

\bibitem[{{Finlator} {et~al.}(2016){Finlator}, {Oppenheimer}, {Dav{\'e}},
  {Zackrisson}, {Thompson}, \& {Huang}}]{finlator16}
{Finlator}, K., {Oppenheimer}, B.~D., {Dav{\'e}}, R., {et~al.} 2016, \mnras,
  459, 2299, \dodoi{10.1093/mnras/stw805}

\bibitem[{{Ford} {et~al.}(2013){Ford}, {Oppenheimer}, {Dav{\'e}}, {Katz},
  {Kollmeier}, \& {Weinberg}}]{ford13}
{Ford}, A.~B., {Oppenheimer}, B.~D., {Dav{\'e}}, R., {et~al.} 2013, \mnras,
  432, 89, \dodoi{10.1093/mnras/stt393}

\bibitem[{Foreman-Mackey {et~al.}(2019)Foreman-Mackey, Farr, Sinha, Archibald,
  Hogg, Sanders, Zuntz, Williams, Nelson, de~Val-Borro, Erhardt, Pashchenko, \&
  Pla}]{Foreman_Mackey_2019}
Foreman-Mackey, D., Farr, W., Sinha, M., {et~al.} 2019, Journal of Open Source
  Software, 4, 1864, \dodoi{10.21105/joss.01864}

\bibitem[{{F{\"o}rster Schreiber} \& {Wuyts}(2020)}]{forster20}
{F{\"o}rster Schreiber}, N.~M., \& {Wuyts}, S. 2020, \araa, 58, 661,
  \dodoi{10.1146/annurev-astro-032620-021910}

\bibitem[{{Fumagalli} {et~al.}(2016){Fumagalli}, {O'Meara}, \&
  {Prochaska}}]{fumagalli16}
{Fumagalli}, M., {O'Meara}, J.~M., \& {Prochaska}, J.~X. 2016, \mnras, 455,
  4100, \dodoi{10.1093/mnras/stv2616}

\bibitem[{{Giroux} \& {Shull}(1997)}]{giroux97}
{Giroux}, M.~L., \& {Shull}, J.~M. 1997, \aj, 113, 1505, \dodoi{10.1086/118367}

\bibitem[{{Gnedin} \& {Madau}(2022)}]{gnedin22}
{Gnedin}, N.~Y., \& {Madau}, P. 2022, Living Reviews in Computational
  Astrophysics, 8, 3, \dodoi{10.1007/s41115-022-00015-5}

\bibitem[{{Hasan} {et~al.}(2020){Hasan}, {Churchill}, {Stemock}, {Mathes},
  {Nielsen}, {Finlator}, {Doughty}, {Croom}, {Kacprzak}, \& {Murphy}}]{hasan20}
{Hasan}, F., {Churchill}, C.~W., {Stemock}, B., {et~al.} 2020, \apj, 904, 44,
  \dodoi{10.3847/1538-4357/abbe0b}

\bibitem[{{Hennawi} {et~al.}(2021){Hennawi}, {Davies}, {Wang}, \&
  {O{\~n}orbe}}]{hennawi21}
{Hennawi}, J.~F., {Davies}, F.~B., {Wang}, F., \& {O{\~n}orbe}, J. 2021,
  \mnras, 506, 2963, \dodoi{10.1093/mnras/stab1883}

\bibitem[{{Hu} {et~al.}(1995){Hu}, {Kim}, {Cowie}, {Songaila}, \&
  {Rauch}}]{hu95}
{Hu}, E.~M., {Kim}, T.-S., {Cowie}, L.~L., {Songaila}, A., \& {Rauch}, M. 1995,
  \aj, 110, 1526, \dodoi{10.1086/117625}

\bibitem[{{Hu} {et~al.}(2022){Hu}, {Khaire}, {Hennawi}, {Walther}, {Hiss},
  {Alsing}, {O{\~n}orbe}, {Lukic}, \& {Davies}}]{hu22}
{Hu}, T., {Khaire}, V., {Hennawi}, J.~F., {et~al.} 2022, \mnras, 515, 2188,
  \dodoi{10.1093/mnras/stac1865}

\bibitem[{{Huang} {et~al.}(2021){Huang}, {Chen}, {Shectman}, {Johnson},
  {Zahedy}, {Helsby}, {Gauthier}, \& {Thompson}}]{huang21}
{Huang}, Y.-H., {Chen}, H.-W., {Shectman}, S.~A., {et~al.} 2021, \mnras, 502,
  4743, \dodoi{10.1093/mnras/stab360}

\bibitem[{{Kacprzak} \& {Churchill}(2011)}]{ggk-cwc11}
{Kacprzak}, G.~G., \& {Churchill}, C.~W. 2011, \apjl, 743, L34,
  \dodoi{10.1088/2041-8205/743/2/L34}

\bibitem[{{Kacprzak} {et~al.}(2010){Kacprzak}, {Churchill}, {Ceverino},
  {Steidel}, {Klypin}, \& {Murphy}}]{kacprzak10}
{Kacprzak}, G.~G., {Churchill}, C.~W., {Ceverino}, D., {et~al.} 2010, \apj,
  711, 533, \dodoi{10.1088/0004-637X/711/2/533}

\bibitem[{{Kacprzak} {et~al.}(2019){Kacprzak}, {Vander Vliet}, {Nielsen},
  {Muzahid}, {Pointon}, {Churchill}, {Ceverino}, {Arraki}, {Klypin},
  {Charlton}, \& {Lewis}}]{kacprzak19}
{Kacprzak}, G.~G., {Vander Vliet}, J.~R., {Nielsen}, N.~M., {et~al.} 2019,
  \apj, 870, 137, \dodoi{10.3847/1538-4357/aaf1a6}

\bibitem[{{Keating} {et~al.}(2014){Keating}, {Haehnelt}, {Becker}, \&
  {Bolton}}]{keating14}
{Keating}, L.~C., {Haehnelt}, M.~G., {Becker}, G.~D., \& {Bolton}, J.~S. 2014,
  \mnras, 438, 1820, \dodoi{10.1093/mnras/stt2324}

\bibitem[{{Keller} {et~al.}(2020){Keller}, {Kruijssen}, \&
  {Wadsley}}]{keller20}
{Keller}, B.~W., {Kruijssen}, J.~M.~D., \& {Wadsley}, J.~W. 2020, \mnras, 493,
  2149, \dodoi{10.1093/mnras/staa380}

\bibitem[{{Kinman} \& {Burbidge}(1967)}]{kinman67}
{Kinman}, T.~D., \& {Burbidge}, E.~M. 1967, \apjl, 148, L59,
  \dodoi{10.1086/180015}

\bibitem[{{Kulkarni} {et~al.}(2019){Kulkarni}, {Worseck}, \&
  {Hennawi}}]{kulkarni19}
{Kulkarni}, G., {Worseck}, G., \& {Hennawi}, J.~F. 2019, \mnras, 488, 1035,
  \dodoi{10.1093/mnras/stz1493}

\bibitem[{{Lanzetta} {et~al.}(1987){Lanzetta}, {Turnshek}, \&
  {Wolfe}}]{lanzetta87}
{Lanzetta}, K.~M., {Turnshek}, D.~A., \& {Wolfe}, A.~M. 1987, \apj, 322, 739,
  \dodoi{10.1086/165769}

\bibitem[{{Lauroesch} {et~al.}(1996){Lauroesch}, {Truran}, {Welty}, \&
  {York}}]{jtl1996}
{Lauroesch}, J.~T., {Truran}, J.~W., {Welty}, D.~E., \& {York}, D.~G. 1996,
  \pasp, 108, 641, \dodoi{10.1086/133780}

\bibitem[{{Lehner} {et~al.}(2016){Lehner}, {O'Meara}, {Howk}, {Prochaska}, \&
  {Fumagalli}}]{lehner16}
{Lehner}, N., {O'Meara}, J.~M., {Howk}, J.~C., {Prochaska}, J.~X., \&
  {Fumagalli}, M. 2016, \apj, 833, 283, \dodoi{10.3847/1538-4357/833/2/283}

\bibitem[{{Lehner} {et~al.}(2022){Lehner}, {Kopenhafer}, {O'Meara}, {Howk},
  {Fumagalli}, {Prochaska}, {Acharyya}, {O'Shea}, {Peeples}, {Tumlinson}, \&
  {Hummels}}]{lehner22}
{Lehner}, N., {Kopenhafer}, C., {O'Meara}, J.~M., {et~al.} 2022, \apj, 936,
  156, \dodoi{10.3847/1538-4357/ac7400}

\bibitem[{{Lochhaas} {et~al.}(2021){Lochhaas}, {Tumlinson}, {O'Shea},
  {Peeples}, {Smith}, {Werk}, {Augustin}, \& {Simons}}]{lochhaas21}
{Lochhaas}, C., {Tumlinson}, J., {O'Shea}, B.~W., {et~al.} 2021, \apj, 922,
  121, \dodoi{10.3847/1538-4357/ac2496}

\bibitem[{{Lu} {et~al.}(1996){Lu}, {Sargent}, {Womble}, \&
  {Takada-Hidai}}]{lu96}
{Lu}, L., {Sargent}, W. L.~W., {Womble}, D.~S., \& {Takada-Hidai}, M. 1996,
  \apj, 472, 509, \dodoi{10.1086/526756}

\bibitem[{{Lundgren} {et~al.}(2009){Lundgren}, {Brunner}, {York}, {Ross},
  {Quashnock}, {Myers}, {Schneider}, {Al Sayyad}, \& {Bahcall}}]{lundgren09}
{Lundgren}, B.~F., {Brunner}, R.~J., {York}, D.~G., {et~al.} 2009, \apj, 698,
  819, \dodoi{10.1088/0004-637X/698/1/819}

\bibitem[{{Lyu} {et~al.}(2023){Lyu}, {Peng}, {Jing}, {Yang}, {Ho}, {Renzini},
  {Wang}, {Wang}, {Xu}, {Zhao}, {Dou}, {Gu}, {Maiolino}, {Mannucci}, \&
  {Yuan}}]{lyu23}
{Lyu}, C., {Peng}, Y., {Jing}, Y., {et~al.} 2023, \apj, 959, 5,
  \dodoi{10.3847/1538-4357/ad036b}

\bibitem[{{Madau} \& {Dickinson}(2014)}]{madau14}
{Madau}, P., \& {Dickinson}, M. 2014, \araa, 52, 415,
  \dodoi{10.1146/annurev-astro-081811-125615}

\bibitem[{{Maller} \& {Bullock}(2004)}]{maller04}
{Maller}, A.~H., \& {Bullock}, J.~S. 2004, \mnras, 355, 694,
  \dodoi{10.1111/j.1365-2966.2004.08349.x}

\bibitem[{{Matejek} \& {Simcoe}(2012)}]{matejek12}
{Matejek}, M.~S., \& {Simcoe}, R.~A. 2012, \apj, 761, 112,
  \dodoi{10.1088/0004-637X/761/2/112}

\bibitem[{{Mathes} {et~al.}(2017){Mathes}, {Churchill}, \& {Murphy}}]{mathes17}
{Mathes}, N.~L., {Churchill}, C.~W., \& {Murphy}, M.~T. 2017, arXiv e-prints,
  arXiv:1701.05624.
\newblock \doarXiv{1701.05624}

\bibitem[{{McCourt} {et~al.}(2018){McCourt}, {Oh}, {O'Leary}, \&
  {Madigan}}]{mccourt18}
{McCourt}, M., {Oh}, S.~P., {O'Leary}, R., \& {Madigan}, A.-M. 2018, \mnras,
  473, 5407, \dodoi{10.1093/mnras/stx2687}

\bibitem[{{McQuinn}(2016)}]{quinn16}
{McQuinn}, M. 2016, \araa, 54, 313, \dodoi{10.1146/annurev-astro-082214-122355}

\bibitem[{{McQuinn} {et~al.}(2009){McQuinn}, {Lidz}, {Zaldarriaga},
  {Hernquist}, {Hopkins}, {Dutta}, \& {Faucher-Gigu{\`e}re}}]{mcquinn09}
{McQuinn}, M., {Lidz}, A., {Zaldarriaga}, M., {et~al.} 2009, \apj, 694, 842,
  \dodoi{10.1088/0004-637X/694/2/842}

\bibitem[{{M{\'e}nard} {et~al.}(2008){M{\'e}nard}, {Nestor}, {Turnshek},
  {Quider}, {Richards}, {Chelouche}, \& {Rao}}]{menard08}
{M{\'e}nard}, B., {Nestor}, D., {Turnshek}, D., {et~al.} 2008, \mnras, 385,
  1053, \dodoi{10.1111/j.1365-2966.2008.12909.x}

\bibitem[{{Mshar} {et~al.}(2007){Mshar}, {Charlton}, {Lynch}, {Churchill}, \&
  {Kim}}]{mshar07}
{Mshar}, A.~C., {Charlton}, J.~C., {Lynch}, R.~S., {Churchill}, C., \& {Kim},
  T.-S. 2007, \apj, 669, 135, \dodoi{10.1086/520792}

\bibitem[{{Narayanan} {et~al.}(2005){Narayanan}, {Charlton}, {Masiero}, \&
  {Lynch}}]{narayanan05}
{Narayanan}, A., {Charlton}, J.~C., {Masiero}, J.~R., \& {Lynch}, R. 2005,
  \apj, 632, 92, \dodoi{10.1086/432750}

\bibitem[{{Narayanan} {et~al.}(2008){Narayanan}, {Charlton}, {Misawa}, {Green},
  \& {Kim}}]{narayanan08}
{Narayanan}, A., {Charlton}, J.~C., {Misawa}, T., {Green}, R.~E., \& {Kim},
  T.-S. 2008, \apj, 689, 782, \dodoi{10.1086/592763}

\bibitem[{{Narayanan} {et~al.}(2007){Narayanan}, {Misawa}, {Charlton}, \&
  {Kim}}]{narayanan07}
{Narayanan}, A., {Misawa}, T., {Charlton}, J.~C., \& {Kim}, T.-S. 2007, \apj,
  660, 1093, \dodoi{10.1086/512852}

\bibitem[{{Nath} \& {Sethi}(1996)}]{nath96}
{Nath}, B.~B., \& {Sethi}, S.~K. 1996, \mnras, 279, 275,
  \dodoi{10.1093/mnras/279.1.275}

\bibitem[{{Nelson} {et~al.}(2018){Nelson}, {Kauffmann}, {Pillepich}, {Genel},
  {Springel}, {Pakmor}, {Hernquist}, {Weinberger}, {Torrey}, {Vogelsberger}, \&
  {Marinacci}}]{nelson18}
{Nelson}, D., {Kauffmann}, G., {Pillepich}, A., {et~al.} 2018, \mnras, 477,
  450, \dodoi{10.1093/mnras/sty656}

\bibitem[{{Nestor} {et~al.}(2011){Nestor}, {Johnson}, {Wild}, {M{\'e}nard},
  {Turnshek}, {Rao}, \& {Pettini}}]{nestor11}
{Nestor}, D.~B., {Johnson}, B.~D., {Wild}, V., {et~al.} 2011, \mnras, 412,
  1559, \dodoi{10.1111/j.1365-2966.2010.17865.x}

\bibitem[{{Nestor} {et~al.}(2005){Nestor}, {Turnshek}, \& {Rao}}]{nestor05}
{Nestor}, D.~B., {Turnshek}, D.~A., \& {Rao}, S.~M. 2005, \apj, 628, 637,
  \dodoi{10.1086/427547}

\bibitem[{{Nestor} {et~al.}(2006){Nestor}, {Turnshek}, \& {Rao}}]{nestor06}
---. 2006, \apj, 643, 75, \dodoi{10.1086/501498}

\bibitem[{Newville {et~al.}(2014)Newville, Stensitzki, Allen, \&
  Ingargiola}]{lmfit}
Newville, M., Stensitzki, T., Allen, D.~B., \& Ingargiola, A. 2014, {LMFIT:
  Non-Linear Least-Square Minimization and Curve-Fitting for Python} (Zenodo),
  \dodoi{10.5281/zenodo.11813}

\bibitem[{{Nielsen} {et~al.}(2013){Nielsen}, {Churchill}, \&
  {Kacprzak}}]{nielsen13-magiicat2}
{Nielsen}, N.~M., {Churchill}, C.~W., \& {Kacprzak}, G.~G. 2013, \apj, 776,
  115, \dodoi{10.1088/0004-637X/776/2/115}

\bibitem[{{Nielsen} {et~al.}(2020){Nielsen}, {Kacprzak}, {Pointon}, {Murphy},
  {Churchill}, \& {Dav{\'e}}}]{nielsen20}
{Nielsen}, N.~M., {Kacprzak}, G.~G., {Pointon}, S.~K., {et~al.} 2020, \apj,
  904, 164, \dodoi{10.3847/1538-4357/abc561}

\bibitem[{{Nielsen} {et~al.}(2022){Nielsen}, {Kacprzak}, {Sameer}, {Murphy},
  {Nateghi}, {Charlton}, \& {Churchill}}]{nielsen22}
{Nielsen}, N.~M., {Kacprzak}, G.~G., {Sameer}, {et~al.} 2022, \mnras, 514,
  6074, \dodoi{10.1093/mnras/stac1824}

\bibitem[{{Noterdaeme} {et~al.}(2012){Noterdaeme}, {Petitjean}, {Carithers},
  {P{\^a}ris}, {Font-Ribera}, {Bailey}, {Aubourg}, {Bizyaev}, {Ebelke},
  {Finley}, {Ge}, {Malanushenko}, {Malanushenko}, {Miralda-Escud{\'e}},
  {Myers}, {Oravetz}, {Pan}, {Pieri}, {Ross}, {Schneider}, {Simmons}, \&
  {York}}]{noterdaeme12}
{Noterdaeme}, P., {Petitjean}, P., {Carithers}, W.~C., {et~al.} 2012, \aap,
  547, L1, \dodoi{10.1051/0004-6361/201220259}

\bibitem[{{Oh}(2002)}]{oh02}
{Oh}, S.~P. 2002, \mnras, 336, 1021, \dodoi{10.1046/j.1365-8711.2002.05859.x}

\bibitem[{{Oppenheimer} \& {Dav{\'e}}(2006)}]{oppenheimer06}
{Oppenheimer}, B.~D., \& {Dav{\'e}}, R. 2006, \mnras, 373, 1265,
  \dodoi{10.1111/j.1365-2966.2006.10989.x}

\bibitem[{{Oppenheimer} {et~al.}(2009){Oppenheimer}, {Dav{\'e}}, \&
  {Finlator}}]{oppenheimer09}
{Oppenheimer}, B.~D., {Dav{\'e}}, R., \& {Finlator}, K. 2009, \mnras, 396, 729,
  \dodoi{10.1111/j.1365-2966.2009.14771.x}

\bibitem[{{Oppenheimer} {et~al.}(2012){Oppenheimer}, {Dav{\'e}}, {Katz},
  {Kollmeier}, \& {Weinberg}}]{oppenheimer12}
{Oppenheimer}, B.~D., {Dav{\'e}}, R., {Katz}, N., {Kollmeier}, J.~A., \&
  {Weinberg}, D.~H. 2012, \mnras, 420, 829,
  \dodoi{10.1111/j.1365-2966.2011.20096.x}

\bibitem[{{Oppenheimer} {et~al.}(2018){Oppenheimer}, {Schaye}, {Crain}, {Werk},
  \& {Richings}}]{oppenheimer18}
{Oppenheimer}, B.~D., {Schaye}, J., {Crain}, R.~A., {Werk}, J.~K., \&
  {Richings}, A.~J. 2018, \mnras, 481, 835, \dodoi{10.1093/mnras/sty2281}

\bibitem[{{Pandya} {et~al.}(2023){Pandya}, {Fielding}, {Bryan}, {Carr},
  {Somerville}, {Stern}, {Faucher-Gigu{\`e}re}, {Hafen},
  {Angl{\'e}s-Alc{\'a}zar}, \& {Forbes}}]{pandya23}
{Pandya}, V., {Fielding}, D.~B., {Bryan}, G.~L., {et~al.} 2023, \apj, 956, 118,
  \dodoi{10.3847/1538-4357/acf3ea}

\bibitem[{{Peeples} {et~al.}(2019){Peeples}, {Corlies}, {Tumlinson}, {O'Shea},
  {Lehner}, {O'Meara}, {Howk}, {Earl}, {Smith}, {Wise}, \&
  {Hummels}}]{peeples19}
{Peeples}, M.~S., {Corlies}, L., {Tumlinson}, J., {et~al.} 2019, \apj, 873,
  129, \dodoi{10.3847/1538-4357/ab0654}

\bibitem[{{P{\'e}roux} \& {Howk}(2020)}]{peroux20}
{P{\'e}roux}, C., \& {Howk}, J.~C. 2020, \araa, 58, 363,
  \dodoi{10.1146/annurev-astro-021820-120014}

\bibitem[{{Petitjean} \& {Bergeron}(1990)}]{pb90}
{Petitjean}, P., \& {Bergeron}, J. 1990, \aap, 231, 309

\bibitem[{{Petitjean} \& {Bergeron}(1994)}]{pb94}
---. 1994, \aap, 283, 759

\bibitem[{{Planck Collaboration}(2020)}]{planck18}
{Planck Collaboration}. 2020, Astronomy \& Astrophysics, 641, A6,
  \dodoi{10.1051/0004-6361/201833910}

\bibitem[{{Popesso} {et~al.}(2023){Popesso}, {Concas}, {Cresci}, {Belli},
  {Rodighiero}, {Inami}, {Dickinson}, {Ilbert}, {Pannella}, \&
  {Elbaz}}]{popesso23}
{Popesso}, P., {Concas}, A., {Cresci}, G., {et~al.} 2023, \mnras, 519, 1526,
  \dodoi{10.1093/mnras/stac3214}

\bibitem[{{Press} {et~al.}(2002){Press}, {Teukolsky}, {Vetterling}, \&
  {Flannery}}]{press02-numrecipe}
{Press}, W.~H., {Teukolsky}, S.~A., {Vetterling}, W.~T., \& {Flannery}, B.~P.
  2002, {Numerical recipes in C++ : the art of scientific computing} (Zenodo)

\bibitem[{{Prochter} {et~al.}(2006){Prochter}, {Prochaska}, \&
  {Burles}}]{prochter06}
{Prochter}, G.~E., {Prochaska}, J.~X., \& {Burles}, S.~M. 2006, \apj, 639, 766,
  \dodoi{10.1086/499341}

\bibitem[{{Puchwein} {et~al.}(2023){Puchwein}, {Bolton}, {Keating}, {Molaro},
  {Gaikwad}, {Kulkarni}, {Haehnelt}, {Ir{\v{s}}i{\v{c}}},
  {{\v{S}}oltinsk{\'y}}, {Viel}, {Aubert}, {Becker}, \& {Meiksin}}]{puchwein23}
{Puchwein}, E., {Bolton}, J.~S., {Keating}, L.~C., {et~al.} 2023, \mnras, 519,
  6162, \dodoi{10.1093/mnras/stac3761}

\bibitem[{{Rafelski} {et~al.}(2012){Rafelski}, {Wolfe}, {Prochaska},
  {Neeleman}, \& {Mendez}}]{rafelski12}
{Rafelski}, M., {Wolfe}, A.~M., {Prochaska}, J.~X., {Neeleman}, M., \&
  {Mendez}, A.~J. 2012, \apj, 755, 89, \dodoi{10.1088/0004-637X/755/2/89}

\bibitem[{{Raghunathan} {et~al.}(2016){Raghunathan}, {Clowes}, {Campusano},
  {S{\"o}chting}, {Graham}, \& {Williger}}]{raghunathan2016}
{Raghunathan}, S., {Clowes}, R.~G., {Campusano}, L.~E., {et~al.} 2016, \mnras,
  463, 2640, \dodoi{10.1093/mnras/stw2095}

\bibitem[{{Rahmati} {et~al.}(2016){Rahmati}, {Schaye}, {Crain}, {Oppenheimer},
  {Schaller}, \& {Theuns}}]{rahmati16}
{Rahmati}, A., {Schaye}, J., {Crain}, R.~A., {et~al.} 2016, \mnras, 459, 310,
  \dodoi{10.1093/mnras/stw453}

\bibitem[{{Rao} {et~al.}(2017){Rao}, {Turnshek}, {Sardane}, \&
  {Monier}}]{rao17}
{Rao}, S.~M., {Turnshek}, D.~A., {Sardane}, G.~M., \& {Monier}, E.~M. 2017,
  \mnras, 471, 3428, \dodoi{10.1093/mnras/stx1787}

\bibitem[{{Rauch} {et~al.}(1996){Rauch}, {Sargent}, {Womble}, \&
  {Barlow}}]{rauch96}
{Rauch}, M., {Sargent}, W.~L.~W., {Womble}, D.~S., \& {Barlow}, T.~A. 1996,
  \apjl, 467, L5, \dodoi{10.1086/310187}

\bibitem[{{Richter} {et~al.}(2006){Richter}, {Fang}, \& {Bryan}}]{richter06}
{Richter}, P., {Fang}, T., \& {Bryan}, G.~L. 2006, \aap, 451, 767,
  \dodoi{10.1051/0004-6361:20054556}

\bibitem[{{Rigby} {et~al.}(2002){Rigby}, {Charlton}, \& {Churchill}}]{rigby02}
{Rigby}, J.~R., {Charlton}, J.~C., \& {Churchill}, C.~W. 2002, \apj, 565, 743,
  \dodoi{10.1086/324723}

\bibitem[{{Rodr{\'\i}guez Hidalgo} {et~al.}(2012){Rodr{\'\i}guez Hidalgo},
  {Wessels}, {Charlton}, {Narayanan}, {Mshar}, {Cucchiara}, \&
  {Jones}}]{rodriguez12}
{Rodr{\'\i}guez Hidalgo}, P., {Wessels}, K., {Charlton}, J.~C., {et~al.} 2012,
  \mnras, 427, 1801, \dodoi{10.1111/j.1365-2966.2012.21586.x}

\bibitem[{{Roman-Duval} {et~al.}(2022){Roman-Duval}, {Jenkins}, {Tchernyshyov},
  {Clark}, {De Cia}, {Gordon}, {Hamanowicz}, {Lebouteiller}, {Rafelski},
  {Sandstrom}, {Werk}, \& {Yanchulova Merica-Jones}}]{roman-duval22}
{Roman-Duval}, J., {Jenkins}, E.~B., {Tchernyshyov}, K., {et~al.} 2022, \apj,
  935, 105, \dodoi{10.3847/1538-4357/ac7713}

\bibitem[{{Rubin} {et~al.}(2010){Rubin}, {Prochaska}, {Koo}, {Phillips}, \&
  {Weiner}}]{rubin10}
{Rubin}, K. H.~R., {Prochaska}, J.~X., {Koo}, D.~C., {Phillips}, A.~C., \&
  {Weiner}, B.~J. 2010, \apj, 712, 574, \dodoi{10.1088/0004-637X/712/1/574}

\bibitem[{{Sargent} {et~al.}(1988{\natexlab{a}}){Sargent}, {Boksenberg}, \&
  {Steidel}}]{sbs88}
{Sargent}, W. L.~W., {Boksenberg}, A., \& {Steidel}, C.~C. 1988{\natexlab{a}},
  \apjs, 68, 539, \dodoi{10.1086/191300}

\bibitem[{{Sargent} {et~al.}(1988{\natexlab{b}}){Sargent}, {Steidel}, \&
  {Boksenberg}}]{ssb88}
{Sargent}, W. L.~W., {Steidel}, C.~C., \& {Boksenberg}, A. 1988{\natexlab{b}},
  \apj, 334, 22, \dodoi{10.1086/166814}

\bibitem[{{Sargent} {et~al.}(1980){Sargent}, {Young}, {Boksenberg}, \&
  {Tytler}}]{sargent80}
{Sargent}, W.~L.~W., {Young}, P.~J., {Boksenberg}, A., \& {Tytler}, D. 1980,
  \apjs, 42, 41, \dodoi{10.1086/190644}

\bibitem[{{Schaye} {et~al.}(2003){Schaye}, {Aguirre}, {Kim}, {Theuns}, {Rauch},
  \& {Sargent}}]{schaye03}
{Schaye}, J., {Aguirre}, A., {Kim}, T.-S., {et~al.} 2003, \apj, 596, 768,
  \dodoi{10.1086/378044}

\bibitem[{{Schechter}(1976)}]{schechter76}
{Schechter}, P. 1976, \apj, 203, 297, \dodoi{10.1086/154079}

\bibitem[{{Schmidt} {et~al.}(2017){Schmidt}, {Worseck}, {Hennawi}, {Prochaska},
  {Crighton}, {Luki{\'c}}, \& {O{\~n}orbe}}]{schmidt17}
{Schmidt}, T.~M., {Worseck}, G., {Hennawi}, J.~F., {et~al.} 2017, Frontiers in
  Astronomy and Space Sciences, 4, 23, \dodoi{10.3389/fspas.2017.00023}

\bibitem[{{Schneider} {et~al.}(1993){Schneider}, {Hartig}, {Jannuzi},
  {Kirhakos}, {Saxe}, {Weymann}, {Bahcall}, {Bergeron}, {Boksenberg},
  {Sargent}, {Savage}, {Turnshek}, \& {Wolfe}}]{schneider93}
{Schneider}, D.~P., {Hartig}, G.~F., {Jannuzi}, B.~T., {et~al.} 1993, \apjs,
  87, 45, \dodoi{10.1086/191798}

\bibitem[{{Sebastian} {et~al.}(2024){Sebastian}, {Ryan-Weber}, {Davies},
  {Becker}, {Keating}, {D'Odorico}, {Meyer}, {Bosman}, {Cupani}, {Kulkarni},
  {Haehnelt}, {Lai}, {Eilers}, {Bischetti}, \& {Gallerani}}]{sebastian23}
{Sebastian}, A.~M., {Ryan-Weber}, E., {Davies}, R.~L., {et~al.} 2024, \mnras,
  530, 1829, \dodoi{10.1093/mnras/stae789}

\bibitem[{{Seyffert} {et~al.}(2013){Seyffert}, {Cooksey}, {Simcoe}, {O'Meara},
  {Kao}, \& {Prochaska}}]{seyffert13}
{Seyffert}, E.~N., {Cooksey}, K.~L., {Simcoe}, R.~A., {et~al.} 2013, \apj, 779,
  161, \dodoi{10.1088/0004-637X/779/2/161}

\bibitem[{{Shull} {et~al.}(2012){Shull}, {Smith}, \& {Danforth}}]{shull12}
{Shull}, J.~M., {Smith}, B.~D., \& {Danforth}, C.~W. 2012, \apj, 759, 23,
  \dodoi{10.1088/0004-637X/759/1/23}

\bibitem[{{Simcoe} {et~al.}(2004){Simcoe}, {Sargent}, \& {Rauch}}]{simcoe04}
{Simcoe}, R.~A., {Sargent}, W. L.~W., \& {Rauch}, M. 2004, \apj, 606, 92,
  \dodoi{10.1086/382777}

\bibitem[{{Simcoe} {et~al.}(2012){Simcoe}, {Sullivan}, {Cooksey}, {Kao},
  {Matejek}, \& {Burgasser}}]{simcoe12}
{Simcoe}, R.~A., {Sullivan}, P.~W., {Cooksey}, K.~L., {et~al.} 2012, \nat, 492,
  79, \dodoi{10.1038/nature11612}

\bibitem[{{Songaila}(1998)}]{songaila98}
{Songaila}, A. 1998, \aj, 115, 2184, \dodoi{10.1086/300387}

\bibitem[{{Songaila}(2005)}]{songaila05}
---. 2005, \aj, 130, 1996, \dodoi{10.1086/491704}

\bibitem[{{Songaila} \& {Cowie}(1996)}]{songaila96}
{Songaila}, A., \& {Cowie}, L.~L. 1996, \aj, 112, 335, \dodoi{10.1086/118018}

\bibitem[{{Steidel} {et~al.}(1994){Steidel}, {Dickinson}, \& {Persson}}]{sdp94}
{Steidel}, C.~C., {Dickinson}, M., \& {Persson}, S.~E. 1994, \apjl, 437, L75,
  \dodoi{10.1086/187686}

\bibitem[{{Steidel} \& {Sargent}(1992)}]{ss92}
{Steidel}, C.~C., \& {Sargent}, W. L.~W. 1992, \apjs, 80, 1,
  \dodoi{10.1086/191660}

\bibitem[{{Syphers} \& {Shull}(2014)}]{syphers14}
{Syphers}, D., \& {Shull}, J.~M. 2014, \apj, 784, 42,
  \dodoi{10.1088/0004-637X/784/1/42}

\bibitem[{{Tan} {et~al.}(2023){Tan}, {Oh}, \& {Gronke}}]{tan23}
{Tan}, B., {Oh}, S.~P., \& {Gronke}, M. 2023, \mnras, 520, 2571,
  \dodoi{10.1093/mnras/stad236}

\bibitem[{{Tepper-Garc{\'\i}a} {et~al.}(2011){Tepper-Garc{\'\i}a}, {Richter},
  {Schaye}, {Booth}, {Dalla Vecchia}, {Theuns}, \& {Wiersma}}]{tepper-garcia11}
{Tepper-Garc{\'\i}a}, T., {Richter}, P., {Schaye}, J., {et~al.} 2011, \mnras,
  413, 190, \dodoi{10.1111/j.1365-2966.2010.18123.x}

\bibitem[{{Tie} {et~al.}(2024){Tie}, {Hennawi}, {Wang}, {Onorato}, {Yang},
  {Ba{\~n}ados}, {Davies}, \& {O{\~n}orbe}}]{tie23}
{Tie}, S.~S., {Hennawi}, J.~F., {Wang}, F., {et~al.} 2024, \mnras, 535, 223,
  \dodoi{10.1093/mnras/stae2193}

\bibitem[{{Tripp} {et~al.}(1997){Tripp}, {Lu}, \& {Savage}}]{tripp97}
{Tripp}, T.~M., {Lu}, L., \& {Savage}, B.~D. 1997, \apjs, 112, 1,
  \dodoi{10.1086/313031}

\bibitem[{{Tripp} {et~al.}(2008){Tripp}, {Sembach}, {Bowen}, {Savage},
  {Jenkins}, {Lehner}, \& {Richter}}]{tripp08}
{Tripp}, T.~M., {Sembach}, K.~R., {Bowen}, D.~V., {et~al.} 2008, \apjs, 177,
  39, \dodoi{10.1086/587486}

\bibitem[{{Tytler} {et~al.}(1987){Tytler}, {Boksenberg}, {Sargent}, {Young}, \&
  {Kunth}}]{tytler87}
{Tytler}, D., {Boksenberg}, A., {Sargent}, W. L.~W., {Young}, P., \& {Kunth},
  D. 1987, \apjs, 64, 667, \dodoi{10.1086/191213}

\bibitem[{{Vasiliev}(2014)}]{vasiliev14}
{Vasiliev}, E.~O. 2014, Astronomy Reports, 58, 954,
  \dodoi{10.1134/S1063772914120105}

\bibitem[{{Verner} {et~al.}(1996){Verner}, {Ferland}, {Korista}, \&
  {Yakovlev}}]{verner96}
{Verner}, D.~A., {Ferland}, G.~J., {Korista}, K.~T., \& {Yakovlev}, D.~G. 1996,
  \apj, 465, 487, \dodoi{10.1086/177435}

\bibitem[{{Weymann} {et~al.}(1998){Weymann}, {Jannuzi}, {Lu}, {Bahcall},
  {Bergeron}, {Boksenberg}, {Hartig}, {Kirhakos}, {Sargent}, {Savage},
  {Schneider}, {Turnshek}, \& {Wolfe}}]{weymann98}
{Weymann}, R.~J., {Jannuzi}, B.~T., {Lu}, L., {et~al.} 1998, \apj, 506, 1,
  \dodoi{10.1086/306233}

\bibitem[{{Wolfe}(1995)}]{wolfe95}
{Wolfe}, A.~M. 1995, in Astronomical Society of the Pacific Conference Series,
  Vol.~80, The Physics of the Interstellar Medium and Intergalactic Medium, ed.
  A.~{Ferrara}, C.~F. {McKee}, C.~{Heiles}, \& P.~R. {Shapiro}, 478

\bibitem[{{Womble}(1995)}]{womble95}
{Womble}, D.~S. 1995, in QSO Absorption Lines, ed. G.~{Meylan}, 157

\bibitem[{{Worseck} {et~al.}(2019){Worseck}, {Davies}, {Hennawi}, \&
  {Prochaska}}]{worseck19}
{Worseck}, G., {Davies}, F.~B., {Hennawi}, J.~F., \& {Prochaska}, J.~X. 2019,
  \apj, 875, 111, \dodoi{10.3847/1538-4357/ab0fa1}

\bibitem[{{Zhu} \& {M{\'e}nard}(2013)}]{zhu13}
{Zhu}, G., \& {M{\'e}nard}, B. 2013, \apj, 770, 130,
  \dodoi{10.1088/0004-637X/770/2/130}

\bibitem[{{Zibetti} {et~al.}(2005){Zibetti}, {M{\'e}nard}, {Nestor}, \&
  {Turnshek}}]{zibetti05}
{Zibetti}, S., {M{\'e}nard}, B., {Nestor}, D., \& {Turnshek}, D. 2005, \apjl,
  631, L105, \dodoi{10.1086/497424}

\bibitem[{{Zou} {et~al.}(2021){Zou}, {Jiang}, {Shen}, {Wu}, {Ba{\~n}ados},
  {Fan}, {Ho}, {Riechers}, {Venemans}, {Vestergaard}, {Walter}, {Wang},
  {Willott}, {Joshi}, {Wu}, \& {Yang}}]{zou21}
{Zou}, S., {Jiang}, L., {Shen}, Y., {et~al.} 2021, \apj, 906, 32,
  \dodoi{10.3847/1538-4357/abc6ff}

\end{thebibliography}

\end{document}